\documentclass{acmtrans2m}

\usepackage{epsfig,graphics}
\usepackage{enumerate, xspace}
\usepackage{amssymb, amstext,amsmath}
\usepackage{color}
\usepackage{changebar,framed}
\usepackage{xspace}
\usepackage{algorithmic}

\newcommand{\fo}{${\rm FO}$}
\newcommand{\for}[1]{${\rm FO}({#1})$}
\newcommand{\fop}{${\rm FO}(+,\ab \times,\ab <,\ab 0,\ab 1)$}
\newcommand{\fopr}[1]{${\rm FO}(+,\ab \times,\ab <,\ab 0,\ab 1,\ab {#1})$}
\newcommand{\fowr}[1]{${\rm FO}(+,\ab \times,\ab <,\ab 0,\ab 1,\ab {#1})+{\rm While}$}
\newcommand{\foprs}{\fopr{\overline{\sigma}}}
\newcommand{\fow}{${\rm FO+While}$}
\newcommand{\fom}[1]{${\rm FO({#1})}$}
\newcommand{\st}[1]{{\cal S\!\!\!\!T}\!^{#1}}
\newcommand{\sti}[1]{{\cal S\!\!\!\!T}_{#1}}

\newcommand{\R}{{\rm {\bf R}}}
\newcommand{\N}{{\rm {\bf N}}}

\newcommand{\PLss}[1]{{\cal P}(#1)}
\newcommand{\before}[1]{{\bf Before}({#1})}
\newcommand{\betweenm}[2]{{\bf Between}^{#1}({#2})}
\newcommand{\befores}{{\bf Before}}
\newcommand{\betweens}[1]{{\bf Between}^{#1}}

\newcommand{\eqcrst}[1]{{\bf EqCR}^{\bf st}({#1})}
\newcommand{\eqcrsts} {{\bf EqCR}^{\bf st}}
\newcommand{\eqcrs}[1]{{\bf EqCR}^{\bf s}({#1})}
\newcommand{\eqcrss} {{\bf EqCR}^{\bf s}}
\newcommand{\eqcrt}[1]{{\bf EqCR}^{\bf t}({#1})}
\newcommand{\eqcrts} {{\bf EqCR}^{\bf t}}
\newcommand{\equidist}[1]{{\bf EqDist}({#1})}
\newcommand{\equidists}{{\bf EqDist}}
\newcommand{\equidistcotemp}[1]{{\bf EqDist}^{\bf cotemp}({#1})}
\newcommand{\equidistcotemps}{{\bf EqDist}^{\bf cotemp}}
\newcommand{\pos}[2]{{\bf Pos}^{#1}({#2})}
\newcommand{\poss}[1]{{\bf Pos}^{#1}}

\newcommand{\unitdists}{{\bf UnitDist}}
\newcommand{\ab}{\allowbreak }
\newcommand{\unittime}[1]{{\bf UnitTime}({#1})}
\newcommand{\unittimes}{{\bf UnitTime}}

\newcommand{\basishat}{(v_0,\ab v_1,\ab  \ldots,\ab v_{n+1})}
\newcommand{\basis}{(p_0,\ab  p_1,\ab  \ldots,\ab p_{n+1})}
\newcommand{\basisv}{(v_0,\ab  v_1,\ab  \ldots,\ab v_{n+1})}
\newcommand{\basishatdd}{(v_0,\ab  v_1,\ab  v_2,\ab v_3)}
\newcommand{\basisdd}{(p_0,\ab  p_1,\ab  p_2,\ab p_3)}

\newcommand{\nbasisi}{(p_{i,\ab 0},\ab  p_{i,\ab 1},\ab  \ldots \ab,\ab p_{i,\ab n})}

\newcommand{\cosys}{{\bf CoSys}}
\newcommand{\splits}{\hfill\cr{}\hfill}

\newtheorem{definition}{Definition}[section]
\newtheorem{theorem}{Theorem}[section]
\newtheorem{example}{Example}[section]
\newtheorem{lemma}{Lemma}[section]
\newtheorem{remark}{Remark}[section]

\newtheorem{property}{Property}[section]
\newtheorem{proposition}{Proposition}[section]
\newtheorem{corollary}{Corollary}[section]

\def\squareforqed{\hbox{\rlap{$\sqcap$}$\sqcup$}}
\def\qed{\ifmmode\squareforqed\else{\unskip\nobreak\hfil
\penalty50\hskip1em\null\nobreak\hfil\squareforqed
\parfillskip=0pt\finalhyphendemerits=0\endgraf}\fi}

\title{First-order Complete and Computationally Complete Query Languages for \\ Spatio-Temporal Databases~\footnote{An extended abstract appeared in the proceedings of DBPL'01~\cite{onszelf}.}}

\author{ \emph{Floris Geerts}\footnote{Post-doctoral researcher of the FWO-Vlaanderen and visiting researcher at the University of Edinburgh.}, \emph{Sofie Haesevoets}, \emph{Bart Kuijpers}\footnote{Corresponding author: University of Limburg,
Department of Mathematics, Physics and Computer Science, B-3590 Diepenbeek, Belgium, {\tt
bart.kuijpers@luc.ac.be}}\\ University of Limburg  }

\begin{abstract}
We address a fundamental question concerning spatio-temporal database systems: ``What are exactly
spatio-temporal queries?'' We define spatio-temporal queries to be computable mappings that are
also \emph{generic}, meaning that the result of a query may only depend to a limited extent on the
actual internal representation of the spatio-temporal data. Genericity is defined as invariance
under groups of geometric transformations that preserve certain characteristics of spatio-temporal
data (e.g., collinearity, distance, velocity, acceleration, ...). These groups depend on the
notions that are relevant in particular spatio-temporal database applications.
 These
transformations also have the distinctive property that they respect the monotone and
unidirectional nature of time.

We investigate different genericity classes with respect to the constraint database model for
spatio-temporal databases and we identify sound and complete languages for the first-order and the
computable queries in these genericity classes. We distinguish between genericity determined by
time-invariant transformations, genericity notions concerning physical quantities and genericity
determined by  time-dependent transformations.
\end{abstract}

\category{}{}{}

\terms{}


\begin{document}
\maketitle

\section{Introduction}
Since the early 1990s, various database systems have been developed to handle spatial
data~\cite{ssd3,ssd1,ssd4,ssd2,ssd6,voisard-spatial-book,ssd5} and solid theories for such systems
have been proposed and studied~\cite{pvv-pods,cdbook-chap8}. Conceptually, spatial databases
contain possibly infinite sets of points in a real space $\R^n$. In more recent years, we have seen
the emergence of database systems and applications that are dealing with \emph{spatio-temporal
data}~\cite{STDBM99,cz-00,CHOROCHRONOS,ge-00,req}. Conceptually, spatio-temporal data can be
modeled as infinite spatial sets that move or change in time, i.e., sets in $\R^n\times \R$.

\begin{figure}
\begin{center}
\input{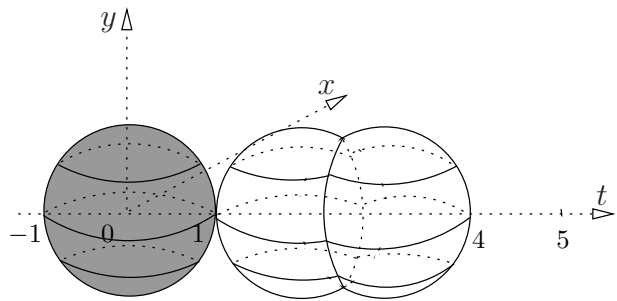}
\end{center}
\caption[dummy]{An example of a spatio-temporal database in $\R^2\times \R$.}\label{movie}
\end{figure}

A recent and much acclaimed method for effectively representing infinite geometrical figures is
provided by the \emph{constraint database model}, that was introduced in 1990 by Kanellakis, Kuper
and Revesz~\cite{kkr90,kkr_cql} (an overview of the area of constraint databases
appeared~\cite{cdbook}; and~\cite{reveszbook} is a textbook on this topic). Until recently this
model has been
 used mainly in the area of spatial databases, but it provides an equally elegant and efficient way to model
spatio-temporal data~\cite{cz-00,cr-97,cr-99,grumbach,kpv-00}. In the setting of the constraint
model, a spatio-temporal relation in $\R^n\times\R$ is finitely represented as a Boolean
combination of polynomial equalities and inequalities. Figure~\ref{movie} depicts the
spatio-temporal database $\{ (x,y;t) \mid x^2+y^2+t^2\leq 1 \lor (x^2+y^2+(t-2)^2=1\land t\leq
5/2)\lor (x^2+y^2+(t-3)^2=1\land t>5/2)\} $ in $\R^2\times \R$. A spatio-temporal database is a
finite collection of such relations and can be finitely represented by the polynomial constraint
formulas that represent its relations.

A number of theoretical studies have appeared on the status of time and its relation with space in
systems that model moving objects. Erwig et al.~\cite{taxo} give a taxonomy of applications ranging
from those that rely on a step-wise constant geometry to applications which need more complete
integration of space and time (like for instance a continuous description of a trajectory).
MOST~\cite{most}, an example of the latter category, relies on a strong interaction of the space
and time components (since the space variables are described by linear polynomials in time) and
provides a query language that is a combination of a spatial query language and a temporal logic.
On the other range of the spectrum, variable independence (defined in terms of orthographic
dimension) gives rise to a less expressive data model which has the advantage of a lower complexity
of query evaluation~\cite{grumbach,libkin}.

We study spatio-temporal queries from the perspective of ex\-pres\-si\-ve po\-wer, and do this
against the background of the full modeling and querying power of the constraint database model and
the first-order and computationally complete languages it offers. We ask which expressions in these
languages may be considered as \emph{reasonable\/} spatio-temporal queries. In database theory it
is usually required that the result of queries should only to a certain limited extent depend on
the actual internal representation of databases and that queries should only ask for properties
that are shared by ``isomorphic'' encodings of the same data. The meaning of ``isomorphic'' may be
influenced by the actual database application and by which notions are relevant to it. In the
context of the relational database model, Chandra and Harel~\cite{ch80} formalized this
independence of the actual encoding in terms of the notion of \emph{genericity}. Paredaens, Van den
Bussche and Van Gucht~\cite{pvv-pods} identified a hierarchy of genericity classes for spatial
database applications. The generic queries in the different classes focus on different geometrical
and topological aspects of the spatial data. On a technical level, generic queries are defined as
being invariant under those transformations of the data that preserve the relevant aspects of the
data. Whereas Chandra and Harel considered the group of the isomorphisms (that possibly fix some
elements of the domain) in the case of relational databases, Paredaens, Van den Bussche and Van
Gucht identified different geometrical and topological transformation groups (affinities,
isometries, translations, homeomorphisms, ...) for spatial database applications.

We define spatio-temporal queries to be computable mappings that are also \emph{generic}, meaning
that the result of a query may only depend to a limited extent on the actual internal
representation of the spatio-temporal data. Genericity is defined as invariance under  some
(application-dependent) group of geometric transformations. These transformations
 preserve certain characteristics of
spatio-temporal data (e.g., collinearity, distance, velocity, acceleration, ...).

We investigate which notions of genericity are appropriate for spatio-tempo\-ral databases and
which transformation groups express them. We observe that the transformations should first and
foremost respect the monotone and unidirectional nature of time, i.e., leave the temporal order of
events unchanged. It follows that the relevant transformation groups are the product of a group of
time-(in)dependent spatial transformations and a group of monotone increasing transformations of
the time-component of the spatio-temporal data. Next, we focus on the former groups and study which
of them leave different spatial and spatio-temporal properties (like collinearity, distance and
orientation) unchanged. We also focus on physical properties of spatio-temporal data (like velocity
and acceleration). The transformation groups that we consider are all subgroups of the
time-dependent or time-independent affinities of $\R^n\times \R$.

We study the notion of spatio-temporal genericity relative to two popular query languages in the
constraint model: first-order logic over the reals (\fo) and an extension of this logic with a
while-loop (\fow). Both languages are known to be effectively computable (given termination in the
case of \fow-programs) and \fow\ is known to be a computationally complete language on
spatio-temporal databases~\cite{cdbook-chap2}. First, we show that all the genericity classes are
undecidable. We show that the considered classes of generic first-order queries are recursively
enumerable, however. Hereto, we define first-order point-based languages in which variables are
assumed to range over points in $\R^n\times \R$ and which contain certain point predicates (such as
${\bf Between}$ and ${\bf Before}$). These point-based languages are shown to be sound and complete
for the first-order queries in the considered genericity classes. We have also shown that
extensions of these point-based logics with a While-loop give sound and complete languages for the
computable queries in the different genericity classes. Our results are inspired by similar results
that were obtained by Gyssens, Van den Bussche and Van Gucht in the context of spatial
databases~\cite{gvv-jcss}. Also, the proof techniques we use for time-independent transformation
groups, are generalisations of techniques introduced in those papers.  However,  our results for
genericity notions described by time-dependent transformations require new proof techniques.

This paper is organized as follows. In Section~\ref{sectie2}, we define spatio-temporal databases,
spatio-temporal queries, and the constraint query languages \fo\ and \fow. In
Section~\ref{sectie3}, we define a number of genericity notions. In Section~\ref{sectie4}, we
present sound and complete first-order query languages for  the different notions of genericity. In
Section~\ref{sectie5}, we present sound and complete languages for the computable queries
satisfying the different notions of genericity. We end with a discussion in Section~\ref{sectie6}.

\section{Definitions and preliminaries}\label{sectie2}

We denote the set of the real numbers by $\R$ and the
$n$-dimensional real space by $\R^n$.

Throughout this paper, we use the following notational convention. Variables that range over real
numbers are denoted by characters $x,\ab  y,\ab   z,\ab   x_1,\ab   y_1,\ab   z_1,\ab x_2,\ab
y_2,\ab   z_2,\ab   \ldots$. When there is the need to distinguish between real variables that
indicate spatial coordinates and time coordinates, we use $x,\ab   y,\ab   z, \ab  x_1,\ab y_1,\ab
z_1, \ab  x_2,\ab y_2, \ab  z_2, \ab  \ldots$ for the former and  use $t,\ab   t_1,\ab t_2, \ab
\ldots$ for the latter. Variables that range over vectors in $\R^{n}$ and that represent spatial
information are denoted by bold characters $\bf{x}, \bf{x_1}, \ab \bf{x_2}, \ldots$. Real constants
are represented by characters $a, \ab  b, \ab  c, a_1, \ab  b_1, c_1,\ab   a_2, \ab  b_2, c_2,
\ldots$. When there is the need to distinguish between real constants that indicate spatial
coordinates and time coordinates, we use $a, \ab  b, \ab  c, \ab  a_1, \ab b_1,\ab   c_1, \ab  a_2,
\ab  b_2, c_2 \ldots$ for the former and  use Greek characters $\tau, \ab \tau_1, \ab  \tau_2, \ab
\ldots$ for the latter.

 Finally, bold characters $\bf {a}, \bf{a}_1, \bf{a}_2, \ldots$ represent constant
$n$-dimensional spatial vectors.

Vectors $(\bf{a},\tau)$ containing mixed spatial and temporal information are denoted $p, q, r,
p_1, \ab q_1, r_1,\ab p_2, q_2, r_2 \ldots$ and variable vectors  $(\mathbf{x},t)$ are represented
by characters $u, \ab v, w, \ab u_1, v_1,\ab  w_1, u_2,\ab  v_2,\ab  w_2,\ab \ldots$.

\subsection{Semi-algebraic and spatio-temporal databases}

We consider $n$-dimensional spatial figures that move or change over time. A moving figure in
$\R^n$ can be described by means of an infinite set of tuples $(a_1,a_2,\ldots, a_n,\ab \tau)$ in
$\R^n\times \R$, where $(a_1,a_2,\ldots, a_n)$ represent the $n$-dimensional spatial coordinates of
$(a_1,a_2,\ldots, a_n, \ab \tau)$ and $\tau$ its time coordinate. Obviously, this infinite
information needs to be represented finitely in order to be stored in the memory of a computer. In
this section, we describe two approaches to model such changing figures, namely the
\emph{semi-algebraic database model} and the \emph{spatio-temporal database model}. Semi-algebraic
databases are based on real numbers, while spatio-temporal databases are based on
$(n+1)$-dimensional points. Both models resort under the constraint database
model~\cite{cdbook,reveszbook}.

\begin{definition}\rm
A \emph{semi-algebraic relation in $\R^n$\/} is a subset of $\R^n$
that can be described as a Boolean combination of sets of the form
$$\{ (x_1, x_2, \ldots, x_n) \in \R^n \mid p(x_1, x_2, \ldots,
x_n) > 0 \},$$ with $p$ a polynomial with integer coefficients in
the real variables $x_1,\allowbreak x_2,\allowbreak \ldots,
\allowbreak x_n$. \qed
\end{definition}

In mathematical terms, semi-algebraic relations are known as \emph{semi-al\-gebra\-ic
sets}~\cite{bcr-real}. In this paper, we will be mainly interested in semi-algebraic relations in
real spaces of the form $\R^{(n+1)\times k}$. These relations can be viewed as $k$-ary relations
over $\R^{n}\times\R$ (i.e., the $n$-dimensional space extended with a time dimension). The next
example illustrates this for $k=1$ and $n=2$.

\begin{example}\rm
Figure \ref{movie} gives an example of a semi-algebraic relation in $\R^3$. This set can be
described as follows: $\{ (x,y,t) \in \R^2\times \R \mid x^2+y^2+t^2\leq 1 \lor
(x^2+y^2+(t-2)^2=1\land t\leq 5/2)\lor (x^2+y^2+(t-3)^2=1\land t>5/2)\}. $ \qed
\end{example}

We call a semi-algebraic relation in $\R^n$ also \emph{a semi-algebraic relation of arity $n$}. A
semi-algebraic database is essentially a finite collection of semi-algebraic relations. We define
this now.

\begin{definition}\label{sad-def}\rm
\medskip\par

A \emph{(semi-algebraic) database schema $\sigma$} is a finite set
of relation names, where each relation name $R$ has a natural
number $ar(R)$, called its \emph{arity}, associated to it.

Let $\sigma$ be a database schema.  A \emph{semi-algebraic
database over $\sigma$} is a structure $\cal{D}$ over $\sigma$
with domain $\R$ such that, for each relation name $R$ of
$\sigma$, the associated relation  $R^{\cal{D}}$ in ${\cal D}$ is
a semi-algebraic relation of arity $ar(R)$. \qed
\end{definition}

\begin{example}\rm
Let $\sigma=\{R,S\}$, with $ar(R)=2$ and $ar(S)=1$  be a
semi-algebraic database schema. Then the structure ${\cal{D}}$
given by $$(\R,R^{\cal{D}}=\{(x,y)\in\R^2\mid x^2+y^2<1\},
S^{\cal{D}}=\{ x \in\R \mid 0\leq x\leq 1 \})$$ is an example of a
semi-algebraic database over $\sigma$ that contains the open unit
disk and the closed unit interval.\qed
\end{example}

We now define spatio-temporal databases. In contrast to
semi-algebraic databases, in which points are described by their
real coordinates, spatio-temporal databases are based on
$(n+1)$-dimensional points. The domain of a spatio-temporal
database is $\R^n \times \R$. We prefer the notation $\R^n \times
\R$ over $\R^{n+1}$ for the domain because it stresses the
distinction between the time coordinate and the $n$ spatial
coordinates of the $(n+1)$-dimensional points.

In the following definition, we work with $\R^n \times \R$ as the domain of the spatio-temporal
databases and we assume that this \emph{underlying dimension $n$} is fixed on before hand. In this
paper, we assume, for technical reasons that will become clear in Section~\ref{sectie4}, that $n
\geq 2$.

Throughout this paper we will often use the  canonical bijection $$can: (\R^n \times
\R)^k\rightarrow \R^{(n+1)\times k}$$ that maps a tuple $((\mathbf{a_1}, \tau_1), \ldots,
(\mathbf{a_k},\tau_k))$ to $(a_{1,1}, \ldots, a_{1,n},\ab \tau_1, \ab \ldots,\ab  a_{k,1}, \ab
\ldots, \ab a_{k,n}, \ab \tau_k)$, where for $1 \leq i \leq k$ and $1 \leq j \leq n$, $a_{i,j}$
denotes the $j$th real coordinate of the vector $\mathbf{a_i}$.

\begin{definition}\label{std-def}\rm
A \emph{(spatio-temporal) database schema $\sigma$} is a finite
set of relation names, where each relation name $R$ has a natural
number $ar(R)$, called its arity, associated to it.

A subset of $(\R^n \times \R)^k$ is a \emph{spatio-temporal
relation of arity $k$} if its image under the canonical
 bijection $can:(\R^n \times \R)^k\rightarrow \R^{(n+1)\times k}$
is a semi-algebraic relation of arity $(n+1)\times k$.

Let $\sigma$ be a spatio-temporal schema.  A \emph{spatio-temporal database over $\sigma$} is a
structure $\st{}$ over $\sigma$ with domain $\R^n \times \R$ such that to each relation name $R$ in
$\sigma$, a spatio-temporal relation $R^{\st{}}$ of arity $ar(R)$ is associated to it.  \qed
\end{definition}

\begin{remark}\label{remark-canon}
\rm A spatio-temporal database $\st{}$ over $\sigma$ can be viewed in a natural way as a
semi-algebraic database $\overline{\st{}}$ over the semi-algebraic schema $\overline{\sigma}$,
which has for each relation name $R$ of $\sigma$, a relation name $\overline{R}$ of arity
$(n+1)\times ar(R)$. For each relation name $R$, $\overline{R}^{\overline{\st{}}}$ is obtained from
$R^{\st{}}$ by applying the canonical bijection $can: (\R^n \times \R)^{ar(R)}\rightarrow
\R^{(n+1)\times ar(R)}$. We will use the notation introduced here, throughout the paper.
\qed\end{remark}

Following this remark, we observe that spatio-temporal relations and databases can be
\emph{finitely} encoded and stored by means of the systems of polynomial equalities and
inequalities (i.e., by means of a quantifier-free formula of first-order logic over the reals with
$+$, $\times$, $<$ and the constants $0$ and $1$) that describe the associated semi-algebraic
relations and databases.

\begin{remark}\label{remark-datastructure}\rm
Throughout this paper, we assume that databases are finitely encoded by systems of polynomial equations
and that  a specific data structure is fixed (possible data structures are dense or sparse representations of polynomials).
The specific choice of data structure is not relevant to the topic of this paper, but we assume that one is fixed. When we
talk about computable queries later on, we mean Turing computable with respect to the chosen encoding and  data structures.
\end{remark}

The model presented here and the results in this paper can be extended straightforwardly to the
situation where spatio-temporal relations are accompanied by classical thematic information, like
the typical alpha-numeric data you find in business applications and also, in combination with
spatial data, in Geographical Information Systems. However, because the problem that is discussed
here is captured by this simplified model, we stick to it for reasons of simplicity of exposition.

\begin{example}\label{vb1} \rm
Figure~\ref{movie} in the Introduction gives an illustration of a
spatio-temporal database over a schema $\sigma = \{ R \}$ with
underlying dimension 2, where $R$ has arity 1. It shows at its
beginning, i.e., at $t=-1$,  a single point in the origin of
$\R^2$. Then it shows a disk whose radius increases and later
decreases and ends in a point at moment $t=1$, followed by a
circle whose radius increases, decreases, increases and then
shrinks to a point. \qed
\end{example}

\begin{definition}\rm
Let $\sigma$ be a spatio-temporal schema and let $\st{}$ be a
spatio-temporal database over $\sigma$ with underlying dimension
$n$. Let $R$ be a relation name in  $\sigma$ and let $\tau_0$ be a
real number representing a moment in time. We call the subset
$$R^{\st{}}\cap (\R^n\times\{\tau_0\})^{ar(R)}$$ of $(\R^n \times
\{\tau_0\})^{ar(R)}$ the \emph{snapshot of $R$ at the moment $\tau_0$}. The snapshot of the
spatio-temporal database $\st{}$ at the moment $\tau_0$ is the finite set of snapshots of all its
relations at $\tau_0$. \qed
\end{definition}

\begin{example}\rm
For the spatio-temporal relation depicted in Figure~\ref{movie},
the snapshot  at ${-1}$ is $\{(0,0,-1) \}$, the snapshot at ${0}$
is the closed unit disk in the plane $t = 0$ and the snapshot at
${5}$ is the empty set. \qed
\end{example}

\subsection{Spatio-temporal and semi-algebraic database queries}

Here, we define spatio-temporal and semi-algebraic  database queries as computable mappings of some
type. In the next section, we will argue that not all such mappings should be regarded as
``reasonable'' queries and that further conditions on the nature of these mappings have to be
imposed.
\medskip

\begin{definition}\label{sa-query}\rm
Let $\sigma$ be a semi-algebraic database schema.  A \emph{$k$-ary semi-algebraic  database query
$Q$ over $\sigma$} is a partial, computable mapping (in the sense of
Remark~\ref{remark-datastructure}) from the set of semi-algebraic databases over $\sigma$ to the
set of $k$-ary semi-algebraic relations. \qed \end{definition}

\begin{definition}\label{st-query}\rm
Let $\sigma$ be a spatio-temporal database schema and let us
consider input spatio-temporal databases over $\sigma$ with
underlying dimension $n$. A \emph{$k$-ary $n$-dimensional
spatio-temporal database query $Q$ over $\sigma$} is a partial,
computable mapping (in the sense of Remark~\ref{remark-datastructure}) from the set of spatio-temporal databases over
$\sigma$ to the set of $k$-ary spatio-temporal relations with
underlying dimension $n$. \qed
\end{definition}

\par\noindent We also call a $k$-ary $n$-dimensional spatio-temporal database
query a \emph{spatio-temporal database query of output type
$(n,k)$}.

 Note that we restrict spatio-temporal database queries to
preserve the underlying dimension of the input database.
\medskip

\begin{example}\rm\label{exq-a}
Let $\sigma = \{R\}$, where $R$ has arity 1 and let the underlying dimension be $2$. The query that
selects those snapshots from the relation $R$ where $R$ shows a circle is a spatio-temporal
database query of output type $(2,1)$. Applied to the database of Example~\ref{vb1} and shown in
Figure~\ref{movie}, this query returns the union of its snapshots in the open time interval
$]1,4[$. \qed
\end{example}

There is a natural way to see spatio-temporal queries as semi-algebraic queries, that is captured
in the following definition of
 equivalence of queries.

 \begin{definition}\label{equiv-queries}\rm
Let $\sigma$ be a spatio-temporal database schema and let us
consider input spatio-temporal databases over $\sigma$ with
underlying dimension $n$. Let $\overline{\sigma}$ be the corresponding semi-algebraic database schema (see Remark~\ref{remark-canon}).
Let $Q$ be a $k$-ary $n$-dimensional
spatio-temporal database query  over $\sigma$ and let
$\overline{Q}$ be a $((n+1)\times k)$-ary semi-algebaric  database query  over $\overline{\sigma}$.
We say that $Q$ and $\overline{Q}$ are \emph{equivalent} if for every  database $\st{}$ over $\sigma$
we have $$\overline{Q(\st{})}=\overline{Q}(\overline{\st{}}).$$
\qed\end{definition}

\subsection{First-order logic and its extension with a while loop as a spatio-temporal query language}\label{2.3}

First-order logic over the field of the real numbers, \fop\ for short, has been well-studied as a
query language for spatial databases~\cite{kkr_cql,pvv-pods,cdbook}. In the setting of
spatio-temporal databases, it can be used as a query language in a similar way. We introduce \fop\
here as a spatio-temporal query language, beit on semi-algebraic databases that represent
spatio-temporal databases.

\begin{definition}\rm
Let $\sigma = \{R_1, R_2, \ldots, R_m \}$ be a spatio-temporal database sche\-ma and let us
consider queries working on input databases over $\sigma$ with underlying dimension $n$. Let
$\overline{R}_i$ ($1\leq i\leq m$) be the corresponding semi-algebraic relation names of arity
$(n+1)\times ar(R_i)$ (we follow the notation of Remark~\ref{remark-canon}) and let
$\overline{\sigma}$ be the semi-algebraic schema $\{\overline{R}_1, \overline{R}_2, \ldots,
\overline{R}_m \}$.

 Let
$\varphi({\bf x_1},t_1,{\bf x_2},t_2,\ldots,{\bf x_k},t_k)$, be a first-order logic formula over
the alphabet $(+,\times,0,1,<,\overline{R}_1, \overline{R}_2, \ldots, \overline{R}_m)$. If ${\bf
x_i}=(x_{i,1},\ldots, x_{i,n})$, then the free variables of $\varphi$ are $x_{1,1},\ldots,
x_{1,n},t_1,x_{2,1},\ldots, x_{2,n},t_2,\ldots,x_{k,1},\ldots, x_{k,n},t_k$. The formula $\varphi$
expresses a semi-algebraic $((n+1)\times k)$-ary query $\overline{Q}$ which is equivalent to a
$k$-ary $n$-dimensional spatio-temporal query $Q$. For each input spatio-temporal database $\st{}$
over $\sigma$, $Q(\st{})$ is defined as the set of points $(({\bf a_1},\tau_1),\ab({\bf
a_2},\tau_2),\ldots,\ab({\bf a_k},\tau_k))$ of $ (\R^n \times \R)^k$ such that
 $$ (\R, +,\times,0,1,<,\overline{R}^{\overline{\st{}}}_1, \overline{R}^{\overline{\st{}}}_2, \ldots,
\overline{R}^{\overline{\st{}}}_m)\models \varphi[{\bf a_1},\tau_1,{\bf a_2},\tau_2,\ldots,{\bf
a_k},\tau_k],$$
where
$\varphi[{\bf a_1},\tau_1,{\bf a_2},\tau_2,\ldots,{\bf a_k},\tau_k]$ denotes the formula
$\varphi({\bf x_1},\ab t_1 ,{\bf x_2} ,\ab t_2 ,\ab\ldots,\ab {\bf x_k},t_k)$ with its free
variables instantiated  by
 ${\bf a_1},\tau_1,{\bf a_2},\tau_2,\ldots,{\bf
a_k},\tau_k$. \qed \end{definition}

We will refer to the first-order query language, introduced here, as \fopr{\overline{R}_1,\ab
\overline{R}_2, \ab \ldots, \ab \overline{R}_m}, or, if the schema is clear from the context, as
\fop.

\begin{example}\rm\label{exq}
As in Example~\ref{exq-a}, let $\sigma = \{R\}$, where $R$ has arity 1 and let the underlying
dimension be $2$. The formula $(\exists x_0)(\exists y_0)(\exists r>0) \mbox{\large (}
(x-x_0)^2+(y-y_0)^2=r^2\Leftrightarrow \overline{R}(x,y,t) \mbox{\large )}$ expresses a
spatio-temporal query of output type $(2,1)$. It selects those snapshots from a spatio-temporal
relation $R$ where $R$ shows a circle. As mentioned, applied to the database of Example \ref{vb1},
this query returns all its snapshots in the time interval $]1,4[$. \qed
\end{example}

We remark that the formalism of semi-algebraic and spatio-temporal databases and the first-order
query language introduced here, fits within the framework of \emph{constraint
databases}~\cite{cdbook,reveszbook}. It is well known that \fop\/-formulas can be effectively
evaluated in the constraint model and therefore also in this context. It is also known that the
output can be represented in the same constraint formalism~\cite{cdbook,cdbook-chap2}, i.e. by a
quantifier-free formula over $(+,\times,0,1,<)$.

\begin{remark}\rm
An arbitrary \fo\/-formula does not necessarily express a spatio-temporal database query as shown
by the following example. The formula $$(\exists t)\overline{R}(x_1,x_2,t)$$ expresses the
projection of the spatio-temporal relation $R$ on the spatial $(x_1,x_2)$-plane. The formula
returns a set of couples $(x_1,x_2)$ in $\R^2$ that form a semi-algebraic set with a purely spatial
meaning. \qed
\end{remark}

We end this section by specifying the programming language \fowr{\overline{\sigma}}\ which is known
to be computationally complete on semi-algebraic databases~\cite{gvv-jcss}. Essentially, this
language is an extension of \foprs\ with assignments and a While loop. The use of similar languages
will be illustrated in Section~\ref{sectie5}. We also refer to~\cite{gvv-jcss,cdbook-chap2} for
illustrations.

\begin{definition}\rm
Let $\sigma$ be a spatio-temporal database schema.
Syntactically, a \emph{program\/} in the language \fowr{\overline{\sigma}}\  is a finite
sequence of \emph{statements\/} and \emph{while-loops\/}. It is assumed there is a sufficient
supply of new relation variables, each with an appropriate arity.

\begin{enumerate}[(i)]
    \item Each statement has the form
$$\overline{R} := \{(x_1,\ldots , x_k)\mid \varphi(x_1,\ldots , x_k)\};.$$ Here, $\overline{R}$ is a new relation variable with assigned arity $k$ (the variables $x_i$ range over $\R$) and
$\varphi$ is  a formula in \fopr{\overline{\sigma'}}, where ${\overline{\sigma'}}$ is the set of relation names
containing the elements of ${\overline{\sigma}}$ together with the relation variables introduced in previous
statements of the program.

    \item A while-loop has the form
$${\bf while}\  \varphi\  {\bf do}\  P\ {\bf ;}$$ where $P$ is a program and $\varphi$ is a
sentence in \fopr{\overline{\sigma'}}, where ${\overline{\sigma'}}$ is again the set of relation names
containing the elements of ${\overline{\sigma}}$ together with the relation variables introduced in previous
statements of the program.

\item One of the relation names occurring in the program is designated as the output relation and
is named $\overline{R}_\text{out}$.
\end{enumerate}\qed
\end{definition}

Semantically, a program in the query language \fowr{\overline{\sigma}}\  expresses a spatio-temporal
query as soon as $\overline{R}_\text{out}$ is assigned a return value. The execution of an \fowr{\overline{\sigma}}-program applied to an input database is performed step-by-step. A statement is executed by
first evaluating the \foprs-formula on the right hand side on the input database together with
the new relations resulting from previous statements. Next, the result of the evaluation of the
right hand side is assigned to the relation variable on the left-hand side. The effect of a while
loop is to execute the body as long as the condition $\varphi$ evaluates to true.

Note that these programs are not guaranteed to halt. For those input databases it
does not, the query represented by the program is not defined on that particular input database.

\section{Spatio-temporal genericity}\label{sectie3}

As stated in the introduction, we are interested in spatio-temporal database queries that are
invariant under the elements of a certain spatio-temporal transformation group (for function
composition) $${\cal F} = \{ f\mid f=(f_1,f_2,\ab \ldots, \ab f_n,\ab f_t): \R^n\times \R
\rightarrow \R^n\times \R  \}.$$ The idea is that the result of spatio-temporal queries should be
largely independent of the particular coordinate system in which the data are presented. In this
section, we formalize this idea by the notion of $\cal F$-genericity.

In the remainder of this section, we look at different types of transformation groups and we impose
two further conditions on these transformations. Firstly, we look at purely temporal conditions.
Secondly, we look at purely spatial or spatio-temporal conditions that reflect the nature of the
queries one is interested in. We also look at transformation groups that are suited for
applications in which physical notions such as velocity and acceleration are of importance.

\subsection{Definition of spatio-temporal genericity}\label{sectie3.1}

Let $f: \R^n\times \R \rightarrow \R^n\times \R $ be a function, let $R$ be a spatio-temporal
relation name of arity $k$ and let $R^{\st{}}$ be a relation instance with underlying dimension
$n$. In the following, we use the notation $f(R^{\st{}})$ to abbreviate the set $\{ (f({\bf
a}_1,\tau_1),\ab f({\bf a}_2,\ab \tau_2),\ab \ldots, \ab f({\bf a}_k,\ab \tau_k))\in (\R^n\times
\R)^k\mid \ab ({\bf a}_1,\tau_1, {\bf a}_2,\tau_2,\ldots, {\bf a}_k,\tau_k)\in R^{\st{}}\}$.

\begin{definition}\label{def-genericity}\rm
Let $Q$ be a spatio-temporal database query that takes databases of signature $\sigma=\{R_1,\ldots,
R_m\}$ with underlying dimension $n$ as input. Let ${\cal F}= \{ f\mid f: \R^n\times \R \rightarrow
\R^n\times \R \}$ be a spatio-temporal transformation group. We say that $Q$ is \emph{${\cal
F}$-generic\/} if, for any $f$ in ${\cal F}$ and for each pair of spatio-temporal databases
$\sti{1}$ and $\sti{2}$ over $\sigma$, the fact that $\sti{{2}} = (R_1^{\sti{{2}}}, \ldots,
R_m^{\sti{{2}}}) = (f(R_1^{\sti{{1}}}), \ldots, f(R_m^{\sti{{1}}}))$ implies that $f(Q(\sti{{1}}))
= Q(\sti{{2}})$.\qed
\end{definition}

This definition will be illustrated in Section~\ref{examples}.

It is clear that if a query is $\cal F$-generic, it is also $\cal F'$-generic for any subgroup $\cal F'$ of $\cal F$.

\subsection{Temporal restrictions on the  transformations}

It is very natural to describe spatio-temporal events with the notions``before'', ``after'' and
``co-temporal''. For instance, when two people arrive shortly after each other, we say \emph{``Mary
arrived before Jane''} rather than \emph{``Mary arrived at 9:31 and Jane at 9:35}''. Another
example is any kind of race. The winner is the one that finishes first. So, foremost the order of
arrival of the participants matters. Exact time moments are only important in very specific
situations.
\medskip
\par\noindent

We start with the definition of a spatio-temporal event.
\begin{definition}\rm
An \emph{event\/} is a subset of $\R^n\times \R$. The projection
of an event $A$ on the time-axis is denoted by $\pi_t(A)$ and
called the \emph{time-domain\/} of $A$. \qed
\end{definition}

Let $A$ and $B$ be events. In the terminology of Allen's interval
calculus~\cite{allen1,allen2}, $A$ and $B$ are called
\emph{co-temporal\/} if $\pi_t(A)=\pi_t(B)$ (we denote this by
$A=_tB$). Allen says $A$ is \emph{before\/} $B$ if $t_A<t_B$ for
all $t_A\in \pi_t(A)$ and all $t_B\in \pi_t(B)$ (we denote this by
$A<_tB$).

Remark that $A\leq_tB:=(A=_tB {\rm \ or \ } A<_tB)$ is a pre-order on events.

\begin{definition}\rm
We say that a transformation $f: \R^n\times \R \rightarrow
\R^n\times \R $ \emph{preserves the order of events\/} if for all
events $A$ and $B$, $A=_tB$ implies $f(A)=_tf(B)$ and $A<_tB$
implies $f(A)<_tf(B)$.\qed\end{definition}

\begin{proposition}\label{orthotime}
A transformation $f=(f_1,f_2,\ldots, f_n,f_t): \R^n\times \R \rightarrow \R^n\times \R:
(\mathbf{x},t)\mapsto (f_1(\mathbf{x},t),\ldots,f_n(\mathbf{x},t),f_t(\mathbf{x},t)) $ preserves
the order of events if and only if $f_t$ is a strictly monotone increasing bijection of $t$ alone.
\end{proposition}

\par\noindent
{\bf Proof.}  The  if-direction is straightforward. To prove the other direction, let $f=(f_1,\ab
f_2,\ab \ldots,\ab f_n,\ab f_t)$ be a transformation of $\R^n\times \R$. Consider any two events
$A=\{(a_1,a_2, \ldots, \ab a_n,\tau)\}$ and $B=\{(a'_1, a'_2, \ldots, a'_n,\tau)\}$. Since $A=_t
B$, then $f_t(a_1,a_2, \ldots, a_n,\tau) = f_t(a'_1,a'_2, \ldots, a'_n,\tau)$. This shows that
$f_t$ is a function of $t$ alone.

Consider any two events  $A=\{(a_1,a_2, \ldots, a_n,\tau_1) \}$ and $B=\{(a_1,a_2, \ldots, \ab
a_n,\tau_2) \}$ with $\tau_1 < \tau_2$. Since $A<_t B$, then   $f_t(\tau_A) < f_t(\tau_B)$. This
shows that $f_t$ is a strictly monotone function of $t$.

The transformation groups that we consider are all groups with
respect to the composition operator $\circ$ of functions.
Therefore, for every transformation $f$ also its inverse exists,
and hence $f$ is a bijection. Given the fact that the component
$f_t$ is a function of $t$ alone, it has to be a bijection too.
\qed
\medskip

\emph{We require that transformations preserve the order of events}. We can therefore  write the
transformation groups of interest as a product of groups, i.e., ${\cal F} =({\cal F}_{st}, {\cal
F}_t)$, where  $$ ({\cal F}_{st}, {\cal F}_t) = \{ (f_{st},f_t) \mid f_{st}=(f_1,f_2,\ldots, f_n):
\R^n\times \R \rightarrow \R^n {\rm \ and\ } f_t: \R\rightarrow \R \}.$$

The particular groups ${\cal F}_t$ that we will consider in this
paper are:
\begin{itemize}
\item ${\cal A}_t = \{t\mapsto a t+b\mid a, b\in \R {\rm \ and \ } a>0\}$, i.e., the monotone
affinities of the time-line; \item ${\cal T}_t = \{t\mapsto t+b\mid b\in \R \}$, i.e., the
translations of the time-line; and \item  ${\it Id}_t = \{{\rm id}\}$, i.e.,  the identity of time.
\end{itemize}

Invariance with respect to this type of transformations of time is
often encountered in physics~\cite{mech}.

\subsection{Spatial and spatio-temporal restrictions on trans\-for\-mations}\label{sectie3.3}

In the following, we consider transformations coming from practical situations where moving objects
are monitored from a fixed position or situations where a fixed object is observed from a moving
position. The frame of reference is therefore changing in a time-dependent way. In real life, this
continuous change of reference system arises in different kinds of situations. For example, when a
moving person is watching an event, his/her description of that event will be related to his/her
position and orientation at each time moment. When this person moves along a straight line at
constant speed, the transformation that describes this continuous change of reference system would
be a time-dependent affinity.

In this paper, we only look at transformations that have an algebraic description. The general form
of the transformation  groups ${\cal F}_{st}$ that we consider have elements of the form:
$$\left( \begin{array}{c} x_1\\  x_2 \\ \vdots\\  x_n\\ t
 \end{array}
 \right)\mapsto
\left(
\begin{array}{cccc }
\alpha_{11}(t) & \alpha_{12}(t) & \cdots & \alpha_{1n}(t)   \\
\alpha_{21}(t) & \alpha_{22}(t) & \cdots & \alpha_{2n}(t)   \\
\vdots&\vdots&\cdots& \vdots
\\ \alpha_{n1}(t) & \alpha_{n2}(t) & \cdots & \alpha_{nn}(t)
 \end{array}
 \right)\cdot  \left(
\begin{array}{c}
x_1\\  x_2 \\ \vdots\\  x_n
 \end{array}
 \right) + \left(
\begin{array}{c}
\beta_1(t)\\  \beta_2(t)\\\vdots\\ \beta_n(t)\\
 \end{array}
 \right),$$
where the $\alpha_{ij}$ and $\beta_i$ are functions from $\R$ to $\R$.
Furthermore, we require that the transformation groups that we consider
are ``semi-algebraic'' (we give a precise definition in  Section~\ref{subsec:53}).

\medskip
\par\noindent
We will consider the following groups ${\cal F}_{st}$ of
transformations:
\begin{itemize}
\item ${\cal A}_{st}$ is the group of transformations of the
above form where the $\alpha_{ij}(t)$ and $\beta_i(t)$ are
arbitrary functions of $t$ such that the matrix of the
$\alpha_{ij}(t)$ has an inverse for each value of $t$, i.e., these
are the \emph{time-dependent affinities};

\item ${\cal A}_{st}^{f}$ is the subgroup of ${\cal A}_{st}$
consisting of transformations for which the functions
$\alpha_{ij}(t)$ and $\beta_i(t)$ only take a finite number of
values, i.e., functions that are piecewise
constant;

\item ${\cal A}_{st}^{c}$ is the subgroup of  ${\cal A}_{st}^{f}$  consisting of transformations
for which the functions $\alpha_{ij}(t)$ are constants and $\beta_i(t)$ are linear functions of
$t$;

\item ${\cal S}_{st}^{}$, ${\cal S}_{st}^{f}$ and ${\cal S}_{st}^{c}$ are subgroups of ${\cal
A}_{st}^{}$, ${\cal A}_{st}^{f}$ and ${\cal A}_{st}^{c}$ respectively, where the matrix of the
$\alpha_{ij}(t)$ represents at each moment a similarity, i.e. the composition of  an isometry
(given by a matrix with determinant $1$) and a scaling (given by a non-zero multiple of the unit
matrix);

\item ${\cal I}_{st}^{}$, ${\cal I}_{st}^{f}$, ${\cal
I}_{st}^{c}$ are the subgroups of the above groups where the
determinant of the matrix consisting of the $\alpha_{ij}(t)$
equals 1 at each moment, i.e., this matrix determines an isometry;

\item ${\cal T}_{st}^{}$, ${\cal T}_{st}^{f}$, ${\cal
T}_{st}^{c}$ are the subgroups of the above groups where the
matrix consisting of the $\alpha_{ij}(t)$ is the identity matrix,
i.e., these are groups of translations.

\end{itemize}

\subsection{Physical transformation groups}\label{sectie3.4}
The following groups are of interest when notions such as velocity, acceleration and force are
important in an application. These transformation groups can be found by solving the differential
equations that express that these physical entities are preserved~\cite{mech}. We consider these
notions for arbitrary and rigid motions, respectively.
 A \emph{rigid motion} is a motion that preserve the shape of a moving body or moving figure, i.e., it is an isometric movement. To study the velocity and acceleration of a
moving  body,  we only consider the movement of the center of mass of that figure and  do not take
into account the changes in shape of  the body.

The transformation groups of interest here are the following.

\begin{itemize}
\item ${\cal V}_{st}^{}$ is the subgroup of ${\cal A}_{st}^{c}$
where the $\beta_i$ are constants. This group of transformations
preserves the \emph{velocity vector} of a moving figure.

\item ${\cal V(R)}_{st}^{}$ is the subgroup of ${\cal
I}_{st}^{c}$ where the $\beta_i$ are constants. This group of
transformations preserves the \emph{velocity vector \/} of a
moving figure in rigid motion.

\item  ${\cal AC}_{st}^{}$ is the group ${\cal A}_{st}^{c}$.
This group of transformations preserves the \emph{acceleration
vector\/} of a moving object.

\item ${\cal AC(R)}_{st}^{}$ is the group ${\cal
I}_{st}^{c}$. This group of transformations preserves the
\emph{acceleration vector\/} of a moving figure in rigid motion.

\end{itemize}

In physics it is customary to consider only translations for what concerns the time dimension,
i.e., the transformations in  the group  ${\cal T}_t$. The group $({\cal AC(R)}_{st}^{}, {\cal
T}_t)$ is also known as \emph{the group of the Galilei transformations\/}~\cite{mech}. It is
particularly useful because all laws of classical mechanics are invariant for this group of
transformations of space-time~\cite{mech}.

\subsection {Examples of generic queries}\label{examples}

We end this section with a number of  examples of queries that are
generic  for some of the genericity classes that we have
introduced above.

Suppose in some city, an experiment is set up to evaluate the traffic situation. A number of probe
cars (for simplicity, we assume two) is continuously driving around the city in a random way. The
trajectories of the cars are stored in a spatio-temporal database, of underlying dimension  2, with
schema $\sigma=\{carA,carB\}$, where the relations $carA$ and $carB$ both have arity  1.
 In these examples, we  assume that time is measured in seconds and distance is measured in meters.
We now give some example queries, and indicate for each the transformation  groups  it is generic
for.

\begin{example}\rm\label{ex-q1}
\emph{$Q_1:$ Does the route followed by car A self-intersect more often than the route followed by car B does?}
\medskip\par\noindent This query is $({\cal V}_{st}, {\cal A}_t)$-generic,  but not $({\cal A}_{st}, {\cal A}_t)$-generic, for instance.   It is not expressible in first-order logic. In Section~\ref{sectie5}, we will give a ``program''   expressing this query. \qed\end{example}

\begin{example}\rm\label{ex-q2}
 \emph{$Q_2:$ Give the places and time moments where it is true for car A that when it reaches them,
it is standing still at that spot for at least $300$ more seconds, (i.e., where and when did car A
encounter a traffic jam?).} \medskip\par\noindent This query is $({\cal V}_{st}, {\cal
T}_t)$-generic. Indeed, the fact that a car has speed zero (when it is standing still) has to be
preserved, which
 requires
 the group ${\cal V}_{st}$, and the length of time intervals has to be preserved, which
 requires
 ${\cal T}_t$.
 This query is expressed by the following \fopr{\overline{carA}}-formula: $$\varphi_2(x, y, t):=
 (\overline{carA}(x, y, t) \land  (\forall t_2)(  (t \leq t_2 \land t_2 \leq t + 300) \rightarrow \overline{carA}(x, y, t_2) ).$$\qed \end{example}

\begin{example}\rm\label{ex-q3}
\emph{$Q_3:$ Was there a collision between car A and car B?}
\medskip\par\noindent This query is $({\cal A}_{st}, {\cal A}_t)$-generic.  This query is expressed by the following \fopr{\overline{carA}, \ab \overline{carB}}-formula:
$$\varphi_3:= (\exists x)(\exists y)(\exists  t)(\overline{carA}(x, y, t) \land \overline{carB}(x, y, t)).$$ \qed \end{example}
\begin{example}\rm\label{ex-q4}
\emph{$Q_4:$ Did car A pass at 500 meters north of car B at time moment $t = 5930$?}
\medskip\par\noindent This query is $({\cal T}_{st}, {\it Id}_t)$-generic.   This query is expressed by the following \fopr{\overline{carA}, \ab \overline{carB}}-formula:
$$\varphi_4 := (\exists x_1)(\exists y_1)(\exists y_2)(\overline{carA}(x_1, y_1, 5930) \land \overline{carB}(x_1, y_2, 5930) \land y_1 = y_2 + 500).$$ \qed \end{example}
\begin{example}\rm\label{ex-q5}
 \emph{$Q_5:$ Did car A encounter any ``empty roads''? (I.e., were there parts of its trajectory where it could drive at constant speed in a straight line for at least 6000 seconds?)}
\medskip\par\noindent This query is $({\cal AC}_{st}, {\cal T}_t)$-generic. The fact that a car
drives at constant speed  (i.e., has an acceleration of zero)   has to be preserved. Note that,
because the car's movement is a polynomial function of time, driving at constant speed means
driving in a straight line. Query $Q_5$ can be
 expressed by the following \fopr{\overline{carA}}-formula:

$$\displaylines{\qquad \varphi_5 := (\exists t_1)(\exists t_2)(\exists  x_1)(\exists  y_1)(\exists  x_2)(\exists  y_2)(\overline{carA}(x_1, y_1, t_1) \land \overline{carA}(x_2, y_2, t_2)\land \splits t_2=t_1 + 6000 \land
(\forall t_3)((t_1 \leq t_3 \land t_3 \leq t_2) \rightarrow (\exists x_3)(\exists y_3)(
\overline{carA}(x_3, y_3, t_3) \land \splits (t_2-t_1)x_3=(t_2-t_3)x_1+(t_3-t_1)x_2\land
(t_2-t_1)y_3=(t_2-t_3)y_1+(t_3-t_1)y_2))).\qquad}$$   \qed\end{example}

 This completes the examples section.
We return to these examples later on, when we have defined point languages.

\section{Sound and complete languages for the ge\-ne\-ric first-order spatio-temporal queries}\label{sectie4}

 In this section, we study the $({\cal F}_{st}, {\cal
F}_t)$-generic queries that are expressible in \fop. To start with,
we give a general undecidability result.
We prove that it is undecidable whether a query is
$({\cal F}_{st}, {\cal F}_t)$-generic, for any nontrivial group $({\cal F}_{st}$, ${\cal F}_t)$.

Next, we show that $({\cal F}_{st}, {\cal F}_t)$-generic \fop-queries are recursive enumerable, however.
We do this by  syntactically specifying languages that capture the $({\cal F}_{st}$, ${\cal F}_t)$-generic queries, for
all groups $({\cal F}_{st}$, ${\cal F}_t)$ listed in Section~\ref{sectie3.3} and
Section~\ref{sectie3.4}.

The strategy to prove the following Theorem was
introduced by Paredaens, Van den Bussche and
Van Gucht~\cite{pvv-pods}.

Let $\N$ denote the set of the natural numbers.

\begin{theorem} \label{theo-undecidable}\rm For all non-trivial groups $({\cal F}_{st}$, ${\cal F}_t)$
mentioned in the previous section, $({\cal F}_{st}, {\cal F}_t)$-genericity of spatio-temporal
\fopr{\sigma}-queries   is undecidable, where $\sigma$ is a non-empty schema. \qed
\end{theorem}

\par\noindent{\bf Proof.}  Let ${\cal F}$ be a group of transformations of $\R^n \times \R$ that contains an element $f_0$
that does not map $(0,0)$ to itself. We show that ${\cal F}$-genericity of spatio-temporal queries
over a certain schema $\sigma = \{R\}$, where $R$ is a one-dimensional unary spatio-temporal
relation, of output type (1,1) is undecidable. For other non-empty schemas the proof
is similar.
 We will do this by reducing
deciding the truth of sentences of the $\forall*$-fragment of number theory
to the genericity question.
The  $\forall*$-fragment of number theory is known to be undecidable since
Hilbert's 10th problem~\cite{hilbert10} can be formulated in it.

We encode a natural number $n$ by the unary one-dimensional
spatio-temporal relation $$enc(n) := \{
(0,0),(1,0),\ldots,(n,0)\}.$$

A ($k$-dimensional) vector of natural numbers  $(n_1,n_2,\ldots,n_k)$ is encoded by the relation
$$enc(n_1,n_2,\ldots,n_k) := enc(n_1) \cup (enc(n_2) + (n_1 + 2,0))
\cup{}$$ $$\ldots \cup (enc(n_k) + (n_1 + 2 + \cdots + n_{k-1} + 2,0) ).$$
 For fixed $k$, the corresponding decoding is expressible in \fop.
We thus associate to the first-order sentence
 $(\forall n_1)\cdots (\forall n_k)\varphi(n_1,\ldots , n_k)$ of number theory to
the following spatio-temporal query $Q_\varphi $ over the input schema $\sigma=\{R\}$:

\begin{quote}
\begin{tabbing}
\quad\=\quad\=\quad\=\kill
if \textit{ $\overline{R}$  encodes a vector $(n_1,\ldots , n_k)\in \N^k$}  \\
then  \\
\> if \textit{$\varphi(n_1,\ldots, n_k)$}  \\
\> then \textit{return $\emptyset$}  \\
\> else \textit{return $\{(0,0)\}$}  \\
else \textit{return $\emptyset$.}
\end{tabbing}
\end{quote}

This query is expressible in \fop.
\medskip

\par\noindent
\emph{Claim.} The query $Q_\varphi$ is ${\cal F}$-generic if and only if the sentence
$(\forall n_1)\cdots (\forall n_k)\varphi(n_1,\ab \ldots ,\ab  n_k)$
 holds in the natural numbers.
\medskip

Now, we  prove this claim.
First, suppose that, for all $(n_1,\ab \ldots ,\ab  n_k) \in \N^k$, $\varphi(n_1,\ab \ldots ,\ab  n_k)$ holds.
Let $R$ be a
one-di\-men\-sional unary spatio-temporal relation and let $f$ be some transformation of ${\cal
F}$. We have to prove that
$$f(Q_\varphi (R)) = Q_\varphi (f(R)).$$

The result of $Q_\varphi(R)$ will always be $\emptyset$: either $R$ does not encode a vector $(n_1,\ab \ldots ,\ab  n_k)$, or
it does and $\varphi(n_1,\ab \ldots ,\ab  n_k)$ holds. For the same reason, $Q_\varphi(f(R))$ also equals $\emptyset$. The
transformation $f$ maps $\emptyset$ to $\emptyset$, hence $f(Q_\varphi(R)) = \emptyset$, which concludes
the first part of the proof.

 Now assume that there exists an $(n^0_1,\ab \ldots ,\ab  n^0_k)$ such that $\varphi(n^0_1,\ab \ldots ,\ab  n^0_k)$ is not true. Let $R$ be
the database that decodes $(n^0_1,\ab \ldots ,\ab  n^0_k)$. The result of $Q_\varphi(R)$ will be the origin $(0,0)$ of $\R
\times \R$. If we now apply $f_0$ to this result, the output is a vector $(y,z) \neq (0,0)$. On the
other side, if we first apply $f_0$ to $R$, there are three possibilities. Either $f_0(R)$ encodes
a vector $(n^1_1,\ab \ldots ,\ab  n^1_k)$ for which $\varphi(n^1_1,\ab \ldots ,\ab  n^1_k)$ is true, then
the result of $Q_\varphi(f_0(R))$ will be
$\emptyset$. Or, $f_0(R)$ encodes a vector $(n^1_1,\ab \ldots ,\ab  n^1_k)$ for which $\varphi(n^1_1,\ab \ldots ,\ab  n^1_k)$ is not true,
and $Q_\varphi(f_0(R))$ returns $(0,0)$. In the last case, $f_0(R)$ does not encode a vector of natural numbers, in which case
the
result of $Q_\varphi(f_0(R))$ will be $\emptyset$ again. In all cases, we have that $Q_\varphi(f_0(R))\neq
f_0(Q_\varphi(R))$. Therefeore,
the query $Q_\varphi$ is not ${\cal F}$-generic.

We can conclude that $Q_\varphi$ is ${\cal F}$-generic if and only if the sentence
$(\forall n_1)\cdots \ab (\forall n_k)\ab \varphi(n_1,\ab \ldots ,\ab  n_k)$ holds in the natural numbers.

Therefore, if $\cal F$-genericity would be decidable, also the truth of sentences in the $\forall^*$-fragment
of number theory would be decidable. This concludes the proof.
\qed
  \medskip

 In the remainder of this section, we show that the first-order queries that are $({\cal
F}_{st}, {\cal F}_t)$-generic are recursively enumerable, however.
We show this by giving sound and complete languages for the
$({\cal F}_{st}, {\cal F}_t)$-generic \fo-queries,  for the groups
$({\cal F}_{st}, {\cal F}_t)$ mentioned in Section~\ref{sectie3}.

We first define these sound and complete languages that are point-based logics.
\begin{definition}\rm\label{pointlangdef}
 Let $\sigma = \{R_1, R_2, \ldots, R_m\}$ be a spatio-tem\-po\-ral database sche\-ma and let
$\Pi$ be a set of predicates. The
 first-order logic over $\sigma$ and $\Pi$, denoted by \for{\Pi,R_1, R_2, \ldots, R_m} or \for{\Pi,\sigma},
can be used as a spatio-temporal query language when variables are interpreted to range over points in $\R^n \times \R$, (we    denote variables by $u, v,
w,\ldots$). The atomic formulas in \for{\Pi,\sigma}\ are equality constraints on point variables,
 the predicates of $\Pi$ applied to point variables, and the relation names $R_1, R_2, \ldots, R_m$ from $\sigma$ applied to point
 variables.\qed \end{definition}

 A \for{\Pi,\sigma}-formula $\varphi(v_1, v_2, \ldots, v_k)$  defines for
each spatio-temporal database $\st{}$ over $\sigma$ a subset
$\varphi(\st{})$ of $(\R^n \times \R)^k$ defined as
$$\{(p_1,\ldots,p_k)\in (\R^n \times \R)^k\mid (\R^n \times \R,\Pi^{\R^n \times \R},R_1^{\st{}},\ldots,R_m^{\st{}})\models \varphi[p_1,...,p_k]\},$$
where $\varphi[p_1,...,p_k]$ is obtained from the formula $\varphi(v_1,...,v_k)$ by instantiating the variables $v_i$ by the constant points $p_i$, $1\leq i\leq k$.

From Definition~\ref{equiv-queries}, it is clear what it means that a \for{\Pi,\sigma}-formula \emph{expresses a
semi-algebraic databases query}.

\begin{definition}\rm
  A query language is said to be
\emph{sound} for the $({\cal F}_{st}, {\cal F}_t)$-generic \fop-queries on spatio-temporal
databases, if that language only expresses $({\cal F}_{st}, {\cal F}_t)$-generic \fop-queries on
spatio-temporal databases.

 A query language is said to be
 \emph{complete} for the $({\cal F}_{st}, {\cal F}_t)$-generic
\fop-queries on spatio-temporal databases,
 if all $({\cal F}_{st}, {\cal F}_t)$-generic \fop-queries on spatio-temporal
databases can be expressed in that language. \qed \end{definition}

\medskip
 In the following, we will omit the dependence on the input
schema when this is clear from
the context, and use the notation \for{\Pi}\ for first-order point languages over the predicate set
$\Pi$.

In the remainder of this section, we first discuss notions of
genericity determined by time-independent transformations (Section~\ref{indep}),
afterwards we discuss applications to physics (Section~\ref{subsec:physics}) and we end with
genericity for the time-dependent transformations (Section~\ref{dependent}).

\subsection{Genericity for time-independent transformations}\label{indep}

In this section, we give a general result concerning $({\cal F}_{st}, {\cal F}_t)$-generic queries
where ${\cal F}_{st}$ is a subgroup of  ${\cal A}_{st}^{c}$, the group of time-independent
affinities of $\R^n\times \R$. First, we introduce the point predicates that we will use for the
different point languages.

To express the temporal order of events, we use the point predicate \befores. Let $p_1$ and $p_2$
be points in $\R^n \times \R$. The expression {\bf Before}($p_1,p_2$) evaluates to true if the time
coordinate $\tau_1$ of $p_1$ is smaller than or equal to the time coordinate $\tau_2$ of $p_2$. In
the following, we will often use the derived binary predicate {\bf Cotemp}, which expresses for two
points $p_1$ and $p_2$ that $\tau_1$ equals $\tau_2$. This predicate can be expressed in terms of
\befores\ as follows:
$${\bf Cotemp}(u,v) := {\bf Before}(u,v) \land {\bf Before}(v,u).$$

There are three more other purely temporal predicates: $\unittimes$, ${\bf 0}_{\bf t}$ and ${\bf
1}_{\bf t}$. The predicate $\unittime{p_1,p_2}$ expresses that the points $p_1, p_2\in
\R^n\times\R$ have time-coordinates $\tau_1$ and $\tau_2$ such that $|\tau_1-\tau_2|=1$. The unary
predicates ${\bf 0}_{\bf t}$ and ${\bf 1}_{\bf t}$ are such that ${\bf 0}_{\bf t}(p)$ and ${\bf
1}_{\bf t}(p)$ respectively express that the time coordinate of the point $p$ equals to zero  and
  to one.

 The following predicates  address spatio-temporal relations between points.  The
  point-predicate $\betweens{n+1}$ is defined such that $\betweenm{n+1}{p_1, \ab p_2, \ab
p_3}$ expresses that the points $p_1$, $p_2$, $p_3$ in $\R^n\times \R$ are collinear (in the space
$\R^n \times \R$) and that $p_2$ is between $p_1$ and $p_3$. The predicates $\leq_{\bf {\it
i}}(p_1, p_2)$ ($1\leq i\leq n$) express that the $i$th spatial coordinate of $p_1$ is less or
equal than the $i$th spatial coordinate of $p_2$. The expression $\equidist{p_1, p_2, p_3, p_4}$ is
true if the distance between the two co-temporal points $p_1$ and $p_2$ equals the distance between
the two co-temporal points $p_3$ and $p_4$. The binary predicate $\unitdists$ applied to two points
$p_1$ and $p_2$ expresses that they are co-temporal and that the (spatial) distance between $p_1$
and $p_2$ equals one. Finally, $\pos{n+1}{p_0, p_1, p_2, \ldots, p_{n+1}}$ expresses that the
$(n+1)$-tuple $(p_0, p_1, p_2, \ldots, p_{n+1})$ of points in $\R^n \times \R$  forms a positively
oriented $(n+1)$-dimensional coordinate system with $p_0$ as origin.
\begin{property}\rm\label{in-fo}
The point predicates \befores, $\betweens{n+1}$, \unittimes, ${\bf 0}_{\bf t}$, ${\bf 1}_{\bf t}$,
$\leq_{\bf {\it i}} (1 \leq i \leq n)$, \equidists, \unitdists\ and $\poss{n+1}$ are all
expressible in \fop. \qed
\end{property}

\par\noindent{\bf Proof.} The \fop-formulas for the different
predicates can be  obtained  by expressing the constraints on the coordinates of the points
satisfying the predicates. We denote the coordinates of a point variable $v_i$ in $\R^n \times \R$
by $(x_{i_1}, x_{i_2}, \ldots, x_{i_n}, t_i)$,  $i=1,2,\ldots$. The translation of the expression
\before{$v_1, v_2$} is $t_1 \leq t_2$ and the translation of \unittime{$v_1, v_2$} equals $\mid t_1
- t_2 \mid = 1$. The translation of ${\bf 0}_{\bf t}$ and ${\bf 1}_{\bf t}$ is straightforward.

It is well known (e.g. \cite{gvv-jcss}) that the predicates $\betweens{n+1}$, $\leq_{\bf {\it i}}
(1 \leq i \leq n)$, \equidists\ and \unitdists\ are expressible in  \fop.  For \equidists\ and
\unitdists\ it is necessary to use $\befores$ to express that their arguments should be
co-temporal.

The expression $\poss{n+1}(v_0, v_1, \ldots, v_{n+1} )$ is translated into \fo\ by expressing that
the vectors $v_1 - v_0, \ldots, v_{n+1} - v_0$ are linearly independent  and  that the
$(n+1)\times(n+1)$-matrix  containing   their coordinates has a strictly positive determinant.
 \qed

\begin{table}[htbp]
  \begin{center}
    \leavevmode
    \begin{tabular}[c]{|l|l|}
\hline\hline
 \ \      $({\cal F}_{st}, {\cal F}_t)$ \ \ & \ \   Sets of point predicates $\Pi{({\cal F}_{st}, {\cal F}_t)}$\\
\hline\hline \ \ $({\cal A}_{st}^{c}, {\cal A}_t)$\ \  & \ \   $\{\betweens{n+1},\befores \}$\\ \ \
$({\cal A}_{st}^{c}, {\cal T}_t)$\ \  & \ \   $\{\betweens{n+1},\befores , \unittimes\}$\\ \ \
$({\cal A}_{st}^{c}, {\rm Id}_t)$\ \ & \ \ $\{\betweens{n+1},\befores,  \unittimes ,{\bf 0}_{\bf
t},{\bf 1}_{\bf t} \}$\\
 \ \ $({\cal S}_{st}^{c}, {\cal F}_t)$\ \  & \ \   $\Pi({\cal A}_{st}^{c}, {\cal F}_t)\cup \{\equidists\}$\\
 \ \ $({\cal I}_{st}^{c}, {\cal F}_t)$\ \  & \ \   $\Pi({\cal A}_{st}^{c}, {\cal F}_t)\cup \{\equidists, \unitdists\}$\\
  \ \ $({\cal T}_{st}^{c}, {\cal F}_t)$\ \  & \ \   $\Pi({\cal A}_{st}^{c}, {\cal F}_t)\cup \{
  \equidists, \unitdists, \leq_{\bf {\it i}} (1\leq i\leq n), \poss{n+1}\}$\\

\hline
    \end{tabular}
\medskip
    \caption{An overview of the different sets of point predicates  for a number of
    spatio-temporal genericity
    notions. In the three last cases ${\cal  F}_t\in \{ {\cal  A}_t,{\cal T}_t,{\it Id}_t \}$.}\label{table-point-predicates-constant}
  \end{center}
\end{table}

\begin{property}\rm\label{invariant}
Let $({\cal F }_{st}, \allowbreak {\cal F}_t)$ be a group   and let $\Pi({\cal F }_{st},
\allowbreak {\cal F}_t)$ be a set of point predicates   as in
Table~\ref{table-point-predicates-constant}. The point predicates in $\Pi({\cal F}_{st}, {\cal
F}_t)$ are invariant under elements of $({\cal F }_{st}, \allowbreak {\cal F}_t)$. \qed
\end{property}

\par\noindent{\bf Proof.}
First, remark that, if we fix ${\cal F}_t$ to be one of  $\{{\cal A}_t, {\cal T}_t, {\it Id}_t \}$,
 then $$({\cal T}^c_{st}, \allowbreak {\cal F}_t) \subset ({\cal I }^c_{st}, \allowbreak {\cal
F}_t) \subset ({\cal S }^c_{st}, \allowbreak {\cal F}_t) \subset ({\cal A }^c_{st}, \allowbreak
{\cal F}_t).$$ Also, if we fix ${\cal F }_{st}$ to be one of $\{{\cal A }^c_{st}, {\cal S }^c_{st}
, {\cal I }^c_{st}, {\cal T }^c_{st}\}$, then  $$({\cal F }_{st}, \allowbreak {\it Id}_t) \subset
({\cal F }_{st}, \allowbreak {\cal T}_t) \subset ({\cal F }_{st}, \allowbreak {\cal A}_t).$$  Also,
all groups $({\cal F}_{st}, {\cal F}_t)$ are subgroups of the affinities of $\R^n \times \R$.
 As we already remarked,  if a point predicate is invariant for a certain
transformation group $({\cal F}_{st}, {\cal F}_t)$, it is also
invariant for all subgroups of $({\cal F}_{st}, {\cal F}_t)$.

We now prove invariance for each of the predicates in the sets $\Pi({\cal F }_{st}, \allowbreak
{\cal F}_t)$ of Table~\ref{table-point-predicates-constant}.

\medskip
\par\noindent
 $\bullet$ \emph{The predicate $\betweens{n+1}$ is invariant under elements of $({\cal A }^c_{st}, \allowbreak
{\cal A}_t)$}. It is well known that affinities preserve  the  betweenness of points.   As all
groups listed in Table~\ref{table-point-predicates-constant} are subgroups of the affinities of
$\R^n \times \R$, the predicate $\betweens{n+1}$ is invariant for all those groups.
\par\noindent  $\bullet$ \emph{The predicate \befores\ is invariant under elements
of $({\cal A }^c_{st}, \allowbreak {\cal A}_t)$}, since the elements of ${\cal A}_t$ are monotone
bijections of time. As shown in Proposition~\ref{orthotime}, the order on time events is preserved
under all strictly monotone increasing bijections of time. The groups ${\cal A}_t$, ${\cal I}_t$,
${\it Id}_t$ are all such bijections.

\par\noindent  $\bullet$ \emph{The predicate \unittimes\ is invariant under elements
of $({\cal A }^c_{st}, \allowbreak {\cal T}_t)$}. It is straightforward that all elements of ${\cal
T}_t$, which are translations in the time direction,  preserve the time difference between any two
points $p_1$ and $p_2$ in $\R^n \times \R$.
\par\noindent
 $\bullet$ \emph{The predicates ${\bf 0}_{\bf
t}$ and ${\bf 1}_{\bf t}$ are invariant under elements of $({\cal A }^c_{st}, \allowbreak {\rm
Id}_t)$}. It is clear that the identity transformation on the time preserves the fact that a point
$p$ in $\R^n \times \R$ has time coordinate zero or one.
\par\noindent  $\bullet$ \emph{The predicate \equidists\ is invariant under elements
of $({\cal S }^c_{st}, \allowbreak {\cal A}_t)$}. It is well known  that isometries and scalings
(and thus similarities) preserve the fact that the distance between one pair of points equals the
distance between a second pair of points. The groups ${\cal A}_t$, ${\cal T}_t$, ${\it Id}_t$ all
preserve co-temporality of  points.
\par\noindent  $\bullet$ \emph{The predicate \unitdists\ is invariant under elements
of $({\cal I }^c_{st}, \allowbreak {\cal A}_t)$}, because isometries are distance preserving transformations.

\par\noindent   $\bullet$ \emph{The predicates $\leq_{\bf {\it i}} (1\leq i\leq n)$ are invariant under elements
of $({\cal T }^c_{st}, \allowbreak {\cal A}_t)$}. It is easy to verify that if for two points $p_1$
and $p_2$ in $\R^n \times \R$, $\leq_{\bf {\it i}}(p_1,p_2)$ is true for some $i$ in $\{1, \ldots,
n\}$, also $\leq_{\bf {\it i}}(f(p_1),f(p_2))$ holds for each $f$ in $({\cal T}^c_{st}, {\cal
F}_t)$, where ${\cal F}_t$ is one of   ${\cal A}_t, {\cal T}_t, {\it Id}_t $.
\par\noindent  $\bullet$ \emph{The predicate $\poss{n+1}$ is invariant under elements
of $({\cal T }^c_{st}, \allowbreak {\cal A}_t)$}, since translations are orientation-preserving transformations.
  \qed

\medskip
\begin{remark}\rm
From now, all results are valid for underlying dimension $n \geq
2$.
\end{remark}

\medskip
The following theorem follows directly from the proof of Theorem
 5.5~\cite{gvv-jcss}.

\begin{theorem}\rm \label{metatheoremsp}
 Let $\sigma$ be a spatio-temporal database schema.  Let ${\cal F}$ be a subgroup of the affinities of $\R^n \times \R$. Let $\Pi$ be a set of
point-predicates that contains $\betweens{n+1}$. If the predicates in $\Pi$ are  \fop-expressible
and invariant under the transformations of ${\cal F}$ and if the fact \emph{``$\basisv$ is the
image of the standard coordinate system of $\R^n\times\R$ under some element $f$ of ${\cal F}$''}
is expressible in $\fo(\Pi)$, then $\fo(\Pi,\sigma)$ is sound and complete for the ${\cal
F}$-generic spatio-temporal database queries that are expressible in \fopr{\overline{\sigma}}.
\qed
\end{theorem}

 We now prove the following theorem.

\begin{theorem}\rm \label{metatheoremst}
 Let $\sigma$ be a spatio-temporal database schema.  Let ${\cal F}_{st}$ be a subgroup of ${\cal A}^c_{st}$ and ${\cal F}_t$ a  subgroup  of ${\cal
A}_t$. Let $\Pi({\cal F}_{st}, {\cal F}_t)$ be a set of point-predicates that contains
$\betweens{n+1}$ and \befores. If the predicates in $\Pi({\cal F}_{st}, {\cal F}_t)$ are
\fo-expressible and invariant under the transformations of $({\cal F}_{st}, {\cal F}_t)$ and if the
fact \emph{``$\basisv$ is the image of the standard coordinate system under some element $f$ of
$({\cal F}_{st}, {\cal F}_t)$''} is expressible in \fom{\Pi({\cal F}_{st}, {\cal F}_t)}, then the
logic \fom{\Pi({\cal F}_{st}, \ab {\cal F}_t),\ab \sigma} is sound and complete for the $({\cal
F}_{st}, {\cal F}_t)$-generic   spatio-temporal database queries that are expressible in
\fopr{\overline{\sigma}}.   \qed
\end{theorem}

\par\noindent{\bf Proof.}
First, we show that the language $\fo(\{\betweens{n+1}, \befores\})$ is sound and complete for the
 $({\cal A}^c_{st}, {\cal A}_t)$-generic \foprs-expressible spatio-temporal database queries  using  Theorem~\ref{metatheoremsp}. Indeed, it is clear that the group $({\cal A}^c_{st}, {\cal A}_t)$ is
a subgroup of the affinities of $\R^n \times \R$. Furthermore, the expression $\befores(u,v)$, is
expressible in \fop\ (see Property~\ref{in-fo}). Also, the predicates $\betweens{n+1}$ and
$\befores$ are both invariant under elements of $({\cal A}^c_{st}, {\cal A}_t)$
 (see  Property~\ref{invariant}).

 To conclude this part of the proof, we need to show that there is  an expression in $\fo(\{\betweens{n+1}, \befores\})$ that, for  $n+2$  arbitrary points $p_0, p_1,
\ab, \ldots, p_{n+1}$ in $\R^n \times \R$, states that $\basis$ is the image of the standard
coordinate system under some element $f$ of $({\cal A}^c_{st}, {\cal A}_t)$. It is  known
(e.g.~\cite{gvv-jcss,schwab-tarski})   that there exists an expression in the language
$\fo(\{\betweens{n+1}\})$ that, for  $n+2$  points $p_0, p_1, \ab, \ldots, p_{n+1}$ of $\R^n \times
\R$, expresses that $\basis$ is the image of the standard $(n+1)$-dimensional coordinate system
under some affinity of $\R^n \times \R$. We refer to this expression as
$${\bf CoSys}_{\cal A}\basishat.$$  Obviously, this formula also belongs to  $\fo(\{\betweens{n+1},
\befores\})$. The expression for the image of the standard coordinate system under some element of
$({\cal A}_{st}^{c}, {\cal A}_t)$ is as follows:

$$\cosys_{({\cal A}_{st}^{c}, {\cal A}_t)}\basishat :=
{\bf CoSys}_{\cal A}\basishat \land$$ $$\bigwedge_{i=1}^n {\bf Cotemp}(v_0, v_i) \ab \land
\neg\befores(v_{n+1}, v_0).$$ It is easy to verify that any coordinate system that is an image of
the standard coordinate system under an element of $({\cal A}_{st}^{c}, {\cal A}_t)$ satisfies this
expression. Also, the reverse is true. For clarity, we show this only for $n = 2$ (the general case
is analogous).

 Any coordinate system $\basisdd$ satisfying the expression $\cosys_{({\cal A}_{st}^{c},
{\cal A}_t)}\ab\basishatdd$ is of the form $p_0 = (a_{0,1},a_{0,2},\tau_0)$, $p_1 =
(a_{1,1},a_{1,2},\tau_0)$, $p_2 = (a_{2,1},a_{2,2},\tau_0)$, $p_3 = (a_{3,1},a_{3,2},\tau_3)$,
where $\tau_0 < \tau_3$ and
 the determinant

  $$\left|\begin{array}{@{}ccc@{}}a_{1,1} - a_{0,1} & a_{1,2} - a_{0,2} & 0\\
a_{2,1} - a_{0,1} & a_{2,2} - a_{0,2} & 0
\\ a_{3,1} - a_{0,1} & a_{3,2} - a_{0,2} & \tau_3 - \tau_0
\end{array}\right| \neq 0.\eqno{(\ast)}$$

Now, we have to show that there exists an element $f$ of $({\cal
A}_{st}^{c}, {\cal A}_t)$ such that the image of the standard
coordinate system under $f$ equals $\basisdd$. As $({\cal
A}_{st}^{c}, {\cal A}_t)$ is a subgroup of the affinities, $f$ is
representable by a matrix. It is straightforward to derive that
$f = (f_{st}, f_t)$, where
 $$f_{st}(x,y,t)=\left(\!\begin{array}{@{}cc@{}}a_{1,1} - a_{0,1} & a_{2,1} -
a_{0,1} \\ a_{1,2} - a_{0,2} & a_{2,2} - a_{0,2}
\end{array}\!\right)\left(\!\begin{array}{@{}c@{}}x \\ y  \end{array}\!\right) + \left(\!\begin{array}{@{}c@{}} (a_{3,1} - a_{0,1})t + a_{0,1}
\\(a_{3,2} - a_{0,2})t + a_{0,2} \end{array}\!\right) \textrm{, and}$$
$$f_t(t) = (\tau_3 - \tau_0)t + \tau_0.$$
\medskip
\par\noindent It is clear that
$(\tau_3 - \tau_0) > 0$  and that,  because of the inequality $(\ast)$, the value of the determinant  $\left|\begin{array}{@{}cc@{}}a_{1,1} - a_{0,1} & a_{2,1} - a_{0,1} \\
a_{1,2} - a_{0,2} & a_{2,2} - a_{0,2}
\end{array}\right|$ differs from zero, hence $f$ is an element of
$({\cal A}_{st}^{c}, {\cal A}_t)$.

 So far, we proved that the language $\fo(\{\betweens{n+1}, \befores\},\sigma)$ is sound and complete for
the $({\cal A}^c_{st}, {\cal A}_t)$-generic queries expressible in \fopr{\overline{\sigma}}. The
fact that any other language \fom{\Pi({\cal F}_{st}, {\cal F}_t),\ab \sigma}, where $\Pi({\cal
F}_{st}, {\cal F}_t)$ contains $\betweens{n+1}$ and \befores, is sound and complete for the $({\cal
F}_{st}, {\cal F}_t)$-generic \fopr{\overline{\sigma}}-queries for each subgroup $({\cal F}_{st},
{\cal F}_t)$ of $({\cal A}_{st}^{c}, {\cal A}_t)$,  under the conditions stated in
Theorem~\ref{metatheoremst}, follows from Theorem~\ref{metatheoremsp} together with the first part
of this proof. \qed

 \begin{theorem}\label{meta-cor-fo}\rm
 Let $\sigma$ be a spatio-temporal database schema.  Let $({\cal F }_{st}, \allowbreak {\cal F}_t)$ be a group
 and  let $\Pi({\cal F }_{st}, \allowbreak {\cal F}_t)$
be as in Table~\ref{table-point-predicates-constant}. The point language $\fo(\Pi({\cal F}_{st},\ab
{\cal F}_t),\ab \sigma )$ is sound and complete for the $({\cal F }_{st}, \allowbreak {\cal F}_t)$-generic
queries expressible in \fopr{\overline{\sigma}}.\qed
\end{theorem}

\smallskip
\par\noindent
{\bf Proof.} We can apply Theorem~\ref{metatheoremst} for all groups in
Table~\ref{table-point-predicates-constant},  because they are all subgroups of $({\cal
A}_{st}^{c}, {\cal A}_t)$. From Properties~\ref{in-fo} and~\ref{invariant}, it follows that all
predicates are expressible in \fop\ and that they are invariant under transformations of the
appropriate groups. The only thing left to prove is that, for all groups $({\cal F }_{st},
\allowbreak {\cal F}_t)$  from Table~\ref{table-point-predicates-constant}, and for  $n+2$  points
$v_0, v_1, \ab, \ldots, v_{n+1}$ in $\R^n \times \R$, the fact ``$\basisv$ is the image of the
standard coordinate system under some element $f$ of $({\cal F}_{st}, {\cal F}_t)$'' is expressible
in $\fo(\Pi({\cal F}_{st}, {\cal F}_t) )$.
 For  each group $({\cal F}_{st}, {\cal F}_t)$

 from Table~\ref{table-point-predicates-constant}, we now give a  formula that
expresses this   fact. The correctness of these formulas is easy to verify.

\par\noindent  $\bullet$ \emph{For the group $({\cal A}_{st}^{c}, {\cal A}_t)$},
we already gave a formula in the proof of Theorem~\ref{metatheoremst}.
 The desired formula is there   denoted by $\cosys_{({\cal A}_{st}^{c}, {\cal A}_t)}$.

\medskip

\par\noindent  $\bullet$ \emph{For the group $({\cal A}_{st}^{c}, {\cal T}_t)$}, we have  $$\displaylines{\qquad \cosys_{({\cal A}_{st}^{c}, {\cal T}_t)}\basishat := \hfill{} \cr  \hfill{}\cosys_{({\cal A}_{st}^{c}, {\cal A}_t)}\basishat \land {\bf UnitTime}(v_0, v_{n+1}).\qquad}$$

\par\noindent  $\bullet$ \emph{For the group $({\cal A}_{st}^{c}, {\it Id}_t)$}, we have $$\displaylines{\qquad\cosys_{({\cal A}_{st}^{c}, {\it Id}_{t})}\basishat := \hfill{} \cr  \hfill{}
\cosys_{({\cal A}_{st}^{c}, {\cal T}_t)}\basishat \land {\bf 0_t}(v_0) \ab  \land {\bf
1_t}(v_{n+1}).\qquad}$$

Let ${\cal  F}_t$ be an element of  $\{ {\cal  A}_t,{\cal T}_t,{\it Id}_t \}$.
\par\noindent  $\bullet$ \emph{For the groups $({\cal S}_{st}^{c}, {\cal F}_t)$}, we have  $$\displaylines{\qquad\cosys_{({\cal S}_{st}^{c}, {\cal F}_t)}\basishat := \hfill{} \cr  \hfill{}\cosys_{({\cal A}_{st}^{c}, {\cal F}_t)}\basishat \land
\bigwedge_{i=1}^n\bigwedge_{j=1}^n\equidists(v_0, v_i, v_0, v_j).\qquad}$$

\par\noindent  $\bullet$ \emph{For the groups $({\cal I}_{st}^{c}, {\cal F}_t)$}, we have $$\displaylines{\qquad\cosys_{({\cal I}_{st}^{c}, {\cal F}_t)}\basishat := \hfill{} \cr  \hfill{}
\cosys_{({\cal S}_{st}^{c}, {\cal F}_t)}\basishat \land \bigwedge_{i=1}^n \unitdists(v_0,
v_i).\qquad}$$

\par\noindent  $\bullet$ \emph{For the groups $({\cal T}_{st}^{c}, {\cal F}_t)$}, we have
$$\displaylines{\qquad\cosys_{({\cal T}_{st}^{c}, {\cal F}_t)}\basishat :=\hfill{} \cr  \hfill{}
 \ab {\bf CoSys}_{({\cal I}_{st}^{c}, {\cal F}_t)}\basishat \land \hfill{} \cr  \hfill{}\poss{n+1}\basishat \ab \land
\bigwedge_{j=1}^n\bigwedge_{i=1}^n \leq_{\bf {\it i}}(v_0, v_j).\qquad}$$  \qed

\subsection{Applications to Physics}\label{subsec:physics}

Here, we focus on the transformation groups $({\cal V}_{st}^{}, {\cal T}_t)$, $({\cal
V(R)}_{st}^{}, {\cal T}_t)$, $({\cal AC}_{st}^{},\ab {\cal T}_t)$ and $({\cal AC(R)}_{st}^{}, \ab
{\cal T}_t)$. To formulate our results we need to define one more point-predicate, namely ${\bf
=}_\textrm{\bf space}$.  If $p_1 = (a_{1,1},\ldots, a_{1,n}, \tau_1)$ and $p_2 =
(a_{2,1},\ldots,a_{2,n},\ab \tau_2)$ are elements of $\R^n\times \R$, then ${\bf =}_\textrm{\bf
space}(p_1,p_2)$ if and only if $a_{1,i}=a_{2,i}$ for  all  $1\leq i\leq n$.

\begin{remark}\rm\label{space-in-fo}
The expression $$\displaylines{\qquad{\bf =}_\textrm{\bf space}(v_1,v_2) := \bigwedge_{i =
1}^n(\leq_i(v_1, v_2) \land \leq_i(v_2, v_1))\qquad}$$ is expressible in  \fop. \qed
\end{remark}

\begin{theorem}\label{phys-th}\rm
 Let $\sigma$ be a spatio-temporal database schema.  Let  the groups $({\cal F }_{st}, \allowbreak {\cal T}_t)$ and the predicate sets $\Pi({\cal F
}_{st}, \allowbreak {\cal T}_t)$  be as in Table~\ref{table-point-predicates-physics}. The point
language $\fo({\Pi({\cal F}_{st}, {\cal T}_t) },\ab \sigma)$ is sound and complete for the $({\cal
F }_{st}, \allowbreak {\cal T}_t)$-generic spatio-temporal queries that are  expressible in
\fopr{\overline{\sigma}}. \qed
\end{theorem}

\begin{table}[htbp]
  \begin{center}
    \leavevmode
    \begin{tabular}[c]{|c|l|}
\hline\hline
 \ \      $({\cal F}_{st}, {\cal T}_t)$ \ \ & \ \   Set of point predicates $\Pi{({\cal F}_{st}, {\cal T}_t)}$\\
\hline\hline \ \ $({\cal V}_{st}^{}, {\cal T}_t)$\ \  & \ \   $\{\betweens{n+1},\befores,
\unittimes,{\bf =}_\textrm{\bf space} \}$\\ \ \ $({\cal V(R)}_{st}^{}, {\cal T}_t)$\ \  & \ \
$\{\betweens{n+1},\befores,\unittimes,{\bf =}_\textrm{\bf space},$ \\ &  \ \ \ \ \ $\equidists,\unitdists\}$\\
\ \ $({\cal AC}_{st}^{}, {\cal T}_t)$\ \ & \ \ $\{\betweens{n+1},\befores, \unittimes\}$\\ \ \
$({\cal AC(R)}_{st}^{}, {\cal T}_t)$\ \ & \ \
$\{\betweens{n+1},\befores, \unittimes,$ \\ &  \ \ \ \ \ $\equidists,\unitdists \}$\\
 \hline
    \end{tabular}
\medskip
    \caption{An overview of the different point-predicate sets for the
    physical transformation groups.}
    \label{table-point-predicates-physics}
  \end{center}
\end{table}

\par\noindent
{\bf Proof.} The transformation groups $({\cal F}_{st}, {\cal T}_t)$ of
 Table~\ref{table-point-predicates-physics}  are all subgroups of the group $({\cal A}_{st}^{c},
{\cal A}_t)$. Furthermore, the predicates of $\Pi{({\cal F}_{st}, {\cal T}_t)}$ are expressible in
 \fop\  (see Property~\ref{in-fo} and Remark~\ref{space-in-fo}).  Straightforward geometrical and physical arguments show  that all
predicates are invariant under the appropriate transformation groups. We can now apply
Theorem~\ref{metatheoremst}. We only have to verify that it is possible to express in the languages
$\fo(\Pi{({\cal F}_{st}, {\cal T}_t)})$ that a coordinate system is the image of the standard
$(n+1)$-dimensional coordinate system under an element of $({\cal F }_{st}, \allowbreak {\cal
T}_t)$. We now give, for each group $({\cal F}_{st}, {\cal F}_t)$ from
Table~\ref{table-point-predicates-physics}, the expression for the fact that $\basisv$ is the image
of the standard coordinate system under some element $f$ of $({\cal F}_{st}, {\cal F}_t)$.

  The correctness
of these expressions is easy to verify.

\par\noindent $\bullet$ \emph{For the group $({\cal V}_{st}^{}, {\cal T}_t)$}, we have  $$\displaylines{\qquad\cosys_{({\cal V}_{st}^{}, {\cal T}_t)}\basishat :=\hfill{} \cr \hfill{}
\cosys_{({\cal A}_{st}^{}, {\cal T}_t)}\basishat \land {\bf =}_\textrm{\bf
space}(v_0,v_{n+1}),\qquad}$$
 because elements of this group map the origin $(0,\ldots,0,0)$ and the unit vector in the time-direction
$(0,\ldots,0,1)$ of the standard coordinate system of $\R^n\times\R$ onto points which have equal
spatial coordinates.

\par\noindent $\bullet$ \emph{For the group $({\cal V(R)}_{st}^{}, {\cal T}_t)$}, we have  $$\displaylines{\qquad\cosys_{({\cal V(R)}_{st}^{}, {\cal T}_t)}\basishat :=\hfill{} \cr \hfill{}
\cosys_{({\cal I}_{st}^{}, {\cal T}_t)}\basishat \land {\bf =}_\textrm{\bf
space}(v_0,v_{n+1}).\qquad}$$

\par\noindent $\bullet$ \emph{For the group $({\cal AC}_{st}^{}, {\cal T}_t)$}, we have  $$\displaylines{\qquad\cosys_{({\cal AC}_{st}^{}, {\cal T}_t)}\basishat := \cosys_{({\cal A}_{st}^{}, {\cal
T}_t)}\basishat.\qquad}$$

\par\noindent $\bullet$ \emph{For the group $({\cal AC(R)}_{st}^{}, {\cal T}_t)$}, we have  $$\displaylines{\qquad\cosys_{({\cal AC(R)}_{st}^{}, {\cal T}_t)}\basishat := \cosys_{({\cal I}_{st}^{}, {\cal
T}_t)}\basishat.\qquad}$$\qed

Next, we illustrate the languages summarized in Table~\ref{table-point-predicates-constant} and
Table~\ref{table-point-predicates-physics} on the appropriate examples of Section~\ref{examples}.

\begin{example}\label{ex-q2between}\rm We give the $\fo(\{\betweens{n+1},\befores, \unittimes, \ab {\bf=}_\textrm{\bf space} \})$ -query $Q'_2$ equivalent to the $({\cal V}_{st}, {\cal T}_t)$-generic query of
Example~\ref{ex-q2}: {\it Give the places and time moments where $car A$ is standing still at that
spot for at least $300$ more seconds. }

 Remember that we  assumed before that time is measured in seconds and distance is measured in meters. We first remark that the fact that one point is a constant number of seconds before another, can be
expressed using \unittimes\ and \befores. We illustrate this for an   easy    example where one
point is 3 seconds after another:

 $$\displaylines{\qquad 3sec(u, v) := (\exists w_1)(\exists  w_2)(\before{u, w_1} \land \before{w_1, w_2} \land \before{w_2, v} \land \splits  \unittime{u, w_1} \land \unittime{w_1, w_2} \land \unittime{w_2, v}).\qquad}$$
Now we give the expression for $Q'_2$: $$\displaylines{\qquad carA(u) \land (\exists v)(300sec(u,
v) \land \splits (\forall w)((\before{u, w} \land \before{w, v}\land carA(w)) \rightarrow
{\bf=}_\textrm{\bf space}(u, w))).\qquad}$$ \qed\end{example}

\begin{example}\label{ex-q5between}\rm We give the $\fo(\{\betweens{n+1},\befores, \unittimes \})$-query $Q'_5$ equivalent to the $({\cal AC}_{st}, {\cal T}_t)$-generic query of
Example~\ref{ex-q5}: {\it Did car A encounter any empty roads? I.e.,  were there  parts of its
trajectory where it could drive at constant speed for at least 6000 seconds.}
$$\displaylines{\qquad (\exists u)(\exists v)(carA(u) \land carA(v) \land 6000sec(u, v) \land
(\forall w)((carA(w) \land \splits \before{u, w} \land \before{w, v}) \rightarrow
(\betweenm{n+1}{u, v, w}))).\qquad}$$ \qed
\end{example}

\subsection{Genericity for time-dependent transformations}\label{dependent}

Here, we focus on notions of genericity determined by time-dependent transformations. Our first
result in this context shows that we can restrict our attention, without loss of generality, to
piece-wise constant transformations.

\begin{proposition}\label{collapse}\rm
Let $Q$ be a spatio-temporal query expressible in \fop\ and let the group ${\cal F}_{st}$ be
${\cal A}_{st}$, ${\cal S}_{st}$, ${\cal
I}_{st}$ or ${\cal T}_{st}$ and the group ${\cal F}_t$ be ${\cal A}_t$, ${\cal T}_t$ or ${\it Id}_t$.
Then $Q$ is $({\cal
F}_{st}, {\cal F}_t)$-generic if and only if it is $({\cal F}_{st}^f, {\cal F}_t)$-generic. \qed
\end{proposition}

 Although we postpone the proof of this proposition until the end of
this section, it allows us to focus on subgroups of $({\cal A}_{st}^f, {\cal A}_t)$.

We first  look at  the group $({\cal A}_{st}^f, {\cal A}_t)$ and next on its subgroups.  It will
become clear later, that the proof strategy for these groups is analogous to that for the group
$({\cal A}_{st}^f, {\cal A}_t)$.

\begin{figure}
  \centerline{\includegraphics[width=300pt]{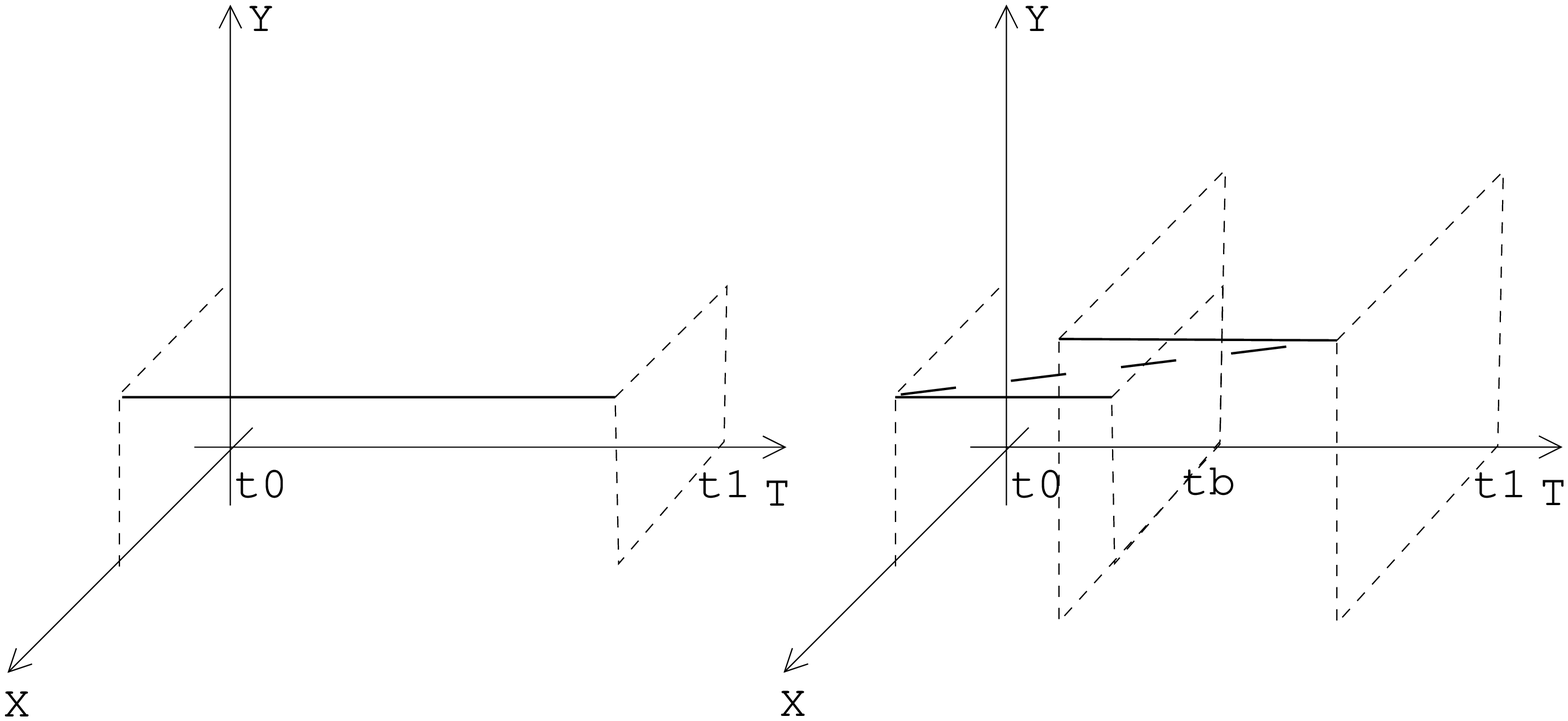}}
  \caption{The elements of $({\cal A}_{st}^{f}, {\cal A}_t)$ do not preserve betweenness of points.}\label{fignoline}
\end{figure}

It is important to note that for $({\cal A}_{st}^f, {\cal A}_t)$ and its subgroups, we cannot apply
 Theorem~\ref{metatheoremst}.   Indeed, it heavily relies on the fact that, using the predicate
$\betweens{n+1}$, it can be expressed that  $n+2$  points form an affine coordinate system for the
space $\R^n \times \R$, and also that some points represent the coordinates of another point,
relative to such an affine coordinate system  (the latter is a straightforward consequence of the
former). When using the transformation group $({\cal A}_{st}^{f}, {\cal A}_t)$ or one of its
subgroups, the predicate $\betweens{n+1}$ is too strong. Indeed, transformations of the group
$({\cal A}_{st}^{f}, {\cal A}_t)$ do not preserve ``betweenness'' in $(n+1)$-dimensional space of
points with different time coordinates.  Therefore,  the notion of collinearity in
$(n+1)$-dimensional space can  no longer be used.  Figure~\ref{fignoline} illustrates this with a
line (left) and the image of the line under some transformation $\alpha = (\alpha_{st}, \alpha_t)$
in $({\cal A}_{st}^{f}, {\cal A}_t)$ for which $\alpha_t$ is the identity function and
$\alpha_{st}$ equals the identity in the time interval $[t_0, t_b[$ and is a constant translation
of space for the interval $[t_b, t_1]$. In the left part of Figure~\ref{fignoline}, it is true that
all points different from the endpoints at time moments $t_0$ and $t_1$ lie between the endpoints.
For the right part of Figure~\ref{fignoline} this is not true (the dashed line connecting the end
points indicates all points between them.)

However, as we want our language to be able to express all first-order $({\cal A}_{st}^{f}, {\cal
A}_t)$-generic queries, somehow there needs to be a link between an  $(n+1)$-dimensional point and
its coordinates. It will become more clear later that, although we cannot express projection along
the time axis, this link can be expressed using the predicates $\betweens{n}, \ab \befores$ and  a
new predicate, $\eqcrsts$. The predicate \befores\ has already been introduced in
Section~\ref{indep}. The expression $\betweenm{n}{p,\allowbreak q,\allowbreak r}$ states, for three
points $p,\allowbreak q,r\allowbreak \in\allowbreak \R^n\times \R$, that they are co-temporal,
collinear in the space $\R^n$ and that $q$ is between $p$ and $r$. We also introduce a new $6$-ary
predicate, $\eqcrsts$. For six points $p_1, p_2, p_3, q_1, q_2, q_3 \in\allowbreak \R^n\times \R$,
$\eqcrst{p_1,\allowbreak p_2,\allowbreak p_3,\allowbreak q_1,\allowbreak q_2,\allowbreak q_3}$
expresses that the cross ratio of the three co-temporal and collinear points $p_1$, $p_2$ and $p_3$
equals the cross ratio of the time coordinates $\tau_{q_1}, \tau_{q_2}$ and $\tau_{q_3}$ of the
points $q_1$, $q_2$ and $q_3$.
 The cross ratio of three collinear points $p$, $q$, $r$ is
$\frac{\mid pq\mid}{\mid pr\mid}$, where $\mid pq\mid$ denotes the length of the line segment
between $p$ and $q$. It is well known that the cross ratio is invariant under affine
transformations.

 For example, in $\R^2\times\R$, $$\eqcrst{(0,\ab 0,\ab 0),\ab (1,\ab 1,\ab 0),\ab (2,\ab 2,\ab 0),\ab (0,\ab 0,\ab 0),\ab (0,\ab 0,\ab 1),\ab (0,\ab 0,\ab 2)}$$ holds, since the former three points have a cross ratio of $\frac{\sqrt{2}}{2\sqrt{2}}$ and the latter three points have a cross ratio of $\frac{1}{2}$.

For ease of use, we will often use the predicates
 $\eqcrss$ for the cross-ratio of spatial coordinates,
and $\eqcrts$ for the cross-ratio of temporal coordinates. Both predicates can be expressed
using $\eqcrsts$:

$$\displaylines{\qquad\eqcrs{u_1, u_2, u_3, v_1, v_2, v_3}
:= (\exists w_1)(\exists w_2)(\exists w_3)\hfill{} \cr \hfill{}(\eqcrst{u_1, u_2, u_3, w_1, w_2,
w_3} \land \eqcrst{v_1, v_2, v_3, w_1, w_2, w_3}),\qquad}$$ and $$\displaylines{\qquad\eqcrt{u_1,
u_2, u_3, v_1, v_2, v_3} := (\exists w_1)(\exists w_2)(\exists w_3)\hfill{} \cr
\hfill{}(\eqcrst{w_1, w_2, w_3, u_1, u_2, u_3} \land \eqcrst{w_1, w_2, w_3, v_1, v_2,
v_3}).\qquad}$$

Next, we present the main theorem of this section. The proof is
composed of three lemmas, as explained below.

\begin{theorem}\label{finite-fo}\rm
 Let $\sigma$ be a spatio-temporal database schema.
The language $\fo(\{\betweens{n}, \ab \befores,\ab \eqcrsts \},\sigma )$   is sound and complete
for the $({\cal A}_{st}^{f}, {\cal A}_t)$-generic spatio-temporal queries that are  expressible in
\fopr{\overline{\sigma}}. \qed
\end{theorem}

For the remainder of this section,  we will assume that $\Pi$ denotes the set $\{\betweens{n},\ab
\befores,\ab \eqcrsts \}$, unless stated otherwise.

We prove this theorem  by  three lemmas. First, the soundness is addressed in
Lemma~\ref{soundness}.   Next,  we prove completeness in two steps: Lemma~\ref{conv-fo} shows that
every  \foprs-formula can be converted into a \fo($\Pi,\sigma$)-formula, parameterized by a set of
 coordinate systems  and  Lemma~\ref{completeness} shows then that every $({\cal A}_{st}^{f},
{\cal A}_t)$-generic spatio-temporal query that is expressible in  \fop\  can be converted into an
equivalent query  expressible  in the language \fo($\Pi$).

\begin{lemma}\label{soundness}\rm
 Let $\sigma$ be a spatio-temporal database schema and let $n$ be the underlying dimension.
The language $\fo(\Pi,\sigma)$ is sound for the $({\cal A}_{st}^{f}, {\cal A}_t)$- generic
spatio-temporal queries expressible in \fopr{\overline{\sigma}}.  \qed
\end{lemma}

\smallskip
\par\noindent
{\bf Proof.} Soundness is proved in two  steps.  First, we show that every
$\fo(\Pi,\sigma)$-formula is  equivalently  expressible in \fopr{\overline{\sigma}}\ and afterwards
that  every $\fo(\Pi,\sigma)$-formula is  invariant  under elements of $({\cal A}_{st}^{f}, {\cal
A}_t)$. Both are proved by induction on the structure of $\fo(\Pi,\sigma)$-formulas.

\medskip
\par\noindent $\bullet$ \emph{Every $\fo(\Pi,\sigma)$-formula is expressible in \fopr{\overline{\sigma}}.} The atomic formulas of $\fo(\Pi,\sigma)$ are equality on point
variables, the pre\-di\-ca\-tes $\betweens{n}$, $\befores$, $\eqcrsts$ and formulas of the type
 $R(v_1, \ldots, v_l)$,  where $R$ is a relation name from $\sigma$, with arity $l$. We now describe,
for each of the above types of atomic formulas, how they can be translated into
 \fopr{\overline{\sigma}}.
 A point variable $v$ occurring in a $\fo(\Pi,\sigma)$-formula is translated into real variables $x_1^v,\ldots, x_n^v,t^v$.
 Equality between two point variables is then expressed in \fopr{\overline{\sigma}}\ by requiring that all
corresponding coordinates of the two
point variables
are equal.

We already know that the predicate $\befores$ is expressible in  \fop.  The pre\-di\-ca\-te
$\betweens{n}$ is translated in a similar way as $\betweens{n+1}$,  with the additional restriction
  that the time coordinates of the variables should be the same.

 The
formula $\eqcrsts{(u_1, u_2, u_3, v_1, v_2, v_3)}$ is translated as the  conjunction of the translation of
the expression {\bf Collinear}$^n(u_1, u_2, u_3)$, which is equal to
$$\betweens{n}(u_1,
u_2,u_3) \vee \betweens{n}(u_2,u_1,u_3) \vee \betweens{n}(u_1,
u_3,u_2)$$
and
 the formula
 $$(t^{v_3}-t^{v_1})^2 \sum_{i=1}^{n}(x_i^{u_1}-x_i^{u_2})^2=(t^{v_2}-t^{v_1})^2 \sum_{i=1}^{n}(x_i^{u_1}-x_i^{u_3})^2.$$
We translate formulas of the type $R(v_1, \ldots, v_l)$, where $R$ is a relation name from $\sigma$
with arity $l$, by the formula
 $\overline{R}(x_{1}^{v_1}, \ldots,x_{n}^{v_1}, t^{v_1},\ldots, x_{1}^{v_l}, \ldots,x_{n}^{v_l}, t^{v_l})$.

Compositions of atomic formulas by logical  connectives  and quantifiers are translated in a
natural way.

\medskip
\par\noindent  $\bullet$ \emph{Every $\fo(\Pi,\sigma)$ formula is invariant for elements of the group $({\cal A}_{st}^{f}, {\cal A}_t)$.} The only non-trivial part  here
is showing that all point predicates are  $({\cal A}_{st}^{f}, {\cal A}_t)$-invariant.   The
predicate $\befores$ is invariant for all transformations $f=(f_1,f_2,\ldots, f_n,f_t)$, that map
$\R^n\times \R$ to $\R^n\times \R $, such that $f_t$ is a strictly monotone increasing bijection of
$t$ alone (Proposition~\ref{orthotime}).  Since all elements of ${\cal A}_t$ are such bijections,
this condition is satisfied for $({\cal A}_{st}^{f}, {\cal A}_t)$.  It is well known that
affinities preserve the cross-ratio of three points. Because the predicate $\betweens{n}$ requires
its parameters to be co-temporal (which is preserved by elements of $({\cal A}_{st}^{f}, {\cal
A}_t)$),  these co-temporal points   will be transformed by the same affinity and hence their
cross-ratio is preserved. Also the predicate $\eqcrsts$ is invariant under elements of $({\cal
A}_{st}^{f}, {\cal A}_t)$, because the group ${\cal A}_{st}^{f}$ preserves the cross-ratio between
the spatial coordinates of co-temporal points and the group ${\cal A}_t$ preserves the cross-ratio
between time coordinates. \qed

 We now show that every  \foprs-formula can be converted into a \fo($\{\betweens{n}, \ab \befores, \ab
\eqcrsts \},\ab \sigma$) formula, which is parameterized by a finite set of coordinate systems.

A coordinate
system in a $n$-dimensional hyperplane of $\R^n\times \R$, orthogonal to the time axis will be referred to as a \emph{spatial
coordinate system} and a coordinate system on the time-axis will be referred to as a \emph{temporal coordinate
system}.

If $p,q$ and $r$ are collinear points in $\R^n\times \R$, then we denote by $\frac{\vec{pq}}{\vec{pr}}$
 the real number $\alpha$ such that $\vec{pq}=\alpha\vec{pr}$.

\begin{lemma}\label{conv-fo}\rm
 Let $\sigma$ be a spatio-temporal database schema and let the underlying dimension be $n$. For every
\fopr{\overline{\sigma}}-formula
$$\overline{\psi}(x_1,x_2,\ldots,x_m,t_1,\ldots, t_l),$$   there exists a $\fo(\Pi,
\sigma)$-formula
$$\psi(u_{t_O},u_{t_E}, u_{0,0}, u_{0,1},\ldots, u_{0,n}, \ldots, u_{l,0}, u_{l,1}, \ldots, u_{l, n},v_1, v_2, \ldots, v_k),$$ where $l$  is the number of variables
 occurring in the formula  that refer to a time dimension and where $k$ is the total number of free variables of $\overline{\psi}$, i.e., $k=m+l$.

Furthermore, for each spatio-temporal
database
 $\st{}$
 over $\sigma$, for each set of spatial coordinate systems
$\nbasisi$, $i=0,\ldots, l$ of the spatial component of $\R^n\times \R$, for each temporal coordinate
system $(p_{t_O},p_{t_E})$ of the temporal component of $\R^n\times \R$,
and for all points $q_1,q_2,\ldots,q_k$ on the line $p_{0,0}p_{0,1}$:
 $$\displaylines{\quad (\R^n\times\R, \Pi^{\R^n\times\R},\st{}) \models \psi[p_{t_O},p_{t_E}, p_{0,0}, p_{0,1},\ldots, p_{0,n}, \ldots, \splits p_{l,0}, p_{l,1}, \ldots, p_{l, n},q_1, q_2, \ldots, q_k]\quad }$$
if and only if $$(\R, +,\times, <,0,1,\alpha(\overline{\st{}}) \models
\overline{\psi}[\frac{\overrightarrow{p_{0,0}q_1}}{\overrightarrow{p_{0,0}p_{0,1}}},\frac{\overrightarrow{p_{0,0}q_2}}{\overrightarrow{p_{0,0}p_{0,1}}},\ldots,\frac{\overrightarrow{p_{0,0}q_k}}{\overrightarrow{p_{0,0}p_{0,1}}}],$$
 where  $\alpha=(\alpha_{st},\alpha_t) $  is an element of $({\cal A}_{st}^{f}, {\cal A}_t)$
such that $(p_{0,0},\ldots,p_{0,n})$ is mapped by $\alpha_{st}$ onto the standard spatial
coordinate system in the  hyperplane $\R^n\times \{(0,\ldots,0,0)\}$ of $\R^n\times \R$, and each
spatial coordinate system $\nbasisi(i=1\ldots,l)$ is mapped on the standard coordinate system in
the hyperplane at time $\R^n\times \{\alpha(\tau_{p_{i,0}})\}$ where the temporal part $\alpha_t$
of $\alpha$ is the unique time-affinity which maps $\tau_{p_O}$ to $0$ and $\tau_{p_E}$ to $1$.
\qed
\end{lemma}
\smallskip
\par\noindent
{\bf Proof.}
 Let $\overline{\psi}$ be a \fopr{\overline{\sigma}}-formula.  We assume that $\overline{\psi}$ is
in prenex normal form. We now describe the translation of $\overline{\psi}$ into a formula $\psi$
of $\fo(\Pi, \sigma)$  (by induction on its structure). In this translation, first the
quantifier-free part of $\overline{\psi}$ is translated  and the quantifiers are later added in the
obvious  way.

 To start with, a 2-dimensional ``computation plane'' is chosen that is used to simulate real variables and constants and all the polynomial equations, polynomial equalities and inequalities.

\par\noindent  $\bullet$ \emph{The choice of a computation plane.} First of all, two moments in  time $u_{t_O}$
and $u_{t_E}$ (time moments are simulated  in $\psi$ by variables  in $\R^n \times \R$) are chosen
such that $\neg \before{u_{t_E},u_{t_O}}$. They form a temporal coordinate system; the formula
describing this is as follows:  $${\bf TCoSys}_{{\cal A}_t}(u_1, u_2) := \neg \before{u_2, u_1}.$$
Next, in the hyperplane of points co-temporal with  $u_{t_O}$, $n+1$ points $u_{0,\ab 0},\ab  \ab
u_{0,\ab 1}, \ab  \ldots \ab,\ab u_{0,\ab n}$ are chosen such that they form an affine coordinate
system for the hyperplane co-temporal with $u_{t_O}$.   The predicate ${\bf CoSys}_{\cal A}^n$,
expressing  this, is similar the the previously introduced predicate ${\bf CoSys}_{\cal A}$ (see
the proof of Theorem~\ref{meta-cor-fo}),  except that some constraints are added that express that
the points should be co-temporal.

 As the variables  $u_{t_O}$, $u_{t_E}$, $u_{0,0}, \ab u_{0,1}, \ldots \ab,u_{0,n}$ represent
arbitrary points (up to the mentioned restrictions), they  parameterise the translation of
$\overline{\psi}$. To start with, $\psi$ will contain the subformula $\psi_{\text{comp}}$, defined
as

$$\displaylines{\qquad\psi_{\text{comp}}(u_{t_O}, u_{t_E}, u_{0,0}, u_{0,1}, \ldots, u_{0,n}) := {\bf
TCoSys}_{\cal A}(u_{t_O}, u_{t_E}) \hfill{} \cr  \hfill{} \land {\bf CoSys}_{\cal A}^n(u_{0,0},
u_{0,1}, \ldots, u_{0,n}) \land {\bf Cotemp}(u_{t_O},u_{0,0}),\qquad}$$
as a conjunct.

 We will use the 2-dimensional plane through the points $u_{0,0},u_{0,1}$ and $u_{0,2}$
as a ``computation plane''. The idea is that we will simulate real variables and constants by
points on the line through $u_{0,0}$ and $u_{0,1}$ and that addition and multiplication of real
terms are simulated by $\fo(\Pi)$ expressions in the plane through  $u_{0,0},u_{0,1}$ and
$u_{0,2}$.
\medskip
\par\noindent  $\bullet$ \emph{The translation of terms and atomic formulas.}
  A quantifier-free \foprs-formula may contain  the following
 terms and atomic subformulas: real variables; the constants $0$ and $1$; polynomial constraints; and
relation predicates where the relation names from $\overline{\sigma}$ are used.  We translate each
separately.
\medskip
\par\noindent  $-$ \emph{The translation of real variables.}
Each real variable $x$ appearing in the formula $\overline{\psi}$ is translated into a spatio-temporal variable
$v$. Also, $\psi$ will contain a conjunct
$$\psi_{\rm var}(v):={\bf Collinear}^n(u_{0,0}, u_{0,1}, v),$$
expressing that $v$ is in the computation plane on the line  connecting $u_{0,0}$ and $u_{0,1}$.
The idea is that a real variable $x$ taking concrete value $a$, is simulated by requiring that
$v$ is such that $\frac{\overrightarrow{u_{0,0}v}}{\overrightarrow{u_{0,0}u_{0,1}}}$ equals $a$.

\par\noindent  $-$ \emph{The translation of the constants $0$ and $1$.} The real constants $0$ and $1$ that may appear in $\overline{\psi}$ are translated into
$u_{0,0}$ and $u_{0,1}$ respectively.

\par\noindent  $-$ \emph{The translation of polynomial constraints.}
The arithmetic operations (addition and multiplication) on real terms will be simulated in the
computation plane $(u_{0,0},u_{0,1},u_{0,2})$ It was shown by Tarski~\cite{schwab-tarski} (the
results of Tarski were also used in~\cite{gvv-jcss}) that all arithmetic operations on points that
are located on the line through $u_{0,0}$ and $u_{0,1}$ can be simulated in the plane
$(u_{0,0},u_{0,1},u_{0,2})$ using only the construct $\betweens{n}$. Hence, a  subformula $p(x_1,
\ldots, x_m)>0$, with $p$ a polynomial with integer coefficients, using the translation of the real
variables $x_1, \ldots, \ab x_m$ in point variables $v_1, \ldots, \ab v_m$, is translated into
$\psi_{\text{poly}}(u_{0,0}, \ab u_{0,1}, \ab u_{0,2}, \ab v_1,\ldots, \ab v_m)$, defined using the
predicate $\betweens{n}$.

 The correctness of the three above translations can be demonstrated as that of the similar translations in
~\cite{gvv-jcss}.

\par\noindent  $-$ \emph{The translation of relation predicates.}
A subformula of $\overline{\psi}$  of type
$\overline{R}(x_{1,1},\ab \ldots, \ab x_{1,n},\ab x_{1,t},\ab
\ldots,\ab x_{m,1},\ab \ldots, \ab x_{m,n},\ab x_{m,t})$, where $\overline{R} \in \overline{\sigma}$ and
where $m$ is the arity of $R$ in $\sigma$,
is translated into a formula $$R(v_1,\ldots, v_m)$$
and  $\psi$ has a conjunct expressing that the point variables
 $v_{1,1},\ab \ldots, \ab v_{1,n},\ab v_{1,t},\ab
\ldots,\ab v_{m,1},\ab \ldots, \ab v_{m,n},\ab v_{m,t}$, that are the translations of $x_{1,1},\ab
\ldots, \ab x_{1,n},\ab x_{1,t},\ab \ldots,\ab x_{m,1},\ab \ldots, \ab x_{m,n},\ab x_{m,t}$, are
the coordinates of $v_1,\ab \ldots, \ab v_m$ respectively.    For the moment, we assume that the
variables $x_{i,t}$ and $x_{j,t}$ are different for $1\leq i<j\leq m$  and later show how to deal
with the general case.
 Indeed, recall
 that each variable $x_{i,j}$ $(1\leq i \leq m, 1\leq j \leq n)$ and
$x_{i,t}$ $(1\leq i \leq m)$ is already translated into a point variable $v_{i,j}$ and $v_{i,t}$,
which are all collinear with $u_{0,0}$ and $u_{0,1}$. To express  the  link between the coordinates
of point variables $v_1,\ldots,v_m$ and the point variables $v_{i,j}$ and $v_{i,t}$, we proceed as
follows. We associate with each point variable $v_i$ ($1\leq i\leq m$) the following set of point
variables:
\begin{enumerate}
\item $n+1$ point variables $u_{i,0},\ldots,u_{i,n}$ representing an $n$-dimensional coordinate
system which is co-temporal with $v_i$; and \item $n$ point variables $v_{i,j}'$ which are
collinear with $u_{i,0}$ and $u_{i,1}$, such that $v_{i,j}'$ represents the $j$th coordinate of
$v_i$ with respect to the coordinate systems specified by $u_{i,0},\ldots,u_{i,n}$, and such that
the coordinate of $v_{i,j}'$, on the line  through $u_{i,0}$ and $u_{i,1}$, gives the same cross
ratio with respect to these points  as the coordinate of $v_{i,j}$, on the line through $u_{0,0}$
and $u_{0,1}$, gives with respect to these points, i.e.,
$\frac{\overrightarrow{u_{i,0}v'_{i,j}}}{\overrightarrow{u_{i,0}u_{i,1}}}=
\frac{\overrightarrow{u_{i,0}v_{i,j}}}{\overrightarrow{u_{0,0}u_{0,1}}}.$ \end{enumerate}

As explained before, the first set of $n+1$ point variables can be defined using the expression
$${\bf
CoSys}_{\cal A}^n(u_{i,0}, u_{i,1}, \ldots, u_{i,n}) \land {\bf Cotemp}(u_{i,0}, v_i).
$$
For the second set of $n$ point variables, we first observe that from~\cite{gvv-jcss}, we know that
we can express, using $\betweens{n}$, that $n$ point variables $v_{i,1}', \ldots, v_{i,n}'$
represent the spatial coordinates of the point variable $v_i$ relative to a chosen spatial
coordinate system (in this case, the  coordinate  system specified by $u_{i,0},\ldots,u_{i,n}$). In
order to establish the link between the point variables $v_{i,j}'$ in the plane specified by
$u_{i,0},\ldots,u_{i,n}$ and the point variables $v_{i,j}$ in the computation plane we need to use
the predicate $\eqcrss$. The predicate $\eqcrss$ performs a transformation between the affine
coordinate systems at two different time moments, and so connects each $v'_{i,j}$ to a $v_{i,j}\ (i
= 1, \ldots, m, j = 1, \ldots, n)$. Remark that all $v'_{i,j}$ are collinear with $u_{i,0}$ and
$u_{i, 1}$, and that all $v_{i,j}$ are collinear with $u_{0,0}$ and $u_{0,1}$.  Therefore,
$\eqcrss$ can be used to express this equality of cross ratios.

Until now, we only considered the spatial coordinates. To link the temporal variables $v_{i,t}$ to
the temporal coordinate of $v_i$, we use the expression $\eqcrsts(u_{0,0},\ab u_{0,1},\ab v_{i,t},
\ab u_{t_O}, \ab u_{t_E},\ab v_i)$. Recall that the predicate $\eqcrsts$  can be used to relate the
cross ratio of points on the time axis to the cross ratio of points, representing coordinates on
the line through $u_{0,0}$ and $u_{0,1}$,   and thus connects each $v_i$ to a $v_{i,t}$ ($i = 1,
\ldots, m$).

Putting everything together results in the expression $\psi_{\text{rel}}$:
\begin{multline*}
(\exists v_1)(\exists v_2)\ldots (\exists v_m) (
R(v_1, v_2, \ldots, v_m)  \wedge \bigwedge_{i=1}^{m}{\bf CoSys}_{\cal
A}^n(u_{i,0}, u_{i,1}, \ldots, u_{i,n}) \\ {}\land
 \bigwedge_{i=1}^{m}{\bf
Cotemp}(u_{i,0}, v_i)  \land (\exists v'_{1,1}) \ldots (\exists v'_{1,n}) \ldots
(\exists v'_{m,1}) \ldots (\exists v'_{m,n})\\(
\bigwedge_{i= 1}^{m}{\bf Coordinates}^n(u_{i,0}, u_{i,1},
\ldots, u_{i,n}, v'_{i,1}, \ldots , v'_{i,n} , v_i )\\
{} \land \bigwedge_{i=1}^m\bigwedge_{j=1}^{n}\eqcrss(u_{0,0}, u_{0,1}, v_{i,j}, u_{i,0}, u_{i,1},
v'_{i,j}) \\
{}\land
\bigwedge_{i=1}^{m}\eqcrsts(u_{0,0}, u_{0,1}, v_{i,t}, u_{t_O}, u_{t_E}, v_i)))
\end{multline*}
where ${\bf Coordinates}^n(u_{i,0},
\ldots, u_{i,n}, v'_{i,1}, \ldots , v'_{i,n} , v_i )$ expresses for each $(1\leq j\leq n)$ that
$v_{i,j}'$ is represents the $j$th coordinate of $v_i$ with respect to the coordinate
systems specified by $u_{i,0},
\ldots, u_{i,n}$.

We now show the correctness of the above  translation  of a relation predicate. We have to prove
that for each spatio-temporal database $\st{}$,  and for any points $p_{t_O}, p_{t_E},\ab
p_{0,0},\ldots,p_{0,n},\ldots,\ab p_{m,0},\ldots,p_{m,n},\ab q_{1,1},\ldots, q_{1,n},q_{1,t},\ab
\ldots, \ab q_{m,1},\ab \ldots,\ab q_{m,n},q_{m,t}$:

 $$
\displaylines{\quad
(\R^n\times\R,\Pi^{\R^n\times\R},\st{}) \models \psi_{\text{rel}}[p_{t_O},
p_{t_E},\ab p_{0,0},\ldots,p_{0,n},\ldots,\ab p_{m,0},\ldots,p_{m,n},\splits q_{1,1},\ldots,
q_{1,n},q_{1,t},\ab \ldots, \ab q_{m,1},\ldots, q_{m,n},q_{m,t}]\quad\cr}
$$

if and only if
$$\displaylines{\quad
(\R,+,\times,0,1,\alpha(\overline{\st{}})) \models \overline{R}[
\frac{\overrightarrow{p_{0,0}q_{1,1}}}{\overrightarrow{p_{0,0}p_{0,1}}},
\ldots,
\frac{\overrightarrow{p_{0,0}q_{1,n}}}{\overrightarrow{p_{0,0}p_{0,1}}},
\frac{\overrightarrow{p_{0,0}q_{1,t}}}{\overrightarrow{p_{0,0}p_{0,1}}} ,
\ldots,\splits
\frac{\overrightarrow{p_{0,0}q_{m,1}}}{\overrightarrow{p_{0,0}p_{0,1}}},
\ldots,
\frac{\overrightarrow{p_{0,0}q_{m,n}}}{\overrightarrow{p_{0,0}p_{0,1}}},
\frac{\overrightarrow{p_{0,0}q_{m,t}}}{\overrightarrow{p_{0,0}p_{0,1}}} ],
\quad\cr}
$$
where  $\alpha=(\alpha_{st},\alpha_t)\in ({\cal A}_{st}^{f}, {\cal A}_t)$  is the affinity which
maps $(p_{0,0},\ldots,p_{0,n})$ to the spatial standard basis at time $\tau_0=0$,
$(p_{i,0},\ldots,p_{i,n})$ to the spatial standard basis at time $\tau_i=\alpha(\tau_{p_{i,0}})$,
where  $\alpha_t$ is uniquely determined on the time axis by $\alpha_t(\tau_{p_O})=0$ and
$\alpha_t(p_{t_E})=1$. Note that by assumption, $x_{i,t}\neq x_{j,t}$ for $(1\leq i < j < m)$ and
hence also $\tau_{p_{i,0}}$ and $\tau_{p_{j,0}}$, and consequently $\tau_i\neq \tau_j$ for $(1\leq
i < j <m)$. This condition is essential  to ensure that $\alpha_t$ exists and is well defined.
Indeed, suppose that there exists an $i$ and $j$ such that $\tau_{p_{i,0}}=\tau_{p_{j,0}}$ and
hence $\tau_i=\tau_j$. Then we would require that $\alpha$ maps two possibly different co-temporal
coordinate systems $(p_{i,0},\ldots,p_{i,n})$ and $(p_{j,0},\ldots,p_{j,n})$ the same standard
basis.  This can clearly not be done by a $({\cal A}_{st}^{f}, {\cal A}_t)$-generic query.

We know that the formula $\psi_\text{rel}$ is true for the points $p_{t_O}, p_{t_E},\ab
p_{0,0},\ldots,p_{0,n},\ab \ldots,\ab p_{m,0},\ldots,\ab p_{m,n},\ab q_{1,1},\ldots,
q_{1,n},q_{1,t},\ab \ldots, \ab q_{m,1},\ab \ldots,\ab q_{m,n},q_{m,t}$ if and only if there exist
points $p_1,\ldots,p_m, \ab q_{1,1}',\ldots, q_{1,n}',\ab \ldots,\ab q_{m,1}',\ab \ldots,\ab
q_{m,n}'$ such that for each $i=1,\ldots,m$:

 \begin{equation}\label{eqproof1} \overrightarrow{p_{0,0}p_i} = \overrightarrow{p_{i,0}p_i}+\overrightarrow{p_{0,0}p_{i,0}}   =
 \sum_{j=1}^n
\frac{\overrightarrow{p_{i,0}q_{i,j}'}}{\overrightarrow{p_{i,0}p_{i,j}}}\overrightarrow{p_{i,0}p_{i,j}} +\overrightarrow{p_{0,0}p_{i,0}},
\end{equation} and the following equations hold:
\begin{equation}\label{eqproof2}
\frac{\overrightarrow{p_{i,0}q_{i,j}'}}{\overrightarrow{p_{i,0}p_{i,1}}} =
\frac{\overrightarrow{p_{0,0}q_{i,j}}}{\overrightarrow{p_{0,0}p_{0,1}}}, \quad 1\leq j \leq n,
\end{equation}
\begin{equation}\label{eqproof4}
\frac{\tau_{p_i} - \tau_{p_{t_O}}}{\tau_{p_{t_E}} - \tau_{p_{t_O}}} =
\frac{\overrightarrow{p_{0,0}q_{i,t}}}{\overrightarrow{p_{0,0}p_{0,1}}}.
\end{equation}

 Using Equation~(\ref{eqproof2}), Equation~(\ref{eqproof1}) is equivalent to
 \begin{equation}\label{eqproof5} \overrightarrow{p_{0,0}p_i} =
\sum_{j=1}^n \frac{\overrightarrow{p_{0,0}q_{i,j}}}{\overrightarrow{p_{0,0}p_{0,1}}}\overrightarrow{p_{i,0}p_{i,j}} +\overrightarrow{p_{0,0}p_{i,0}}.
\end{equation}
Considering the fact
that $\alpha$ is a linear transformation, and using equation~(\ref{eqproof5}), the following holds:

 $$ \alpha(\overrightarrow{p_{0,0}p_i}) = \sum_{j=1}^n
\frac{\overrightarrow{p_{0,0}q_{i,j}}}{\overrightarrow{p_{0,0}p_{0,1}}}\alpha(\overrightarrow{p_{i,0}p_{i,j}})+\alpha(\overrightarrow{p_{0,0}p_{i,0}}).
$$
Moreover, let $e_i(\tau)$ be the $i$th vector of the standard spatial basis at time $\tau$ and denote
by $e_i=e_i(0)$. We then have
 \begin{equation}\label{eqproof6}
\alpha(\overrightarrow{p_{0,0}p_i}) =  \sum_{j=1}^n
\frac{\overrightarrow{p_{0,0}q_{i,j}}}{\overrightarrow{p_{0,0}p_{0,1}}}\overrightarrow{e_0(\tau_i)e_j(\tau_i)}+\overrightarrow{e_0e_0(\tau_i)}.
\end{equation}
As equation~(\ref{eqproof4}) is invariant under elements of $({\cal A}_{st}^{f}, {\cal A}_t)$, we
also have that
\begin{equation}\label{eqproof7}
\frac{\alpha(\tau_{p_i}) - \alpha(\tau_{p_{t_O}})}{\alpha(\tau_{p_{t_E}}) - \alpha(\tau_{p_{t_O}})} =
\frac{\tau_i - 0}{1 - 0} = \tau_i =
\frac{\overrightarrow{p_{0,0}q_{i,t}}}{\overrightarrow{p_{0,0}p_{0,1}}}.
\end{equation}

So we have that:
$$
\alpha(\overrightarrow{p_{0,0}p_i}) =\sum_{j=1}^n
\frac{\overrightarrow{p_{0,0}q_{i,j}}}{\overrightarrow{p_{0,0}p_{0,1}}}\overrightarrow{e_0(\tau_i)e_j(\tau_i)}+\frac{\overrightarrow{p_{0,0}q_{i,t}}}{\overrightarrow{p_{0,0}p_{0,1}}}\overrightarrow{e_0e_{n+1}}.
$$

Since all standard bases  $(e_0(\tau_i),\ldots,e_{n}(\tau_i))$  are parallel along the time axis,
we have that
$$
\alpha(\overrightarrow{p_{0,0}p_i}) =
\sum_{j=1}^n
\frac{\overrightarrow{p_{0,0}q_{i,j}}}{\overrightarrow{p_{0,0}p_{0,1}}}\overrightarrow{e_0e_j} +
\frac{\overrightarrow{p_{0,0}q_{i,t}}}{\overrightarrow{p_{0,0}p_{0,1}}}\overrightarrow{e_0e_{n+1}}.
$$

This completes the correctness proof for the conversion of relational predicates.
\medskip

\par\noindent  $\bullet$ \emph{The translation of composed formulas.}   When all the
atomic subformulas of $\overline{\psi}$ have been translated as described above, the logical
connectives can be added  in a natural way. We assume that two atomic formulas $\overline{\chi}_1$
and $\overline{\chi}_2$ are translated already,  into   $\chi_1$ and $\chi_2$. The translations of
$\overline{\chi}_1\land\overline{\chi}_2$ and $\overline{\chi}_1\vee\overline{\chi}_2$ are
$\chi_1\land\chi_2$ and $\chi_1\vee\chi_2$, respectively. The formula $\neg\overline{\chi}_1$ is
translated into $\neg\chi_1$.

Remember that with the conversion of a formula $\overline{\psi}_{\text{rel}}$ of type
$\overline{R}(x_{1,1},\ldots, \ab x_{1,n},\ab x_{1,t}, \ldots,\ab x_{m,1},\ldots, \ab x_{m,n},\ab
x_{m,t})$ we assumed that $x_{i,t}\neq x_{j,t}$ for any $(1\leq i < j < m)$. The reason is that we
want to have only one affine coordinate system for every different time moment considered in that
formula. Indeed, an element $\alpha$ of $({\cal A}_{st}^{f}, {\cal A}_t)$ is a one-to-one mapping
from the snapshots of a certain input database  $\st{}$ to the snapshots of the output database
$\alpha(\st{})$. Therefore, we cannot map two different co-temporal coordinate systems to the same
standard coordinate system using such an affinity.

Suppose now that $x_{i,t}=x_{j,t}$ for some $(1\leq i < j \leq m)$. Then we adapt the previous
translation  with the extra requirement that $v_{i,k}=v_{j,k}$  for $k=0,\ldots,n$ and we have
unique coordinate system for each point  occurring in time.

When translating an
 \foprs-formula  $\overline{\psi}$, it is in general not known in advance which time coordinates are equal
(this may depend on the input database; and it is undecidable in general which time coordinates are
equal in an \foprs-formula). To circumvent this problem, we consider all possible orders (using
$\befores$) of the time variables of $\overline{\psi}$ (a real variable denoting a time moment is
recognized as it appears on the $i(n+1)$-th place  ($i = 1, \ldots , m$) in the argument list of a
spatio-temporal relation predicate) and take the disjunction over all possible orders of these time
variables. We denote the set of all possible orders by $P$.

For each $\rho \in P$ the formula $\psi_{\rho}$ is the translation of $\overline{\psi}$ taken the
(in)equalities into account  according to the order of the time variables corresponding to $\rho$.
Hence, each $\psi_\rho$ formula can have a different number $\ell_\rho$ of free variables,
depending on $\rho$. We denote by $\ell$ the total number of free variables across all formulas
$\psi_\rho$, $\rho\in P$.

When connecting several subformulas, the same principle has to be used, as arithmetic  subformulas
  can impose equality on different time variables.

When applying the thus obtained translation of the quantifier-free part of $\psi$ to a
spatio-temporal database instance, only some of the disjuncts will apply  (possibly  depending on
the particular input database).

\medskip
\par\noindent  $\bullet$ \emph{Formulas with quantifiers.}  Finally, the quantifier prefix of $\overline{\psi}$ is translated in the natural way.
Suppose that we already translated the quantifier-free formula $\overline{\chi}$ into the formula
$\chi$. Then the translation of $(\exists x)(\overline{\chi})$ is  $(\exists v)( \chi)$, where $v$
is the point variable associated to $x$ for which we have already declared ${\bf
Collinear}^n(u_{0,0},u_{0,1},v)$. \qed

\bigskip
\begin{lemma}\label{completeness}\rm
 Let $\sigma$ be a spatio-temporal database schema. For every $({\cal A}_{st}^{f}, {\cal A}_t)$-generic spatio-temporal query expressible in  \foprs,
there exists an equivalent $\fo(\{\betweens{n}, \befores, \eqcrsts \},\sigma)$-query.   \qed
\end{lemma}

\medskip
\par\noindent
{\bf Proof.} Given a $({\cal A}_{st}^{f}, {\cal A}_t)$-generic spatio-temporal query
 of output type
$(n,k)$,  expressible in \foprs,  $$ \overline{\psi}(x_{1,1},\ldots,x_{1,n},x_{1,t}, \ldots \ab
,x_{k,1}, \ldots \ab ,x_{k,n},x_{k,t}).$$

The conversion procedure,  given in Lemma~\ref{conv-fo}, returns   a formula
  $$\displaylines{\qquad\psi(u_{t_O},u_{t_E},\ab u_{0,0},\ldots, u_{0,n},u_{1,0}, \ldots, u_{1,n}, \ldots, u_{k,0}, \ldots,
 u_{k,n},\hfill{} \cr  \hfill{}
 v_{1,1}, \ldots \ab ,v_{1,n},v_{1,t}, \ldots \ab ,
 v_{k,1},\ldots \ab ,v_{k,n},v_{k,t}),\qquad}$$  parameterized by one temporal and $k$ spatial
 coordinate systems and which is, up to a transformation of the group $({\cal A}_{st}^{f}, {\cal
 A}_t)$,  that depends on the coordinate systems,  equivalent to the original formula $\overline{\psi}$. Since it has additional free variables, the query $\psi$ clearly has the wrong output type. A $\fo(\Pi,\sigma)$-query equivalent to $\overline{\psi}$
should be a formula $$\psi_{\text{final}}(v_1,v_2,\ldots,v_k)$$ having $k$ free variables only. We
obtain the desired  formula   by introducing $k$ new point variables $v_i$, and for each $1\leq
i\leq k$, $n$ new point variables $v_{i,1}',\ldots,v_{i,n}'$ such that $v_{i,j}'$ is collinear with
$u_{i,0}$ and $u_{i,1}$ and  $$ {\bf Coordinates}^n(u_{i,0}, \ldots,, u_{i,n},
v_{i,1}',\ldots,v_{i,n}',v_i).\eqno{(1)}
$$
Moreover, we require that $$\eqcrsts(u_{0,0}, u_{0,1}, v_{i,t}, u_{t_O}, u_{t_E}, v_i)\eqno{(2)}$$
and
$$
\bigwedge_{j=1}^n \eqcrss(u_{0,0}, u_{0,1}, v_{i,j}, u_{i,0}, u_{i,1},
v_{i,j}').\eqno{(3)}
$$
The final formula $\psi_{\text{final}}$ is now obtain by   existentially  quantifying all point
variables, except  for $v_1,\ldots,v_k$ in the conjunction of $\psi$ with the expressions (1), (2)
and (3).

  Now consider the  (partial) output of $\psi_{\text{final}}$ when we  choose   a specific
coordinate system for each set of  variables $u_{i,0},\ldots,u_{i,n}$. By similar reasoning as in
 Lemma~\ref{conv-fo}, we obtain that this partial output equals
 $$\alpha'^{-1}(\overline{\psi}(\alpha(\overline{\st{}})))$$ where $\alpha' = (\alpha'_{st}, \alpha'_t)$ and
$\alpha = (\alpha_{st}, \alpha_t)$ both are transformations as specified in the statement of
Lemma~\ref{conv-fo}.  This means that they both satisfy the same set of constraints, i.e.,
$\alpha'_t = \alpha_t$ and for certain time moments $\tau$, $\alpha'_{st}(\tau) =
\alpha_{st}(\tau)$. In between those time moments $\alpha'_{st}$ and $\alpha_{st}$ can differ.
However, it follows from Lemma~\ref{conv-fo} that  $\overline{\psi}(\alpha(\overline{\st{}})) =
\overline{\psi}(\alpha'(\overline{\st{}}))$,  for any two transformations $\alpha$ and $\alpha'$
satisfying the constraints as described in the statement of Lemma~\ref{conv-fo}. Hence, we can
conclude without loss of generality that the partial output of $\psi_{\text{final}}$ when we fill
in a specific coordinate system for each set of variables $u_{i,0},\ldots,u_{i,n}$ equals
$\alpha^{-1}(\overline{\psi}(\alpha(\overline{\st{}})))$  where $\alpha$ is a transformation as
specified in the statement of Lemma~\ref{conv-fo}.

If we now consider all possible coordinate systems for each set of variables
$u_{i,0},\ldots,u_{i,n}$
 $$\psi_{\text{final}}(\st{}) = \bigcup_{c}\bigcup_{\alpha_c}(\alpha_c^{-1}(\overline{\psi}(\alpha_c(\overline{\st{}})))),$$  where $c$ ranges over all possible coordinate system assignments and $\alpha_c$
ranges over all transformations satisfying the constraints following from this choice of coordinate
systems.

The union
$\bigcup_{c}\bigcup_{\alpha_c}(\alpha_c^{-1}(\overline{\psi}(\alpha_c(\overline{\st{}}))))$  is in
fact the union over all elements $\alpha$ of $({\cal A}_{st}^{f}, {\cal A}_t)$ of
$\alpha^{-1}(\overline{\psi}(\alpha(\overline{\st{}})))$. So,  $$\psi_{\text{final}}(\st{}) =
\bigcup_{\alpha}(\alpha^{-1}(\overline{\psi}(\alpha(\overline{\st{}})))),$$  where $\alpha$ ranges
over all elements of $({\cal A}_{st}^{f}, {\cal A}_t)$.

Since $\psi_{\text{final}}$ is a $({\cal A}_{st}^{f}, {\cal A}_t)$-generic query     and the group
$({\cal A}_{st}^{f}, {\cal A}_t)$ is semi-algebraic (we give a precise definition in
Section~\ref{subsec:53}), we have that for every $\alpha$
$\alpha^{-1}(\overline{\psi}(\alpha(\overline{\st{}}))) = \overline{\psi}(\overline{\st{}})$.

So,  finally, $$\psi_{\text{final}}(\st{}) = \overline{\psi}(\overline{\st{}}).$$ \qed

 \bigskip\par\noindent {\bf Proof of Theorem~\ref{finite-fo}.}
Lemma~\ref{soundness}, Lemma~\ref{conv-fo} and Lemma~\ref{completeness} together prove
Theorem~\ref{finite-fo}.\qed \medskip

We are now ready to prove Proposition~\ref{collapse}:

\bigskip\par\noindent {\bf Proof of Proposition~\ref{collapse}.} Note that we only consider a finite number of
moments in time in the proof of Lemma~\ref{conv-fo} (there are only a finite number of time
variables in any  \foprs-formula $\varphi$). This implies that the transformation groups ${\cal
A}_{st}^f$ and ${\cal A}_{st}$ yield the same results. So, we can use the proof given above for the
group $({\cal A}_{st},{\cal A}_t).$ Indeed, in between the moments of time that are considered, it
is indeed not important which transformation function is used.
 \qed \medskip

Theorem~\ref{finite-fo} has a number of corollaries. We need two extra point predicates, namely
$\equidistcotemps$ and $\poss{n}$ before we can state those corollaries. First,
$\equidistcotemp{\ab p\ab ,q,r,s}$ is true for four co-temporal points $p,q,r$ and $s$ if and only
if the (Euclidean) distance between $p$ and $q$ equals the distance between $r$ and $s$. Second,
the expression $\pos{n}{p_0,p_1,\ldots,p_n}$ is true for $n+1$ co-temporal points
$p_0,p_1,\ldots,p_n$ if and only if $(p_0,\ab p_1,\ab \ldots,\ab p_n)$ forms a positively oriented
coordinate system.

\begin{corollary}\label{basissen}\rm
 Let $\sigma$ be a spatio-temporal database schema. Let $({\cal F }_{st}, \allowbreak {\cal F}_t)$ and $\fo({\cal F
}_{st}, \allowbreak {\cal F}_t))$ be taken from
Table~\ref{table-point-predicates-finite}. The language
 $\PLss{\Pi({\cal F}_{st}, {\cal F}_t),\sigma }$ is sound and complete for
the $({\cal F }_{st}, \allowbreak {\cal F}_t)$-generic spatio-temporal queries that are expressible
in \foprs.\qed
\end{corollary}

\begin{table}[htbp]
  \begin{center}
    \leavevmode
    \begin{tabular}[c]{|l|l|}
\hline\hline
 \ \      $({\cal F}_{st}, {\cal F}_t)$ \ \ & \ \   Sets of point predicates $\Pi{({\cal F}_{st}, {\cal F}_t)}$\\
\hline\hline
 \ \ $({\cal
A}_{st}^{(f)}, {\cal A}_t)$\ \ & \ \   $\{\betweens{n}, \befores, \eqcrsts \}$\\
 \ \ $({\cal A}_{st}^{(f)}, {\cal
T}_t)$\ \  & \ \   $\{\betweens{n}, \befores, \eqcrsts, \unittimes \}$\\
 \ \ $({\cal A}_{st}^{(f)}, {\it
Id}_t)$\ \  & \ \   $\{\betweens{n}, \befores, \eqcrsts,
\unittimes ,{\bf 0}_{\bf t},{\bf
1}_{\bf t} \}$\\

 \ \ $({\cal S}_{st}^{(f)}, {\cal F}_t)$\ \  & \ \   $\Pi({\cal A}_{st}^{(f)}, {\cal F}_t)\cup \{\equidistcotemps\}$\\
 \ \ $({\cal I}_{st}^{(f)}, {\cal F}_t)$\ \  & \ \   $\Pi({\cal A}_{st}^{(f)}, {\cal F}_t)\cup \{\equidistcotemps, \unitdists\}$\\
  \ \ $({\cal T}_{st}^{(f)}, {\cal F}_t)$\ \  & \ \   $\Pi({\cal A}_{st}^{(f)}, {\cal F}_t)\cup \{
  \equidistcotemps, \unitdists, \leq_{\bf {\it i}} (1\leq i\leq n),\poss{n}\}$\\

\hline
    \end{tabular}
\medskip
    \caption{An overview of the different sets of  point predicate for some
    transformation groups. We have ${\cal  F}_t\in \{ {\cal  A}_t,{\cal T}_t,{\it Id}_t \}$.}
    \label{table-point-predicates-finite}
  \end{center}
\end{table}
\medskip

\par\noindent
{\bf Proof.} It follows directly from the proof of Theorem~\ref{finite-fo} that, for each subgroup
$({\cal F}_{st}^{(f)}, {\cal F}_t)$ of $({\cal A}_{st}^{(f)}, {\cal A}_t)$, the language
$\fo(\Pi,\sigma)$ is sound and complete for the $({\cal F}_{st}^{(f)}, {\cal F}_t)-$generic queries
expressible in \foprs\  if and only if the following three conditions are satisfied:

\begin{enumerate}[(i)]
    \item the set $\Pi$ contains the elements $\betweens{n}$, $\befores$ and $\eqcrsts$;
    \item all   elements of $\Pi$ are  \foprs-expressible and invariant
 under the transformations of $({\cal F}_{st}^{(f)}, {\cal F}_t)$;
    \item the facts  ``$(v_0, v_1, \ldots, v_n)$ is the image of the standard coordinate system in the hyperplane co-temporal with $v_{t_O}$ under an element of $({\cal F}_{st}^{(f)}, {\cal F}_t)$'' and
  ``$(v_{t_O}, v_{t_E})$ is the image of the standard temporal coordinate system under an element of $({\cal F}_{st}^{(f)}, {\cal F}_t)$'', where $v_0, v_1, \ldots, v_n, v_{t_O}$ and $v_{t_E}$ are points in $(n+1)$ dimensional real space, are expressible in $\fo(\Pi)$. \end{enumerate}

All groups listed in Table~\ref{table-point-predicates-finite} are
subgroups of $({\cal A}_{st}^{(f)}, {\cal A}_t)$ and satisfy the
first condition. It is also straightforward to verify that they satisfy
the second condition.

For the third condition, we list for every group mentioned in
Table~\ref{table-point-predicates-finite} the expressions for the
spatial and temporal coordinate system. The proof that these
 expressions are correct are straightforward.

\medskip
\par\noindent  $\bullet$ \emph{For the group $({\cal A}_{st}^{(f)}, {\cal A}_t)$,}
the expressions  for ${\bf TCoSys}_{\cal A}(\ab u_1, \ab u_2)$ and ${\bf CoSys}_{\cal A}^n(\ab u_0,
\ab u_1, \ab \ldots, \ab u_n)$ were given in Lemma~\ref{conv-fo}.

\medskip
\par\noindent $\bullet$ \emph{For the group $({\cal A}_{st}^{(f)}, {\cal I}_t)$,}
the  expression for the spatial coordinate system does not change, but $${\bf TCoSys}_{\cal T}(u_1,
u_2):= {\bf TCoSys}_{\cal A}(u_1, u_2) \land \unittimes(u_1, u_2).$$

\medskip
\par\noindent $\bullet$ \emph{For the group $({\cal A}_{st}^{(f)}, {\it Id}_t)$,}
the  expression for the spatial coordinate system does again not change, but
$${\bf TCoSys}_{ Id}(u_1, u_2):= {\bf TCoSys}_{\cal
T}(u_1, u_2) \land  {\bf 0}_{\bf t}(u_1) \land {\bf 1}_{\bf t}(u_2).$$

For the following groups, we only list the expression for the spatial coordinate system.  The
temporal coordinate system depends on the groups ${\cal F}_t$ and is completely analogous to the
previous cases.
\medskip
\par\noindent $\bullet$ \emph{For the group $({\cal S}_{st}^{(f)}, {\cal F}_t)$,}
 we have   $$\displaylines{\qquad {\bf CoSys}_{\cal S}(u_0,u_1, \ldots,u_n):= \hfill{} \cr
\hfill{} {\bf CoSys}_{\cal A}^n( u_0, u_1, \ldots,u_n) \land \bigwedge_{i=1}^n \bigwedge_{j=1}^n
\equidistcotemps(u_0, u_i, u_0, u_j).\qquad}$$

\medskip
\par\noindent $\bullet$ \emph{For the group $({\cal I}_{st}^{(f)}, {\cal F}_t)$,} we have  \par\noindent
  $${\bf CoSys}_{\cal
I}(u_0,u_1,\ldots,u_n):= {\bf CoSys}_{\cal S}(u_0,u_1,\ldots,u_n) \ab \land \bigwedge_{i=1}^n
\unitdists(u_0,u_i).$$

\medskip
\par\noindent $\bullet$ \emph{For the group $({\cal T}_{st}^{(f)}, {\cal F}_t)$,} we have  \par\noindent $$\displaylines{ \qquad{\bf CoSys}_{\cal T}(u_0, u_1,\ldots,u_n):= \hfill{} \cr \hfill{} {\bf CoSys}_{\cal
I}(u_0,u_1,\ldots,u_n) \land  \bigwedge_{i=1}^n \bigwedge_{j=1}^n \leq_{\bf {\it i}}(u_0,u_j) \land
\poss{n}(u_0,u_1,\ldots,u_n).\qquad}$$ \qed

Next, we illustrate the languages summarized in Table~\ref{table-point-predicates-finite} with the
appropriate examples of  Section~\ref{examples}.

\begin{example}\label{ex-q3between}\rm We give the $\fo(\{\betweens{n}, \befores, \eqcrsts \})$-expression  $\varphi'_3$ e\-qui\-valent to the $({\cal I}_{st}, {\cal A}_t)$-generic query of Example~\ref{ex-q3}: {\it Was there a collision between car A and car B?}: $$\displaylines{\qquad
\varphi'_3 := (\exists u)(carA(u) \land carB(u)).\qquad}$$ Remark that this query can be expressed
without the use of the point predicates from $\Pi$. \qed
\end{example}

\begin{example}\label{ex-q4between}\rm We give the $\fo(\{ \betweens{n}, \befores, \eqcrsts, \equidistcotemps, \ab \unitdists, \leq_{\bf {\it i}} (1\leq i\leq n),\poss{n}, \unittimes ,{\bf 0}_{\bf t},{\bf
1}_{\bf t}\})$  expression $\varphi'_4$  equivalent to the $({\cal T}_{st}, {\it id}_t)$-generic
query of Example~\ref{ex-q4}: {\it Did car A pass at 500 meters north of car B at time moment $t =
5930$?}

The fact that a point has time coordinate 5930 can be expressed using $\unittimes$, ${\bf 0}_{\bf
t}$, and ${\bf 1}_{\bf t}$. We illustrate this with a predicate expressing the fact that a point
has time coordinate $3$:
 $$\displaylines{\qquad eq3t(u) := (\exists v_1)(\exists v_2)({\bf 1}_{\bf t}(v_1) \land \before{v_1, v_2} \land \unittime{v_1, v_2} \land \splits \before{v_2, u} \land \unittime{v_2, u}).\qquad}$$
 The fact that the distance between two points is 500 can be expressed using $\unitdists$ in a way
comparable to the construction of the predicate $3sec$ of Example~\ref{ex-q2between}.

Now we give the  expression $\varphi'_4$:
$$\displaylines{\qquad (\exists u)(\exists v)(\exists w)(carA(u) \land carB(v) \land eq5930t(u) \land eq5930t(v) \land \splits (\leq_{\bf {\it 1}}(u, w) \land  \leq_{\bf {\it 1}}(w, u)) \land (\leq_{\bf {\it 2}}(v, w) \land  \leq_{\bf {\it 2}}(w, v))
\land 500meters(u,w)).\qquad}$$\qed \end{example}

 \section{Sound and complete languages for the computable ge\-ne\-ric   spatio-temporal queries}\label{sectie5}
 In this section, we show that the languages  $\fo(\Pi({\cal F}_{st}, {\cal F}_t),\sigma)$ of
the previous section, when extended with assignment statements and a ${\rm While}$ loop, yield
languages that are computationally sound and complete  for the computable queries that are $({\cal
F}_{st}, {\cal F}_t)$-generic. To start with, we explain in more detail how point-based logics are
extended with assignment statements and  a while loop.  Afterwards, this section is organized in
the same way as Section~\ref{sectie4}. We first discuss sound and complete languages for the
queries generic for  time-independent transformation groups.  Then we focus on genericity   for
groups related to physical notions.   Finally, we address sound and complete languages for the
queries  that are generic  for the  time-dependent transformations.

We start with extending the point-based logics described in Definition~\ref{pointlangdef} with
while loops.
\medskip
\begin{definition}\rm
Let $\Pi$ be a finite set of point predicates, and let $\sigma$ be a database schema.
Syntactically, a \emph{program\/} in the language $\fo(\Pi,\sigma)+ {\rm While}$ is a finite
sequence of \emph{statements\/} and \emph{while-loops\/}. It is assumed there is a sufficient
supply of new relation variables, each with an appropriate arity.

\begin{enumerate}[(i)]
    \item Each statement has the form
$$R := \{(u_1,\ldots , u_k)\mid \varphi(u_1,\ldots , u_k)\};.$$ Here, $R$ is a new relation variable with assigned arity $k$ (the variables $u_i$ range over $\R^n\times \R$) and
$\varphi$ is  a formula in $\fo(\Pi,\sigma')$, where $\sigma'$ is the set of relation names
containing the elements of $\sigma$ together with the relation variables introduced in previous
statements of the program.

    \item A while-loop has the form
$${\bf while}\  \varphi\  {\bf do}\  P\ {\bf ;}$$ where $P$ is a program and $\varphi$ is a
sentence in $\fo(\Pi, \sigma')$,  where $\sigma'$ is again the set of relation names containing the
elements of $\sigma$ together with the relation variables introduced in previous statements of the
program.

\item One of the relation names occurring in the program is designated as the output relation
 and is named $R_\text{out}$. \end{enumerate}\qed
\end{definition}
Semantically, a program in the query language $\fo(\Pi,\sigma)+ {\rm While}$ expresses a
spatio-temporal query as soon as $R_\text{out}$ is assigned a return value. The execution of a
$\fo(\Pi,\sigma)+ {\rm While}$-program applied to an input database is performed step-by-step. A
statement is executed by first evaluating the  $\fo(\Pi,\sigma)$-formula  on the right hand side on
the input database together with the  newly created relations  resulting from previous statements.
Next, the result of the evaluation of the right hand side is assigned to the relation variable on
the left-hand side. The effect of a while loop is to execute the body as long as the condition
$\varphi$ evaluates to true.

Note that  a $\fo(\Pi,\sigma )+ {\rm While}$-program  is not guaranteed to halt. For those input
databases it does not, the query represented by the program is not defined on that particular input
database.

Consider the following example which will be used later on to express the query from Example~\ref{ex-q1}.

\begin{example}\label{while-ex}\rm
Suppose that we have a spatio-temporal  database with schema $\sigma=\{R, \ab S\}$, where the
underlying dimension is two and both $R$ and $S$ have arity one. We assume that all points in $R$
and $S$ have disjoint time coordinates. This means that we can sort all points according to their
time coordinates. We also assume that $R$ and $S$ both contain a finite number of points.

The query $Q$ we want to answer is the following: {\it Does $R$ contain more points than $S$?}. It
is well known that we cannot express this query in first-order  logic~\cite{grumbach-su}.
 The $\fo(\Pi,\sigma)+ {\rm While}$-program expressing $Q$ is:

 \begin{quote}
\begin{tabbing}
\quad\=\quad\=\quad\=\kill
$R_{\textrm Not}$ := $\{\}$;\\
$S_{\textrm Not}$ := $\{\}$;\\
$R_{\textrm Smallest}$ :=  $\{$\= $(u)| R(u) \land \neg R_{\textrm Not}(u) \land $ \\
\> $(\forall v)((R(v) \land \neg R_{\textrm Not}(v)) \rightarrow (\before{u, v}))\}$;\\
$S_{\textrm Smallest}$ := $\{$\=  $(u)| S(u) \land \neg S_{\textrm Not}(u) \land $ \\
\> $(\forall v)((S(v) \land \neg S_{\textrm Not}(v)) \rightarrow (\before{u, v}))\}$;\\
{\bf While} \= $(\exists u)(R_{\textrm Smallest}(u)) \land (\exists v)(S_{\textrm Smallest}(v))$
{\bf do}\\
\> $R_{\textrm Not}$ := $\{(u)| R_{\textrm Not}(u) \vee R_{\textrm Smallest}(u)\}$;\\
\> $S_{\textrm Not}$ := $\{(u)| S_{\textrm Not}(u) \vee S_{\textrm Smallest}(u)\}$;\\
\> $R_{\textrm Smallest}$ :=  $\{$\= $(u)| R(u) \land \neg R_{\textrm Not}(u) \land $ \\
\> \> $(\forall v)((R(v) \land \neg R_{\textrm Not}(v)) \rightarrow (\before{u, v}))\}$;\\
\> $S_{\textrm Smallest}$ := $\{$\=  $(u)| S(u) \land \neg S_{\textrm Not}(u) \land $ \\
\> \>  $(\forall v)((S(v) \land \neg S_{\textrm Not}(v)) \rightarrow (\before{u, v}))\}$;\\
$R_{\rm out} := \{()| (\exists u)(R_{\textrm Smallest}(u))\}$;\\
\end{tabbing}
\end{quote}

Intuitively, this program repeatedly takes the earliest point from both $R$ and $S$ until they do
not both contain unvisited points anymore. When the while loop terminates and $R$ still contains
unvisited points, true  is returned. \qed\end{example}

\subsection{Genericity for time-independent transformations}

In this section, we prove a general result concerning computable $({\cal F}_{st}, {\cal
F}_t)$-generic queries where $({\cal F}_{st}, {\cal F}_t)$ is a time-independent affinity of
$\R^n\times \R$,  i.e., a group from Table~\ref{table-point-predicates-constant}. The following
theorem follows directly from the proof of Theorem 6.1~\cite{gvv-jcss}.

\begin{theorem}\label{meta-compl}\rm  Let $\sigma$ be a spatio-temporal database schema. Let ${\cal F}$ be a subgroup of the affinities of $\R^n \times \R$, let $\Pi$ be a set of point
predicates and let  $\fo(\Pi,\sigma)$ be a point language that is sound and complete for the ${\cal
F}$-generic queries expressible in \foprs. Then the language $\fo(\Pi)+ {\rm While}$ is sound and
complete for the ${\cal F}$-generic computable queries. \qed
\end{theorem}

From this, we can derive the following result:

\begin{corollary}\rm   Let $\sigma$ be a spatio-temporal database schema. Let $({\cal F }_{st}, \allowbreak {\cal F}_t)$ be a group  and let $\Pi({\cal F }_{st}, \allowbreak
{\cal F}_t)$ be as in Table~\ref{table-point-predicates-constant}. The point language
$\fo({\Pi({\cal F}_{st}, {\cal F}_t) },\sigma)\ab +{\rm While}$ is sound and complete for the
computable $({\cal F }_{st}, \allowbreak {\cal F}_t)$-generic
 queries over $\overline{\sigma}$. \end{corollary}

\medskip
\par\noindent{\bf Proof.} The correctness follows from Theorem~\ref{meta-cor-fo} and Theorem~\ref{meta-compl}.\qed

\subsection{Applications to Physics}

Here, we focus again on the transformation groups $({\cal V}_{st}^{}, {\cal T}_t)$, $({\cal
V(R)}_{st}^{}, {\cal T}_t)$, $({\cal AC}_{st}^{}, \ab {\cal T}_t)$ and $({\cal AC(R)}_{st}^{},\ab {\cal
T}_t)$. As they are all subgroups of the affinities of $\R^n \times \R$, we can apply
Theorem~\ref{meta-compl} again.

\begin{corollary}\rm
  Let $\sigma$ be a spatio-temporal database schema. Let $({\cal F }_{st}, \allowbreak {\cal T}_t)$ be a group from
Table~\ref{table-point-predicates-physics} and let $\Pi({\cal F }_{st}, \allowbreak {\cal T}_t)$ be
as in Table~\ref{table-point-predicates-physics}. The point language $\fo({\Pi({\cal F}_{st}, {\cal
T}_t) , \sigma })+{\rm While}$ is sound and complete for the computable spatio-temporal queries
over $\overline{\sigma}$  that are $({\cal F }_{st}, \allowbreak {\cal I}_t)$-generic.
\end{corollary}

\medskip
\par\noindent{\bf Proof.} The correctness follows from Theorem~\ref{phys-th} and Theorem~\ref{meta-compl}.\qed

\begin{example}\label{ex-q1between}\rm We now give the $\fo(\{\betweens{n+1},\befores,
{\bf =}_\textrm{\bf space})+{\rm While}$-program expressing query
$Q_1$ of Example~\ref{ex-q1}: {\it Does the route followed by car
A self-intersect more often than the route followed by car B
does?}.

If a car is standing still at a certain position, this will result in an infinite number of points
in  $\R^{n}\times\R$  with the same spatial coordinates.  However,  one would not consider this
situation to be an infinite number of self-intersections. Therefore, when such a situation happens,
we only consider the last moment of the interval during which the car is at that specific location.

Intuitively, the program first computes the relations containing all self-intersections of the
trajectories of both cars, and then determines whether the route of car A self-intersects the most.
The program of Example~\ref{while-ex}  can be used to perform   this last task. We slightly adapt
it such that it expresses query $Q_1$:

\begin{quote}
\begin{tabbing}
\quad\=\quad\=\quad\=\kill
$A_{\cap}$ := $\{$ \=  $(u)| carA(u)
\land (\exists v)(carA(v) \land \before{u, v} \land {\bf
=}_\textrm{\bf space}(u, v) \land$ \\
\>$(\forall w)($\= $(carA(w) \land \before{u, w} \land \before{w,
v}\land u\not= w\land v\not= w )$ \\
\> \> $\rightarrow \neg({\bf =}_\textrm{\bf space}(w,v))) ) \}$ ;\\
$B_{\cap}$ := $\{$ \=  $(u)| carB(u) \land (\exists v)(carB(v)
\land \before{u, v} \land {\bf
=}_\textrm{\bf space}(u, v) \land$ \\
\>$(\forall w)($\= $(carB(w) \land \before{u, w} \land \before{w,
v}\land u\not= w\land v\not= w )$ \\
\> \> $\rightarrow \neg({\bf =}_\textrm{\bf space}(w,v))) ) \}$ ;\\

$A_{\textrm Not}$ := $\{\}$;\\
$B_{\textrm Not}$ := $\{\}$;\\
$A_{\textrm Smallest}$ :=  $\{$\= $(u)| A_{\cap}(u) \land \neg A_{\textrm Not}(u) \land (\forall v)($ \\
\> $(A_{\cap}(v) \land \neg A_{\textrm Not}(v)) \rightarrow (\before{u, v}))\}$;\\
$B_{\textrm Smallest}$ := $\{$\=  $(u)| B_{\cap}(u) \land \neg B_{\textrm Not}(u) \land (\forall v)($ \\
\> $(B_{\cap}(v) \land \neg B_{\textrm Not}(v)) \rightarrow (\before{u, v}))\}$;\\
{\bf While} \= $(\exists u)(A_{\textrm Smallest}(u)) \land
(\exists v)(B_{\textrm Smallest}(v))$
{\bf do}\\
\> $A_{\textrm Not}$ := $\{(u)| A_{\textrm Not}(u) \vee A_{\textrm Smallest}(u)\}$;\\
\> $B_{\textrm Not}$ := $\{(u)| B_{\textrm Not}(u) \vee B_{\textrm Smallest}(u)\}$;\\
\> $A_{\textrm Smallest}$ :=  $\{$\= $(u)| A_{\cap}(u) \land \neg A_{\textrm Not}(u) \land (\forall v)($ \\
\> \> $(A_{\cap}(v) \land \neg A_{\textrm Not}(v)) \rightarrow (\before{u, v}))\}$;\\
\> $B_{\textrm Smallest}$ := $\{$\=  $(u)| B_{\cap}(u) \land \neg B_{\textrm Not}(u) \land (\forall v)($ \\
\> \>  $(B(v) \land \neg B_{\textrm Not}(v)) \rightarrow (\before{u, v}))\}$;\\
 $R_{\rm out} := \{()| (\exists u)(A_{\textrm Smallest}(u))\}$;\\
\end{tabbing}
\end{quote}\qed
\end{example}

\subsection{Genericity for time-dependent transformations}\label{subsec:53}

Finally, we study notions of genericity determined by
 groups of  time-dependent transformations. Here, we only show results for the
groups of arbitrary time-dependent transformations ${\cal F}_{st}$. We concentrate on the group
$({\cal A }_{st}^{}, \allowbreak {\cal A}_t)$. The other time-dependent transformation groups will
be addressed afterwards (Corollary~\ref{cor-complete}). For the groups ${\cal F}_{st}^f$ the
problem of identifying sound and complete languages is open, we will  discuss the problems
concerning this at the end of this section.

We introduce some definitions first. Recall that we introduced, in Section~\ref{sectie3.1}, the
abbreviation $f(R^{\st{}})$ for the  formula $\{ (f({\bf a}_1,\tau_1),\ab f({\bf a}_2,\ab
\tau_2),\ab \ldots, \ab f({\bf a}_k,\ab \tau_k))\mid \ab ({\bf a}_1,\tau_1, \ab {\bf a}_2,\tau_2,
\ab \ldots, \ab {\bf a}_k,\tau_k)\ab \in R^{\st{}}\},$ where $R$ is a relation name and $\st{}$ a
spatio-temporal database over a schema $\sigma$ that contains $R$.

\begin{definition}\rm \label{iso-db-def}
Let $\sti{1}$ and $\sti{2}$ be spatio-temporal databases   over the schema $\sigma = \{R_1, \ldots,
R_m \}$ with underlying dimension $n$.  The databases $\sti{1}$ and $\sti{2}$ are called
\emph{$({\cal F}_{st},{\cal F}_t)$-isomorphic} if and only if there exists a $f=(f_{st},f_{t})\in
({\cal F}_{st}, {\cal F}_t)$ such that for all $R_i$ in $\sigma$,  $
f(R_i^{\sti{1}})=R_i^{\sti{2}}.$  \qed
\end{definition}

 Recall that a representation of a spatio-temporal database $\st{}$ over a schema
$\sigma = \{R_1, \ldots, R_m\}$ is a tuple $(\varphi_1, \ldots, \varphi_m)$ of quantifier-free formulas in \fop,
such that  $\varphi_i$ describes $R_i^{\st{}}$.

Assuming some order on the characters or symbols that may appear in a \fop-formulas, we can
lexicographically order the \fop-formulas.
\begin{definition}\rm
 The \emph{$({\cal F}_{st},{\cal F}_t)$-canonization} of a spatio-temporal database $\st{}$ over a schema $\sigma=\{R_1,\ldots,R_m\}$, denoted
by \emph{${\bf Canon}_{({\cal F}_{st},{\cal F}_t)}(\st{})$}, is the spatio-temporal database
$\st'{}$, which is $({\cal F}_{st},{\cal F}_t)$-isomorphic to $\st{}$ and has a representation  by
quantifier-free \fop-formulas $$(\varphi_{{\bf Canon}_{({\cal F}_{st},{\cal F}_t)}(R_1)},\ldots,
\varphi_{{\bf Canon}_{({\cal F}_{st},{\cal F}_t)}(R_1)})$$ that  occurs lexicographically first
among the representations of spatio-temporal databases $({\cal F}_{st},{\cal F}_t)$-isomorphic to
$\st{}$. \qed
\end{definition}

\begin{definition}\rm Let $\st{}$ be a spatio-temporal database. The \emph{$({\cal F}_{st},{\cal
F}_t)$-type of $\st{}$}, denoted \emph{$Type_{({\cal F}_{st},{\cal F}_t)}(\st{})$}, equals $$\{ f
\in ({\cal F}_{st},{\cal F}_t) | f(\st{}) = {\bf Canon}_{({\cal F}_{st},{\cal F}_t)}(\st{})\}.$$
\qed
\end{definition}

We can derive directly from a similar proposition of Gyssens, Van den Bussche and Van
Gucht~\cite{gvv-jcss} that, for a spatio-temporal database $\st{}$, a representation of ${\bf
Canon}_{({\cal F}_{st},{\cal F}_t)}(\st{})$ can be computed  if and only if   $({\cal F}_{st},{\cal
F}_t)$ is a \emph{
 transformation group}.

A transformation group ${\cal G}$ of $\R^n\times \R$ is semi-algebraic  if and only if   there
exists a semi-algebraic subset  of $\R^l$, described by a \fop-formula $\varphi_{\cal G}$, for some
fixed $l$, representing all elements of ${\cal G}$, such that the set
$$\displaylines{\qquad \{(g_1, \ldots, g_l, x_1, \ldots, x_n, t, x'_1, \ldots, x'_n, t')|\splits \varphi_{\cal G}(g_1, \ldots, g_l) \land
\varphi_{{\cal G}-img}(g_1, \ldots, g_l,x_1, \ldots, x_n, t, x'_1, \ldots, x'_n, t')\},\qquad}$$
also called the \emph{graph} of ${\cal G}$, is a semi algebraic subset of $\R^{l + 2(n+1)}$. The
formula $\varphi_{{\cal G}-img}$ expresses that, for the element of ${\cal G}$ represented by the
tuple $(g_1, \ldots, g_l)$, the tuple $(x_1, \ldots, x_n, t)$ is mapped to $(x'_1, \ldots, x'_n,
t')$.

\begin{remark}\rm\label{semialg-transf-ex}
The transformation group $({\cal A}_{st},{\cal A}_t)$ is semi-algebraic  if and only if  the
transformation groups ${\cal A}_{st}$ and ${\cal A}_t$  are both semi-algebraic.  The
semi-algebraic set for the group ${\cal A}_t$ is $\varphi_{{\cal A}_t} \equiv\{(\alpha,
\beta)\in\R^2| \alpha >
 0\}$. The semi-algebraic set associated to the group ${\cal A}_{st}$ is  given by the \fop-formula
$\varphi_{{\cal A}_{st}}(\alpha_{1,1}, \ldots, \alpha_{1,n}, \ab \ldots, \ab \alpha_{n,1}, \ldots,
\alpha_{n,n}, \ab \beta_1, \ldots, \beta_n, t)$ expressed by
$$\left|\begin{array}{cccc }
\alpha_{1,1} &  \cdots & \alpha_{1,n}   \\
\vdots&\cdots& \vdots
\\ \alpha_{n,1}  & \cdots & \alpha_{n,n}
 \end{array}
 \right| \neq 0.$$

We now give the formulas for $\varphi_{{\cal A}_t-img}$ and $\varphi_{{\cal A}_{st}-img}$.  The
formula $\varphi_{{\cal A}_t-img}(\ab \alpha, \ab \beta, \ab x_1, \ab \ldots, \ab x_n, \ab x_t, \ab
x'_1, \ab \ldots, \ab x'_n, \ab x'_t)$ can be given as
$$x'_t = \alpha x_t+\beta \land \bigwedge_{i = 1}^n x'_i = x_i.$$
The formula
$\varphi_{{\cal A}_{st}-img} (\alpha_{1,1}, \ab \ldots, \ab \alpha_{1,n},\ab  \ldots,
\ab \alpha_{n,1}, \ab \ldots,\ab  \alpha_{n,n}, \ab \beta_1, \ab \ldots, \ab \beta_n, \ab t,\ab  x_1, \ab \ldots, \ab x_n, \ab x_t,
\ab x'_1,\ab
\ldots,\ab x'_n, \ab x'_t)$ on the other hand is $$  t=t_x \land t_x =
t'_x \land
 \bigwedge_{i = 1}^n \alpha_{i,1}x_1 + \cdots + \alpha_{i,n}x_n + \beta_i = x'_i .$$

 The graph of the group $({\cal A}_{st},{\cal A}_t)$ is now a semi-algebraic subset of
$\R^{l + 2(n + 1)}$, where $l = n^2 + n + 3$, described as follows:

 $$\displaylines{ \qquad \gamma_{({\cal A}_{st},{\cal A}_t)}(\alpha_{1,1}, \ldots, \alpha_{1,n}, \ldots,
\alpha_{n,1}, \ldots, \alpha_{n,n}, \beta_1, \ldots, \beta_n, t, \splits \alpha, \beta, x_1,
\ldots, x_n, t_x, x'_1, \ldots, x'_n, t'_x):= \splits \varphi_{{\cal A}_{st}-img}(\alpha_{1,1},
\ldots, \alpha_{1,n}, \ldots, \alpha_{n,1}, \ldots, \alpha_{n,n}, \beta_1, \ldots, \beta_n,
t,\splits  x_1, \ldots, x_n, x_t, x'_1, \ldots,x'_n, x_t) \splits \land  \varphi_{{\cal
A}_t-img}(\alpha, \beta, x'_1, \ldots, x'_n, x_t, x'_1, \ldots, x'_n, x'_t)  .\qquad}$$\qed
\end{remark}

\bigskip
We now prove the main theorem of this section. The proof technique used here was introduced by
Gyssens, Van den Bussche and Van Gucht~\cite{gvv-jcss}. We first sketch the proof technique, but
only   give  details about  the aspects of the proof that need modifications in the context of
spatio-temporal databases. These modifications are based on proof techniques introduced in
Section~\ref{sectie4}.

\begin{theorem}\label{main-compl}\rm
 Let $\sigma$ be a spatio-temporal database schema.
The point language $\fo(\{\betweens{n},\allowbreak \befores, \allowbreak \eqcrsts\},\sigma) +
\ab{\rm While}$ is sound and complete for the $({\cal A }_{st}^{}, \allowbreak {\cal A}_t)$-generic
computable spatio-temporal queries over $\overline{\sigma}$.\qed
\end{theorem}

\par\noindent
{\bf Proof.} It suffices to show that an $({\cal A }_{st}^{}, \allowbreak {\cal A}_t)$-generic
computable query $Q$ over  $\overline{\sigma}$ can be simulated in the language
$\fo(\{\betweens{n},\allowbreak \befores, \allowbreak \eqcrsts\},\sigma ) + \allowbreak {\rm
While}$.  We first briefly sketch the proof strategy, including the conversion procedure and the
encoding and decoding step, that appear in it. Later the coding and decoding will be  explained in
more detail.  For the remainder of this proof, ${\Pi}$ will denote the set
$\{\betweens{n},\befores,\eqcrsts\}$.

We start with the encoding that will be used to convert formulas that represent spatio-temporal
relations into natural numbers
\medskip
\par\noindent $\bullet$ \emph{The encoding mechanism.}
Let $\st{}$ be a spatio-temporal database over $\sigma$. Let $K$ be the maximum of the arities of
all relations in $\sigma$ and the query $Q$. Let $n$ be the underlying dimension. Then each
relation of $\st{}$ can be represented by a quantifier-free \fop-formula using only the variables
$x_1, \ldots, x_{(n + 1)K}$, the symbols $\leq, +, \times, (, ), \vee$ and $\neg$, and the
constants $0$ and $1$.

 We denote these $9+(n+1)K$ by  $s_1, \ldots, s_{9+(n+1)K}$.
Hence, we can encode a quantifier-free \fop-formula as a string $s = s_{i_1}\ldots s_{i_k}$ as the
natural number $N=p_1^{{i_1}}\ldots p_k^{{i_k}}$, where $p_j$ is the $j$-th prime number. And we
denote $N$ by $Encode(s)$.

\medskip

\par\noindent $\bullet$ \emph{Proof strategy.}  Given a spatio-temporal database $\st{}$ over a schema  $\sigma =
\{R_1, \ab \ldots,\ab  R_m \}$,  the simulation of a $({\cal A }_{st}^{}, \allowbreak {\cal
A}_t)$-generic $k$-ary computable query on input $\st{}$ is broken up into three steps:
\begin{enumerate}[(i)]

    \item{\it The encoding step:}   The database
 $\st{}$ is encoded as
a tuple of natural numbers $(N_{R_1},\ldots,N_{R_m})$, one for each relation of the database. Here,
$N_{R_i} = Encode(s_i)$, where ($s_1, \ldots ,s_m$) are the string representation of the
quantifier-free formulas $\varphi_{{\bf Canon}_{({\cal A }_{st}^{},{\cal A}_t)}(R_i)}$
($i=1,\ldots,m$) of the database  ${\bf Canon}_{({\cal A }_{st}^{},{\cal A}_t)}(\st{})$.   It will
be shown below that this encoding can be performed in the language  $\fo({\bf \Pi},\sigma)$ $+{\rm
While}$. The set $Type_{({\cal A}_{st},{\cal A}_t)}(\st{})$ is also computed, to be used  in the
decoding step.

    \item {\it The
computing step:}  It was shown  that the language $\fo({\Pi}) + {\rm While}$ has full computational
power on the natural numbers, by simulating a counter machine~\cite{gvv-jcss}.

More specifically, one can simulate a counter machine $M$ in $\fo({\bf \Pi}) + {\rm While}$ such
that on input $(N_{R_1}, \ldots, N_{R_m})$, $M$ halts if  and only if  $Q$ is defined on the
corresponding $\st{}$ and $M$ will output a natural number $N_q$ which is the encoding of
$Q(\st{})$.

    \item {\it The
decoding step:}   If $M$ terminates on input $(N_{R_1}, \ldots, N_{R_m})$ then it outputs a natural
number $N_q$. Using $Type_{({\cal A}_{st},{\cal A}_t)}(\st{})$, the decoding algorithm computes the
point set of which $N_q$ is the encoding.   This can be implemented in the language $\fo({\Pi})$
$+{\rm While}$. \end{enumerate}

We show next the details in the encoding and decoding algorithms that are different for $({\cal
A}_{st},{\cal A}_t)$-generic queries, as compared to the affine-generic queries considered in
~\cite{gvv-jcss}. For  ease of exposition, we will assume for the remainder of this proof that the
input spatio-temporal database  $\st{}$ has only one relation, with arity one, i.e.,
$\sigma=\{R\}$.  For relations with arity greater than one, the encoding algorithm has to consider
more variables. If the input database contains more relations, each relation has to be encoded
separately.

\medskip
\par\noindent $\bullet$ \emph{The encoding algorithm can be expressed in $\fo({\Pi}) +{\rm While}$.}
Roughly speaking, the encoding procedure enumerates all natural numbers and meanwhile stores the
evaluation of the  terms and formulas that are encoded by those numbers in relations that are
called $T$ and $F$, respectively. This enumeration continues until one natural number is found that
encodes a relation that is $({\cal A }_{st}^{}, \allowbreak {\cal A}_t)$-isomorphic to $R$. This
relation, for which the evaluation is stored in $F$, corresponds to ${\bf Canon}_{({\cal
A}_{st},{\cal A}_t)}(R)$. The set $Type_{({\cal A}_{st},{\cal A}_t)}(\st{})$ is also computed, to
use in the decoding step.

 First, we explain the role of the relations $T$ and $F$ in more detail, as well as the way they are
built during the encoding process.

The encoding program builds up terms and formulas until the formula is found that encodes ${\bf
Canon}_{({\cal A}_{st},{\cal A}_t)}(\st{})$. The terms and formulas are stored in the relations $T$ and
$F$. In general, the arity of $T$ is $(n+1) + 2 + 2 + l\times (n+1)$, where $n$ is the underlying
dimension and $l=ar(R)$. Under the assumption that $ar(R)=1$
and the underlying dimension is $2$, each tuple in $T$ is of the form
$$(u_{t_O}, u_{t_E},
u_0,u_1,u_2 ,u_t,p_1, p_2, p_t, v ),$$ where $(u_{t_O}, u_{t_E})$ is a temporal coordinate system,
$(u_0,u_1,u_2)$ a spatial coordinate system, $u_t$ the encoding of a term which only uses the
variables $x_1$, $x_2$, $x_t$  (which are translated into $v_1$, $v_2$, $v_t$), and $v$ the value
of the term when evaluated under the valuation $v_1 \mapsto p_1, v_2 \mapsto p_2, v_t \mapsto p_t$.
The arity of $F$ is $(n+1) + 2 + 1 + l\times (n+1)$. Under the same assumptions, each tuple in $F$ is of the form
$$(u_{t_O}, u_{t_E},
u_0,u_1,u_2 ,u_f,p_1, p_2, p_t),$$ where $(u_{t_O}, u_{t_E})$ and $(u_0,u_1,u_2)$ are as before,
$u_f$ the encoding of a formula $\overline{\varphi}$ which only uses the variables $x_1, x_2, x_t$,
and where $\varphi(p_1, p_2, p_t)$ is true.

In Figure~\ref{alg-encoding}, we give the structure of the encoding program in
$\fo[\{\betweens{n},$ $\befores,$ $\eqcrsts\}]$ $+{\rm While}$.  In this algorithm, it is assumed
that substrings $s'$ of a string $s$ is encountered in the enumeration before $s$ is encountered.

\begin{figure}
\begin{algorithmic}\small \rm
    \STATE $m := 0$
    \STATE $T :=  \emptyset$
    \STATE $F:= \emptyset$
    \STATE found :=False
    \WHILE{not found}
        \STATE $m := m + 1$
        \IF{$m$ encodes $x_1$}
           \STATE $T := T \cup \{(u_{t_O}, u_{t_E},
u_0,u_1,u_2 ,m,p_1, p_2, p_t, p_1) | p_1, p_2, p_t \text{ collinear with }u_0\text{ and }u_1\}$
        \ELSIF{$m$ encodes $x_2$}
            \STATE $T := T \cup \{(u_{t_O}, u_{t_E},
u_0,u_1,u_2 ,m,p_1, p_2, p_t, p_2) | p_1, p_2, p_t \text{ collinear with }u_0\text{ and }u_1\}$
        \ELSIF{$m$ encodes $x_t$}
            \STATE $T := T \cup \{(u_{t_O}, u_{t_E},
u_0,u_1,u_2 ,m,p_1, p_2, p_t, p_t) | p_1, p_2, p_t \text{ collinear with }u_0\text{ and }u_1\}$
        \ELSIF{$m$ encodes $0$}
            \STATE $T := T \cup \{(u_{t_O}, u_{t_E},
u_0,u_1,u_2 ,m,p_1, p_2, p_t, u_0) | p_1, p_2, p_t \text{ collinear with }u_0\text{ and }u_1\}$
        \ELSIF{$m$ encodes $1$}
            \STATE $T := T \cup \{(u_{t_O}, u_{t_E},
u_0,u_1,u_2 ,m,p_1, p_2, p_t, u_1) | p_1, p_2, p_t \text{ collinear with }u_0\text{ and }u_1\}$
        \ELSIF{$m$ encodes $(s + t)$}
            \STATE $T := T \cup \{(u_{t_O}, u_{t_E},
u_0,u_1,u_2 ,m,p_1, p_2, p_t, p_e) |\ab T(u_{t_O}, u_{t_E}, u_0,u_1,u_2 , enc(s), p_1, p_2, p_t,
p_c) \ab\land \ab T(u_{t_O}, u_{t_E}, u_0,u_1,u_2 , enc(t), p_1, p_2, p_t, p_d) \ab\land {\bf
Plus}(p_c, p_d, p_e)\}$
        \ELSIF{$m$ encodes $(s\times t)$}
            \STATE $T := T \cup \{(u_{t_O}, u_{t_E},
u_0,u_1,u_2 ,m,p_1, p_2, p_t, p_e) |\ab T(u_{t_O}, u_{t_E}, u_0,u_1,u_2 , enc(s), p_1, p_2, p_t,
p_c) \ab \land T(u_{t_O}, u_{t_E}, u_0,u_1,u_2 , enc(t), p_1, p_2, p_t, p_d) \ab\land {\bf
Times}(p_c, p_d, p_e)\}$
        \ELSIF{$m$ encodes $(s \leq t)$}
            \STATE $F := F \cup \{(u_{t_O}, u_{t_E},
u_0,u_1,u_2 ,m,p_1, p_2, p_t) | (\exists c)(\exists d)(\ab T(u_{t_O}, u_{t_E}, u_0,u_1,u_2 ,
enc(s), p_1, p_2, p_t, p_c) \ab \land T(u_{t_O}, u_{t_E}, u_0,u_1,u_2 , enc(t), p_1, p_2, p_t, p_d)
\ab \land  {\bf Less}(p_c, p_d))\}$
        \ELSIF{$m$ encodes $(\neg\varphi)$}
            \STATE $F := F \cup \{(u_{t_O}, u_{t_E},
u_0,u_1,u_2 ,m,p_1, p_2, p_t) | \neg F(u_{t_O}, u_{t_E}, u_0,u_1,u_2 , enc(\varphi),p_1, p_2,
p_t)\}$
        \ELSIF{$m$ encodes $(\varphi \vee \psi)$}
            \STATE $F := F \cup \{(u_{t_O}, u_{t_E},
u_0,u_1,u_2 ,m,p_1, p_2, p_t) | F(u_{t_O}, u_{t_E}, u_0,u_1,u_2 , enc(\varphi),p_1, p_2, p_t) \vee
 F(u_{t_O}, u_{t_E}, u_0,u_1,u_2 , enc(\psi),p_1, p_2, p_t)\}$
        \ENDIF{}
        \STATE found:= $m$ encodes a formula which represents ${\bf Canon}_{({\cal A}_{st},{\cal A}_t)}(R)$
    \ENDWHILE{}
    \STATE $N_{{\bf Canon}_{({\cal A}_{st},{\cal A}_t)}(R)} := m$
    \STATE $Type_{({\cal A}_{st},{\cal A}_t)} := \{ a \in ({\cal A}_{st},{\cal A}_t) \mid a(R) = {\bf Canon}_{({\cal A}_{st},{\cal A}_t)}(R)\}$
\end{algorithmic}
\caption{The encoding program. The input is a \fopr{\overline{R}}-sentence.}\label{alg-encoding}
\end{figure}

We will discuss in detail \begin{enumerate}[(i)]
\item the representation of natural numbers (as we only have
point-variables), \item the expression that checks whether a certain natural number encodes a formula
which represents ${\bf Canon}_{({\cal A}_{st},{\cal A}_t)}(\st{})$, and \item  the computation of the set
$Type_{({\cal A}_{st},{\cal A}_t)}(R)$.\end{enumerate}
 All other elements of the encoding can be adopted from the
proof of~\cite{gvv-jcss} with  only slight modifications. For ease of exposition, we give the
formulas for $n = 2$.
\begin{enumerate}[(i)]
    \item Natural numbers can be represented by $(n+1)$-dimensional points using the computation plane
technique introduced in Section~\ref{sectie4}.  Further on, in de encoding and decoding algorithm,
we need to simulate assignments such as $m:=0$ and $m:=m+1$ (since we have to run through all
natural numbers in those algorithms). As an illustration, we explain here how thet are simulated in
$\fo({\Pi})$ $+{\rm While}$. The expression $m := 0$, for example,  is translated in $\fo({\Pi})$
$+{\rm While}$ by assigning to a spatio-temporal relation a point that is the origin of  the chosen
computation plane. The translated expression is

$$\displaylines{ \qquad N:= \{(u_{t_O}, u_{t_E},
u_0,u_1,u_2,v) \mid  {\bf TCoSys}_{\cal A}(u_{t_O},u_{t_E}) \land \splits {\bf CoSys}_{\cal A}^n(
u_0,u_1,u_2) \land {\bf Collinear}(u_0, u_1, v)\land v = u_0\}.\qquad}$$

For the assignment $m := m + 1$, we have:
$$\displaylines{ \qquad N:= \{(u_{t_O}, u_{t_E},
u_0,u_1,u_2, v) \mid  (\exists w) ( N(u_{t_O}, u_{t_E}, u_0,u_1,u_2, w) \splits \land {\bf
Plus}(u_{t_O}, u_{t_E}, u_0,u_1,u_2, w, u_1, v))\}.\qquad}$$

The predicate {\bf Plus}, which expresses that, relative to a computation plane, a certain point
represents the sum of two other points can be written in $\fo({\bf \Pi})$ $+{\rm While}$ because of
 Theorem~\ref{finite-fo}.

\item We now give the expression $\varphi$ that checks whether a certain natural number $m$ encodes
a formula which represents $\varphi_{{\bf Canon}_{({\cal A}_{st},{\cal A}_t)}(R_i)}$
($i=1,\ldots,m$). Remember that the evaluation of the formula encoded by $m$ is stored in the
relation $F$. This relation has arity  $n + 3 + K$, where $K$ is the maximal arity in the input
database schema. Let $(p_1, p_2, \ldots, p_{n+3+K})$ be a tuple of points satisfying $F$. The
points $p_1$ and  $p_2$ are a temporal coordinate system, $p_{n+3}$ represents the natural number
$m$ encoding the formula and $p_3, p_4, \ldots, p_{n+2}$  form a hyperplane of which the plane
through $p_3$, $p_4$ and $p_5$ will be use as a computation plane. The last $K$  points are the
translation of the free variables in the formula encoded by $m$.

\smallskip

 Let the formula $\psi_{({\cal A}_{st},{\cal A}_t)}$ be the translation of the formula $\varphi_{({\cal
A}_{st},{\cal A}_t)}$ from Example~\ref{semialg-transf-ex}. Intuitively, the next formula checks,
for a natural number $m$, whether there exists an element of  the group $({\cal A}_{st},{\cal
A}_t)$ that maps each point in $R$ to a point in the set of points satisfying the formula encoded
by $m$, the evaluation  of which is
 stored
in $F$.

The following formula  $\psi$ checks  whether the right quantifier-free formula has been found. It
reflects the stop condition of the While-loop that runs through the natural numbers.  This formula
$\psi$ can be written as

$$\displaylines{(\forall u_{t_O})(\forall  u_{t_E})(\forall  u_0)(\forall  u_1)(\forall  u_2) ({\bf TCoSys}_{\cal A}(u_{t_O},u_{t_E})
\land \splits {\bf CoSys}_{\cal A}^n(u_0, u_1, u_2)) \rightarrow  (\exists v_{\alpha})(\exists
v_{\beta})(\exists w)(\forall u_t) (\exists v_{a_{1,1}})(\exists v_{a_{1,2}})(\exists
v_{a_{2,1}})(\exists  v_{a_{2,2}})\splits (\exists  v_{b_1})(\exists  v_{b_2}) ( N(u_{t_O},u_{t_E},
u_0, u_1, u_2, w) \land \splits (\forall v_x)(\forall  v_y)(\forall v_t)(F(u_{t_O}, u_{t_E}, u_0,
u_1, u_2, w, v_x, v_y, v_t) \leftrightarrow \splits (\exists v)(\exists v'_x)(\exists v'_y)(R(v)
\land \mbox{\bf comp-coord}(u_{t_O},u_{t_E}, u_0, u_1, u_2, v, v'_x, v'_y, u_t) \land \splits
\psi_{({\cal A}_{st},{\cal A}_t)}(u_{t_O},u_{t_E}, u_0, u_1, u_2, v_{a_{1,1}}, v_{a_{1,2}},
v_{a_{2,1}}, v_{a_{2,2}}, v_{b_1}, v_{b_2},  u_t, v_{\alpha}, v_{\beta},\splits v'_x, v'_y, u_t,
v_x, v_y, v_t ) ) )) .\qquad}$$

In the above formula, we omitted, for all point variables except $v$, the sub formulas expressing
collinearity with $v_0$ and $v_1$. Also, the predicate {\bf comp-coord} is an abbreviation for the
fact that the translation of $v$'s coordinates to the computation plane are $v'_x$, $v'_y$ and
$u_t$. The exact formula expressing this can be found in the proof of Lemma~\ref{conv-fo}, when the
translation of relation predicates is explained.

    \item For the set $Type_{({\cal A}_{st},{\cal A}_t)}(R) = \{ \alpha \in ({\cal A}_{st},{\cal A}_t) \mid \alpha(R) =
{\bf Canon}_{({\cal A}_{st},{\cal A}_t)}(R)\}$, we compute two separate relations storing the
${\cal A}_{t}$-type, respectively ${{\cal A}_{st}}$-type of the encoded relation. In the previous
formula, it was checked whether there exists a transformation mapping all points in $R$ to points
in the formula coded by $m$  (i.e., in $F$). Here, we compute that transformation:

$$\displaylines{\qquad  T_{{\cal A}_t} := \{(u_{t_O}, u_{t_E}, u_0, u_1, u_2, v_{\alpha}, v_{\beta})| ({\bf TCoSys}_{\cal A}(u_{t_O},u_{t_E})
\land \splits {\bf CoSys}_{\cal A}^n(u_0, u_1, u_2)) \rightarrow
(\exists w)(N(u_{t_O}, u_{t_E}, u_0, u_1, u_2, w) \land \splits
(\forall v_x)(\forall  v_y)(\forall  v_t)( F(u_{t_O}, u_{t_E}, u_0, u_1, u_2, w,
v_x, v_y, v_t) \leftrightarrow \splits (\exists v)(\exists  v'_x)(\exists  v'_y)(\exists
v'_t)(\exists  v''_x)(\exists  v''_y)(\exists  v_{\alpha_{0,0}})(\exists  v_{\alpha_{0,1}})(\exists
v_{\alpha_{1,0}})(\exists  v_{\alpha_{1,1}})(\exists  v_{\beta_{0}})
(\exists v_{\beta_{1}})\splits (  R(v) \land \mbox{\bf
comp-coord}(u_{t_O},u_{t_E}, u_0, u_1, u_2, v, v'_x, v'_y, v'_t)
\land \splits \psi_{{\cal A}_{st}}(u_{t_O},u_{t_E}, u_0, u_1, u_2,
v_{\alpha_{0,0}}, v_{\alpha_{0,1}}, v_{\alpha_{1,0}},
v_{\alpha_{1,1}}, v_{\beta_{0}}, v_{\beta_{1}}, \splits v'_t, v'_x, v'_y,
v'_t, v''_x, v''_y, v'_t) \land \splits \psi_{{\cal
A}_t}(u_{t_O},u_{t_E}, u_0, u_1, u_2, v_{\alpha}, v_{\beta},
v''_x, v''_y, v'_t, v_x, v_y, v_t))))\} \qquad}$$

and

$$\displaylines{\qquad  T_{{\cal A}_{st}}:= \{(u_{t_O}, u_{t_E}, u_0, u_1, u_2, v_{\alpha_{0,0}}, v_{\alpha_{0,0}}, v_{\alpha_{0,1}}, v_{\alpha_{1,0}},
v_{\alpha_{1,1}}, v_{\beta_{0}}, v_{\beta_{1}}, u_t)| \splits
({\bf TCoSys}_{\cal A}(u_{t_O},u_{t_E}) \land  {\bf CoSys}_{\cal
A}^n(u_0, u_1, u_2)) \rightarrow \splits (\exists w)(  N(u_{t_O},
u_{t_E}, u_0, u_1, u_2, w) \land  (\forall v_x)(\forall  v_y)(\forall  v_t)(\splits
F(u_{t_O}, u_{t_E}, u_0, u_1, u_2, w, v_x, v_y, v_t)
\leftrightarrow (\exists v)(\exists  v'_x)(\exists  v'_y)(\exists  v'_t)(\exists  v_{\alpha})(\exists
v_{\beta})\splits  (R(v) \land \mbox{\bf comp-coord}(u_{t_O},u_{t_E},
u_0, u_1, u_2, v, v'_x, v'_y, v'_t) \land \splits T_{{\cal
A}_t}(u_{t_O},u_{t_E}, u_0, u_1, u_2, v_{\alpha}, v_{\beta}) \land
\splits \psi_{{\cal A}_t}(u_{t_O},u_{t_E}, u_0, u_1, u_2,
v_{\alpha}, v_{\beta}, v'_x, v'_y, v'_t, v'_x, v'_y, v_t) \land
\splits \psi_{{\cal A}_{st}}(u_{t_O},u_{t_E}, u_0, u_1, u_2,
v_{\alpha_{0,0}}, v_{\alpha_{0,1}}, v_{\alpha_{1,0}},
v_{\alpha_{1,1}}, v_{\beta_{0}}, v_{\beta_{1}}, \splits v_t, v'_x, v'_y,
v_t, v_x, v_y, v_t))))\} .\qquad}$$

\end{enumerate}

\medskip
\par\noindent $\bullet$ \emph{The decoding algorithm can be expressed in $\fo({\bf \Pi}) +{\rm While}$.}
Input databases are encoded by natural numbers. A counter machine simulates the query on this
natural number and returns a natural number that encodes the output.
 In the decoding algorithm, again all natural numbers are enumerated and the evaluation of the terms and
formulas they encode are stored in relations called $T$ and $F$. When the number that is the output
of the counter machine is encountered, the relation $F$ contains all points of the result, up to
the transformation stored in Type$_{({\cal A}_{st},{\cal A}_t)}$  (because the query is $({\cal
A}_{st},{\cal A}_t)-$generic).  The result corresponds to the set
  $Q({\bf Canon}_{({\cal A}_{st},{\cal A}_t)}(\st{}))$.  As $Q$ is assumed to be a $({\cal A}_{st},{\cal
A}_t)-$generic query, we have that for all $f \in Type_{({\cal A}_{st},{\cal
A}_t)}(\st{})$
$$ Q({\bf Canon}_{({\cal A}_{st},{\cal A}_t)}(\st{})) =
Q(f(\st{})) = f(Q(\st{})),$$ so $Q(\st{})$ is computed as $$\bigcup_{f \in Type_{({\cal
A}_{st},{\cal A}_t)}(\st{})}f^{-1}(Q({\bf Canon}_{({\cal A}_{st},{\cal A}_t)}(\st{}))) = \bigcup_{f
\in Type_{({\cal A}_{st},{\cal A}_t)}(\st{})}f^{-1}(f(Q(\st{}))).$$
 For completeness, we give a program Decode that, when applied to the encoding $N_{\varphi}$ of
a formula $\varphi$, computes in a relation variable Dec the spatio-temporal relation defined by
$\varphi$. Thereto it suffices to modify the encode program as shown in
Figure~\ref{alg-decoding}.

\begin{figure}\rm
\label{alg-decoding}
\begin{algorithmic}
 \STATE $m := 0$
    \STATE $T :=  \emptyset$
    \STATE $F:= \emptyset$
    \STATE found :=False
    \WHILE{not found}
        \STATE $m := m + 1$
        \STATE build relations $T$ and $F$
        \STATE found:= $m = N_{\varphi}$
    \ENDWHILE
    \STATE Dec := all points which are the image under the
transformation stored in Type$_{({\cal A}_{st},{\cal A}_t)}$ of the points with coordinates
(represented as points on the line $u_0u_1$) $p_x,p_y,p_t$ such that $F(u_{t_O}, u_{t_E},
u_0,u_1,u_2 , m,p_1, p_2, p_t)$.
\end{algorithmic}
\caption{The decoding program. The input is a natural number encoding a relation.}
\end{figure}

The formula constructing the output, using the above, only differs slightly from the formulas we
gave when explaining the encoding algorithm. In the encoding phase, it had to be checked, for some
natural number $m$, whether there existed a transformation mapping all points of $R$ to the points
satisfying the formula encoded by $m$. Also, that transformation was computed. Here, we have the
transformation stored in Type$_{({\cal A}_{st},{\cal A}_t)}$, and we know we have the right natural
number $m$, so all points mapped by the transformation in Type$_{({\cal A}_{st},{\cal A}_t)}$ to
points satisfying the formula encoded by $m$, are returned.

\bigskip
\par\noindent To conclude we summarize the conversion procedure. Given a $k$-ary computable query Q over a schema $\sigma = \{R_1, \ldots, R_m \}$,
 there exists a counter program $M$ such that for each database $\st{}$ over $\sigma$, if
$(n_{R_1},\ldots,n_{R_m})$ are the results of applying the program Encode to $\st{}$ then
$M(n_{R_1},\ldots,n_{R_m})$ is the encoding of the quantifier-free formula defining $Q(\st{})$,
using the variables $x_1^1,\ab \ldots,\ab x_1^{n+1},\ab \ldots,\ab x_K^1,\ab \ldots,\ab x_K^{n+1}$.
If $Q(\st{})$ is not defined, then M does not halt on this input. As already noted above, we can
simulate  $M$ by a program $P$ in $\fo({\bf \Pi}) +\ab {\rm While}$. Hence, the query Q is
expressed by the program

 \begin{tabular}{l}Encode; \\ $P$;\\ Decode; \end{tabular}\qed

\bigskip The reason that the problem of identifying sound and
complete languages for the groups ${\cal F}_{st}^f$ is still open,
is that for those groups, there is no first-order logic formula
expressing their graph. Indeed, it is not possible to express that
there should exist a finite number of time moments for which there
is a different affinity, when describing the groups ${\cal
F}_{st}^f$. Hence, we cannot use the above proof technique.

\bigskip
The previous theorem has a number of corollaries.

 \begin{theorem}\label{cor-complete} \rm  Let $\sigma$ be a database
schema. Let $({\cal F }_{st}, \allowbreak {\cal F}_t)$ be one of the groups  $({\cal A}_{st}^{ },
\ab {\cal A}_t)$, $({\cal A}_{st}^{ }, \ab {\cal I}_t)$, $({\cal A}_{st}^{ },  \ab {\it Id}_t)$,
$({\cal S}_{st}^{ },  \ab {\cal F}_t)$, $({\cal I}_{st}^{ },  \ab {\cal F}_t)$, or $({\cal
T}_{st}^{ }, \ab {\cal F}_t)$ with ${\cal F}_t\in\{{\cal A}_t,\ab {\cal T}_t,\ab {\it Id}_t\}$  and
let $\Pi({\cal F }_{st}, \allowbreak {\cal F}_t)$ be as in
Table~\ref{table-point-predicates-finite}. The point language $\fo(\Pi({\cal F}_{st}, {\cal
F}_t),\sigma )+{\rm While}$ is sound and complete for the $({\cal F }_{st}, \allowbreak {\cal
F}_t)$-generic computable spatio-tempo\-ral queries over $\overline{\sigma}$.\qed
\end{theorem}

\par\noindent{\bf Proof.} The proof of this corollary is similar to the proof of Theorem~\ref{main-compl}. The encoding and decoding
programs for the various transformation groups only differ where
the transformation in Type$_{({\cal A}_{st},{\cal A}_t)}$ is
described, and where a coordinate system needs to be defined. The
rest of the proof is the same, regardless of the transformation
groups considered. The descriptions of the coordinate systems for
the various transformation groups can be found in the proof of
Corollary~\ref{basissen}\qed

\section{Conclusion and discussion}\label{sectie6}

We have investigated different genericity classes relative to the constraint database model for
spatio-temporal databases and we have identified sound and complete languages for the  \foprs,
respectively the computable, queries in (most of) these genericity classes. Some results were
obtained by techniques introduced by Gyssens, Van den Bussche and Van Gucht~\cite{gvv-jcss}, but
for time-dependent transformations we have introduced new proof techniques.

For what concerns computationally complete languages these
techniques seem to be insufficient to deal with the genericity
 notions that are expressed by the groups $({\cal A}_{st}^{f },
{\cal A}_t)$, $({\cal A}_{st}^{f }, {\cal I}_t)$, $({\cal A}_{st}^{ f}, {\it Id}_t)$, $({\cal
S}_{st}^{f }, {\cal F}_t)$, $({\cal I}_{st}^{ f}, {\cal F}_t)$, and $({\cal T}_{st}^{f }, {\cal
F}_t)$ with ${\cal F}_t\in\{{\cal A}_t,{\cal T}_t,{\it Id}_t\}$.  The problem in adapting the proof
technique of Theorem~\ref{main-compl} to these groups is that it is not clear how we can express in
the respective point-based logics that two spatio-temporal databases can be mapped to each other by
some piece-wise constant affinity. Indeed, since the number of pieces is not defined \emph{a
priori}, this might not be expressible. This would imply that yet another new proof technique would
be required to deal with the remaining cases.

\bibliographystyle{acmtrans}

\end{document}